\documentclass[12pt, twocolumn, trackchanges, twocolappendix, ]{aastex631}

\usepackage{color}
\usepackage{amsmath}
\usepackage{float}
\usepackage{gensymb}
\usepackage{upgreek}
\usepackage{longtable}
\usepackage{threeparttablex}
\usepackage{booktabs}
\usepackage{placeins}
\usepackage{url}
\usepackage{enumitem} 

\submitjournal{ApJ}

\begin{document}

\title{Polarization properties of 28 repeating fast radio burst sources with CHIME/FRB}

\correspondingauthor{Cherry Ng}
\email{cherry.ng-guiheneuf@cnrs-orleans.fr}

\author[0000-0002-3616-5160]{Cherry Ng}
\affiliation{Laboratoire de Physique et Chimie de l'Environnement et de l'Espace - Université d'Orléans/CNRS, 45071, Orléans Cedex 02, France}
\author[0000-0002-8897-1973]{Ayush Pandhi}
\affiliation{David A. Dunlap Department of Astronomy and Astrophysics, University of Toronto, 50 St. George Street, Toronto, ON M5S 3H4, Canada}
\affiliation{Dunlap Institute for Astronomy and Astrophysics, University of Toronto, 50 St. George Street, Toronto, ON M5S 3H4, Canada}
\author[0000-0001-7348-6900]{Ryan Mckinven}
\affiliation{Department of Physics, McGill University, 3600 rue University, Montréal, QC H3A 2T8, Canada}
\affiliation{Trottier Space Institute, McGill University, 3550 rue University, Montréal, QC H3A 2A7, Canada}
\author[0000-0002-8376-1563]{Alice P. Curtin}
\affiliation{Department of Physics, McGill University, 3600 rue University, Montréal, QC H3A 2T8, Canada}
\affiliation{Trottier Space Institute, McGill University, 3550 rue University, Montréal, QC H3A 2A7, Canada}
\author[0000-0002-6823-2073]{Kaitlyn Shin}
\affiliation{MIT Kavli Institute for Astrophysics and Space Research, Massachusetts Institute of Technology, 77 Massachusetts Ave, Cambridge, MA 02139, USA}
\affiliation{Department of Physics, Massachusetts Institute of Technology, 77 Massachusetts Ave, Cambridge, MA 02139, USA}
\author[0000-0001-8384-5049]{Emmanuel Fonseca}
\affiliation{Department of Physics and Astronomy, West Virginia University, P.O. Box 6315, Morgantown, WV 26506, USA}
\affiliation{Center for Gravitational Waves and Cosmology, West Virginia University, Chestnut Ridge Research Building, Morgantown, WV 26505, USA}
\author[0000-0002-3382-9558]{Bryan~M.~Gaensler}
\affiliation{Department of Astronomy and Astrophysics, University of California Santa Cruz, 1156 High Street, Santa Cruz, CA 95064, USA}
\affiliation{Dunlap Institute for Astronomy and Astrophysics, University of Toronto, 50 St. George Street, Toronto, ON M5S 3H4, Canada}
\affiliation{David A. Dunlap Department of Astronomy and Astrophysics, University of Toronto, 50 St. George Street, Toronto, ON M5S 3H4, Canada}
\author[0000-0003-3236-8769]{Dylan L.	Jow}
\affiliation{Kavli Institute for Particle Astrophysics \& Cosmology, P.O. Box 2450, Stanford University, Stanford, CA 94305, USA}
\author[0000-0001-9345-0307]{Victoria Kaspi}
\affiliation{Department of Physics, McGill University, 3600 rue University, Montréal, QC H3A 2T8, Canada}
\affiliation{Trottier Space Institute, McGill University, 3550 rue University, Montréal, QC H3A 2A7, Canada}
\author[0000-0001-7931-0607]{Dongzi	Li}
\affiliation{Department of Astrophysical Sciences, Princeton University, Princeton, NJ 08544, USA}
\author[0000-0002-7164-9507]{Robert	Main}
\affiliation{Department of Physics, McGill University, 3600 rue University, Montréal, QC H3A 2T8, Canada}
\affiliation{Trottier Space Institute, McGill University, 3550 rue University, Montréal, QC H3A 2A7, Canada}
\author[0000-0002-4279-6946]{Kiyoshi W.~Masui}
\affiliation{MIT Kavli Institute for Astrophysics and Space Research, Massachusetts Institute of Technology, 77 Massachusetts Ave, Cambridge, MA 02139, USA}
\affiliation{Department of Physics, Massachusetts Institute of Technology, 77 Massachusetts Ave, Cambridge, MA 02139, USA}\author[0000-0002-2551-7554]{Daniele Michilli}
\affiliation{MIT Kavli Institute for Astrophysics and Space Research, Massachusetts Institute of Technology, 77 Massachusetts Ave, Cambridge, MA 02139, USA}
\affiliation{Department of Physics, Massachusetts Institute of Technology, 77 Massachusetts Ave, Cambridge, MA 02139, USA}
\author[0000-0003-0510-0740]{Kenzie	Nimmo}
\affiliation{MIT Kavli Institute for Astrophysics and Space Research, Massachusetts Institute of Technology, 77 Massachusetts Ave, Cambridge, MA 02139, USA}
\author[0000-0002-4795-697X]{Ziggy	Pleunis}
\affiliation{Anton Pannekoek Institute for Astronomy, University of Amsterdam, Science Park 904, 1098 XH Amsterdam, The Netherlands}
\affiliation{ASTRON, Netherlands Institute for Radio Astronomy, Oude Hoogeveensedijk 4, 7991 PD Dwingeloo, The Netherlands}
\author[0000-0002-7374-7119]{Paul	Scholz}
\affiliation{Department of Physics and Astronomy, York University, 4700 Keele Street, Toronto, ON MJ3 1P3, Canada}
\affiliation{Dunlap Institute for Astronomy and Astrophysics, University of Toronto, 50 St. George Street, Toronto, ON M5S 3H4, Canada}
\author[0000-0001-9784-8670]{Ingrid	Stairs}
\affiliation{Department of Physics and Astronomy, University of British Columbia, 6224 Agricultural Road, Vancouver, BC V6T 1Z1 Canada}
\author[0000-0002-3615-3514]{Mohit	Bhardwaj}
\affiliation{McWilliams Center for Cosmology \& Astrophysics, Department of Physics, Carnegie Mellon University, Pittsburgh, PA 15213, USA}
\author[0000-0002-1800-8233]{Charanjot	Brar}
\affiliation{National Research Council of Canada, Herzberg Astronomy and Astrophysics Research Centre, 5071 W. Saanich Rd. Victoria, BC, V9E 2E7, Canada}
\author[0000-0003-2047-5276]{Tomas	Cassanelli}
\affiliation{Department of Electrical Engineering, Universidad de Chile, Av. Tupper 2007, Santiago 8370451, Chile}
\author[0000-0003-3457-4670]{Ronniy~C.~Joseph}
\affiliation{Department of Physics, McGill University, 3600 rue University, Montréal, QC H3A 2T8, Canada}
\affiliation{Trottier Space Institute, McGill University, 3550 rue University, Montréal, QC H3A 2A7, Canada}
\author[0000-0002-8912-0732]{Aaron~B.~Pearlman}
\affiliation{Department of Physics, McGill University, 3600 rue University, Montréal, QC H3A 2T8, Canada}
\affiliation{Trottier Space Institute, McGill University, 3550 rue University, Montréal, QC H3A 2A7, Canada}
\author[0000-0001-7694-6650]{Masoud	Rafiei-Ravandi}
\affiliation{Department of Physics, McGill University, 3600 rue University, Montr\'eal, QC H3A 2T8, Canada}
\author[0000-0002-2088-3125]{Kendrick	Smith}
\affiliation{Perimeter Institute for Theoretical Physics, 31 Caroline Street N, Waterloo, ON N25 2YL, Canada}



\begin{abstract}
As part of the Canadian Hydrogen Intensity Mapping Experiment Fast Radio Burst (CHIME/FRB) project, we report 41 new Rotation Measures (RMs) from 20 repeating Fast Radio Bursts (FRBs) obtained between 2019 and 2023 for which no previous RM was determined.
We also report 22 additional RM measurements for eight further repeating FRBs.
We observe temporal RM variations in practically all repeating FRBs.
Repeaters appear to be separated into two 
categories:  those with dynamic and those with stable RM environments, 
differentiated by the ratios of RM standard deviations over the averaged RM magnitudes. 
Sources from stable RM environments likely have little RM contributions from the interstellar medium (ISM) of their host galaxies, whereas sources from dynamic RM environments share some similarities with Galactic pulsars in eclipsing binaries but appear distinct from Galactic centre solitary pulsars.
We observe a new stochastic, secular, and again stochastic trend in the temporal RM variation of FRB~20180916B, which does not support binary orbit modulation being the reason for this RM changes. 
We highlight two more repeaters that show RM sign change, namely FRBs~20290929C and 20190303A. 
We perform an updated comparison of polarization properties between repeating and non-repeating FRBs, which show a 
marginal dichotomy in their distribution of electron-density-weighted parallel-component line-of-sight magnetic fields.

\end{abstract}




\section{Introduction} \label{sec:intro}

Fast Radio Bursts (FRBs) are bright, millisecond bursts of radio waves from extragalactic distances. First discovered by \cite{Lorimer2007}, the provenance of these energetic bursts remains a mystery. 
Polarization analysis of FRBs is a way to obtain observational constraints that can provide important clues on their  emission mechanisms
and their host environments, hence improving our understanding of the progenitor theories.

In the literature, polarimetric information is still relatively rare for FRBs. At the time of writing, polarimetric information is only available for over 100 sources out of the 700+ currently published FRB sources. This is because most of the FRB search programs thus far are conducted using intensity data so polarization analysis is limited to search projects equipped with the capability to trigger raw voltage callback or to repeating FRBs where subsequent follow-up observations collecting Stokes information are possible.  

The detection of linear polarization implies the possibility of constraining the Faraday rotation measure (RM), which represents the amount of rotation 
of the polarization position angle of linearly polarized light as it 
passes through a magneto-ionic medium due to the influence of the magnetic field and the column density of the free electrons along the line-of-sight between the FRB and the observer.
Thus far, FRBs have mostly been observed with high linear polarization fraction, although values from 0 to 100\% have been reported \citep[see, e.g.][]{Sherman2024,Pandhi2024}. FRBs have also been seen with a wide range of RM magnitudes, from almost zero to tens of thousands of rad\,m$^{-2}$ \citep{Michilli2018}. 
The observed RM values can shed light on the local environment of the FRB, given the intergalactic medium RM contribution is expected to be small \citep[see, e.g.][]{Amaral2021}. 
For example, the high RM magnitude of the order of 10$^{5}$\,rad\,m$^{-2}$ and large temporal RM variations observed for FRB~20121102 may be associated with a turbulent supernova remnant \citep{Marcote2017}. 
On the other end of the scale, the almost-zero RM observed in FRB\,20200120E has been attributed to the clean environment of the host globular cluster \citep{Bhardwaj2021,Nimmo2022,Kirsten2021}.
Small observed RM values can also be used as an upper bound to constrain the cosmic contribution to the line-of-sight RM and help to probe the magnetization of the cosmic web \citep{Hackstein2019}.

The polarization angle (PA; $\psi$) is the emission angle (geometry) 
of the linearly polarized component of the light from the 
FRB source and is typically presented as a function of time during the burst.
Diverse PA variations during bursts (for example, sweeping up, down, up and down) have been observed 
in some FRBs \citep[e.g. FRB~20180301;][]{Luo2020}, with at least one recently reported one-off FRB displaying PA evolution that closely matches the rotating vector model often applied to pulsar measurements \citep{Mckinven2024}.
It has been suggested that such diverse variations are consistent with a magnetospheric origin for the FRB radio emission associated with different open magnetic field line regions as the central compact object, possibly a magnetar, rotates,
and disfavour synchrotron maser models that predict constant PAs attributed to ordered magnetic fields. 
However, the fact that both varying and constant PAs have been observed in repeaters and non-repeaters \citep{Luo2020,Michilli2018,Masui2015} in different environments \citep{Feng2022} also suggests that the PA is intrinsic to the radiation physics of the FRB and might not be a meaningful classifier to distinguish between different repetition behaviors and host environments.

Currently, $\sim$3\% of all published FRBs have been seen to repeat\footnote{https://blinkverse.alkaidos.cn} \citep{RN3}. Repeating FRBs allow for the study of temporal variations in the polarization properties. RM variations on timescales of days to years have been detected, potentially pointing to dynamic magneto-ionic environments \citep{Hilmarsson2021b,Mckinven2023a}. 
A temporal RM sign change in FRB~20190520B indicated a reversal in the direction of the magnetic field lines, where the orbit of a binary star has been proposed as an explanation for the observed polarization properties \citep{Dai2022,AnnaThomas2023}.

The Canadian Hydrogen Intensity Mapping Experiment (CHIME) is a radio telescope located at the Dominion Radio Astrophysical Observatory (DRAO). As a transit telescope, CHIME has no moving parts and scans each sky position within its $\sim$200\,deg$^{2}$ instantaneous field-of-view with a daily cadence.
This makes the FRB search backend of CHIME \citep{chime/frb2018} particularly well suited for discovering and subsequently monitoring repeating FRB behavior. At the time of writing, 60 repeating FRBs are reported on the CHIME repeating FRB database\footnote{https://www.chime-frb.ca/repeaters} \citep{RN1,Fonseca2020,RN3}. 
Furthermore, CHIME is equipped with a complex voltage recording system (herein referred to as the baseband buffer system) \citep{Michilli2021}, which holds data from the last $\sim$20\,s.
FRBs above the detection threshold trigger a call-back data dump from the baseband ring buffer, enabling a polarimetric study of FRBs with high time and frequency resolution.

This paper presents the first polarization information for 20 repeating FRBs whose discoveries were initially reported by \citet{RN3} (hereafter CHIME23), which increases the number of repeaters with polarization properties from 18 to 38. 
We also present updated polarization information for eight repeating FRBs with previously published polarization data. 
In Section~\ref{sec:analysis}, we describe the data set and the analysis techniques. In Section~\ref{sec:results}, we present the polarization results and discuss the astrophysics implications in Section~\ref{sec:discussion}, before concluding in Section~\ref{sec:conclusion}.

\section{Observations, data reduction \& analysis} \label{sec:analysis}

This work presents the first polarization data for 20 repeating FRBs whose discoveries were initially reported in CHIME23. 
During the period April 2019 to July 2023, triggered baseband data \citep{Michilli2021} were captured for about half of the bursts CHIME/FRB detected from these repeaters. These 53 events are listed in Table~\ref{tab:RN3pol}.
These bursts met the baseband triggering signal-to-noise threshold that was set at 10.
Five out of the 25 sources from the CHIME23 sample (FRBs~20190226B, 20200420A, 20181226F, 20191114A and 20200913C) have no baseband data in the time period. These sources are not discussed further in this paper. 
In addition, eight repeaters whose polarization properties have previously been reported by  \citet{Kumar2019,Bhardwaj2021,Hilmarsson2021b,Kumar2022,Mckinven2023a,Mckinven2023b} continued to be detected by CHIME. During the period January 2019 to May 2024, 22 events with new polarization results from these sources were detected and are presented in Table~\ref{tab:RN1-2pol}. Note that all but one FRB in this group were initially discovered by CHIME/FRB \citep{RN1,Fonseca2020,Bhardwaj2021}. The exception is FRB~20171019A, which was discovered using the ASKAP radio telescope \citep{shannon2018}.

\begin{table*}
\begin{center}
\caption{Galactic DM and RM estimates and the coordinates for the 28 repeating FRBs presented in this paper. The sky coordinates are taken from \citet{RN1,Fonseca2020,RN3}, whereas the DMs are estimated using the \textsc{PyGEDM} package. } 
\begin{tabular}{l | c c |c c| c }
\toprule
Source & $\alpha$ & $\delta$ & \multicolumn{2}{c}{$\rm{DM_{MW}}\; \rm{[pc \, cm^{-3}]}$} & $\rm{RM_{MW}}\; \rm{[rad \, m^{-2}]}$ \\
TNS Name & (J2000) & (J2000)  & NE2001 & YMW16 &   \\
\hline
FRB 20171019A & 334.3(18) & $-$8.66(11) & $\sim$37 & $\sim$26 & $-$24(7) \\
FRB 20180910A & 354.8(9) & 89.01(1)    & $\sim$57 & $\sim$55 & $-$15(9)\\
FRB 20180916B & 29.170(7) & 65.740(18) & $\sim$199& $\sim$325& $-$94(45) \\
FRB 20181119A & 190.50(12) & 65.13(15) & $\sim$34 & $\sim$26 & $+$22(5)\\
FRB 20190110C & 249.33(1) & 41.445(9)  & $\sim$37 & $\sim$30 & 14(1)\\
FRB 20190117A &  331.75(3) & 17.383(4) & $\sim$48 & $\sim$40 & $-$26(8) \\
FRB 20190208A &  283.75(5) & 46.966(4) & $\sim$65 & $\sim$71 & $-$8(13) \\
FRB 20190303A &  208.25(5) & 48.250(4) & $\sim$30 & $\sim$22 & $+$21(5)\\
FRB 20190430C & 277.210(9) & 24.770(9) & $\sim$99 & $\sim$84 & 92(26)\\
FRB 20190604A &  218.75(4) & 53.283(3) & $\sim$32 & $\sim$24 & $+$12(2) \\
FRB 20190609C & 73.32(1) & 24.068(6) & $\sim$113 & $\sim$153 & $-$32(13)\\
FRB 20190804E & 261.34(2) & 55.069(8) & $\sim$43 & $\sim$37 & 11(6)\\
FRB 20190915D & 11.78(3) & 46.86(2) & $\sim$89 & $\sim$88 & 10(6)\\
FRB 20191013D & 40.42(2) & 13.63(3)    & $\sim$43 & $\sim$36 & $-$7(4) \\
FRB 20191106C & 199.58(1) & 43.002(9) & $\sim$25 & $\sim$21 & $-$1(3)\\
FRB 20200118D & 106.91(1) & 42.837(9)  & $\sim$77 & $\sim$91 & 2(6)\\
FRB 20200120E &  149.486(9) & 68.82(2) & $\sim$40 & $\sim$35 & $-$11(8) \\
FRB 20200127B & 119.2(1) & 86.609(8)  & $\sim$54 & $\sim$51 & $-$23(10)\\
FRB 20200202A & 25.93(2) & 44.290(7) & $\sim$83 & $\sim$84 & $-$61(14)\\
FRB 20200223B & 8.265(8) & 28.831(7) & $\sim$46 & $\sim$37 & $-$54(6)\\
FRB 20200619A & 272.6(1)  & 55.56(6) & $\sim$51 & $\sim$45 & 32(12)\\
FRB 20200809E & 20.0(1)  & 82.89(2) & $\sim$72 & $\sim$81 & $-$5(10)\\
FRB 20200926A & 283.2(2) & 53.9(3)         & $\sim$64 & $\sim$59 & 8(9) \\
FRB 20200929C & 17.04(2) & 18.47(1) & $\sim$38 & $\sim$29 & $-$18(3)\\
FRB 20201114A & 221.59(8) & 71.79(3)   & $\sim$38 & $\sim$31 & 8(12)\\
FRB 20201124A & 77.0152(4) & 26.0610(2) & $\sim$123 & $\sim$197 & $-$57(33)\\
FRB 20201130A & 64.39(1) & 7.94(1) & $\sim$56 & $\sim$69 & 32(21)\\
FRB 20201221B & 124.20(3) & 48.78(2) & $\sim$51 & $\sim$46 & $-$1(3)\\
 \hline
\end{tabular}
\label{tab:dmrmgal}
\end{center}
\end{table*}

The key features of the CHIME baseband polarization processing are summarized below. 
First, the baseband data are localized and beamformed at the best sky position of each repeater using the CHIME/FRB localization pipeline presented by \citet{Michilli2021}. These data are coherently de-dispersed at the Signal-to-Noise (S/N)-optimizing DM determined by the search pipeline. The output of the baseband pipeline is data consisting of dual polarization information along N--S and E--W directions, with 1024 frequency channels at a frequency resolution of 390\,kHz with a time resolution of 2.56\,$\mu$s.
From these input data, 
we search for the structure-optimized DM (DM$_{\mathrm{struct}}$) using  the \textsc{dm\_phase}\footnote{\url{https://github.com/danielemichilli/DM_phase}} package \citep[see][for more details]{Curtin2024}. It is important to properly align the FRB signal before conducting the polarization analysis, since an incorrect DM value could impact the RM and $\psi$ measurements. 

The polarization pipeline of \citet{Mckinven2021} computes the four Stokes parameters ($I$ for total intensity, $Q$ and $U$ for linearly polarized components, and $V$ for the circularly polarized component) as well as the polarization angle.
Two RM detection methods are implemented in the polarization pipeline. The RM synthesis technique \citep{Burn1966,Brentjens2005} steps through trial Faraday depths ($\phi$) and calculates the corresponding average linearly polarized intensity across CHIME's 400--800\,MHz bandpass. 
The range of $|\phi|$ searched is inversely proportional to the bandwidth of the event and is set to be the point where intra-channel depolarization is at maximum 50\%.
Since it is assumed that the FRB emission region is small and has little Faraday complexity (the ``Faraday thin regime"), if the FRB signals are polarized, we expect to see a single peak at a specific Faraday depth in the resulting Faraday dispersion function (FDF). By fitting a parabolic curve to the FDF, we obtain the Rotation Measure (RM$_{\mathrm{FDF}}$) at the peak of the fit while the uncertainty in RM$_{\mathrm{FDF}}$ is taken to be the full-width-half-maximum divided by $2\times${S/N}.

The FDF method does not correct for the cable delay between the two linear polarizations, which means we can lose some polarized signal. 
We have attempted to extract the RM using the QU-fitting technique \citep{O'Sullivan2012}, which models the oscillations in Stokes $Q$ and $U$ introduced by Faraday rotation. 
This method allows for a simultaneous fitting of the instrumental leakage in Stokes $U$--$V$ and the polarization model ($Q$, $U$, RM, $\psi$).
However, we notice an inconsistency in the sign of the RM values between the FDF and QU-fitting methods. It appears that QU-fitting does not perform well for narrow or low S/N bursts and can show the wrong RM sign. 
For this reason, we choose to report only RM values derived using the FDF method in Tables~\ref{tab:RN3pol} and~\ref{tab:RN1-2pol}.

Intra-channel depolarization could lead to a non-detection if the intrinsic RM is high. To mitigate this issue, a semi-coherent RM search \citep{Mckinven2021} is performed for events where no significant RM peaks were found in the RM synthesis. Coherent de-rotation is conducted at a sparse grid of trial RMs in the range of $-10^{6} \le \mathrm{RM} \le 10^{6}$\,rad\,m$^{-2}$, followed by an incoherent RM search like the regular pipeline around neighbouring RM values. We note that in this work, the semi-coherent search did not detect any RM with large magnitudes.

To study the extragalactic DM and RM contributions for each FRB (see Section~\ref{sec:Ayush}), 
we compute and subtract the Milky Way RM contribution using the map of \cite{Hutschenreuter2022} and the DM contribution using the thermal electron density maps of \cite{Yao2017} and \cite{NE2001} by using a python wrapper code \textsc{PyGEDM} package\footnote{https://github.com/FRBs/pygedm}.
See Table~\ref{tab:dmrmgal} for the Galactic DM and RM estimates for the 28 repeaters presented in this paper. 
The Galactic DM estimates do not include the DM contribution of the Galactic halo, which is expected to contribute an average value of $\sim30-50\,\rm{rad\, m^{-2}}$, depending on the halo model assumed \citep[e.g.,][]{Dolag2015, Yamasaki2020,Cook2023}.
We have not taken into account the RM contribution from the Earth's ionosphere, since conveniently for CHIME, most FRBs were detected close to the zenith during meridian transients. This means that the ionospheric RM variations due to pointings at different line-of-sight elevations are expected to be minimal (no more than a few rad\,m$^{-2}$) and less than the uncertainty of the RM values. 

\section{Results} \label{sec:results}
\begin{figure*}[ht!]
\begin{center}
    \includegraphics[width=0.188\textwidth]{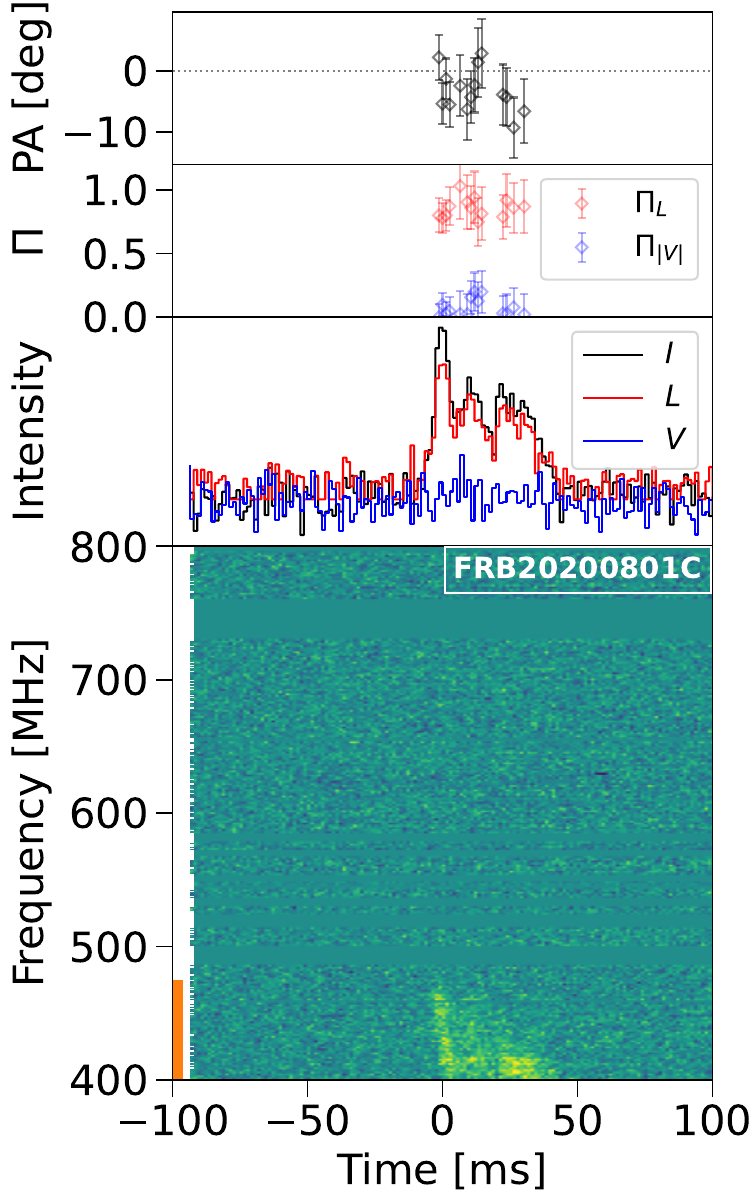}
    \includegraphics[width=0.188\textwidth]{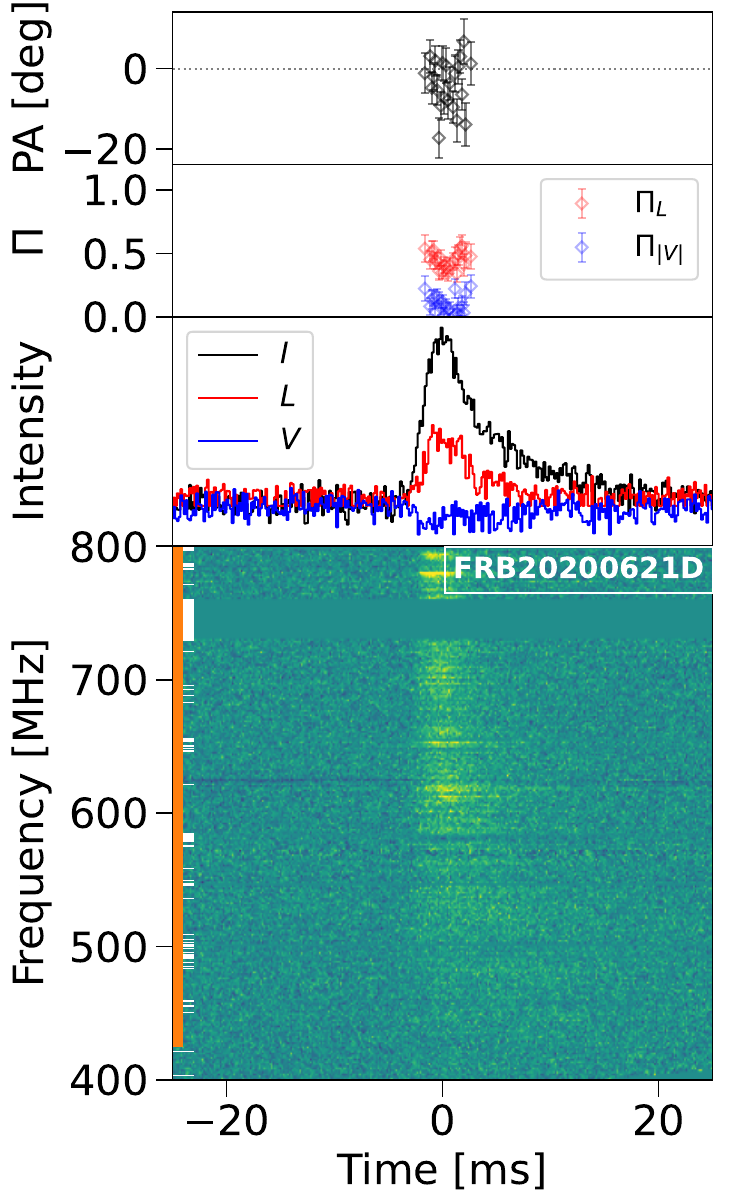}
    \includegraphics[width=0.188\textwidth]{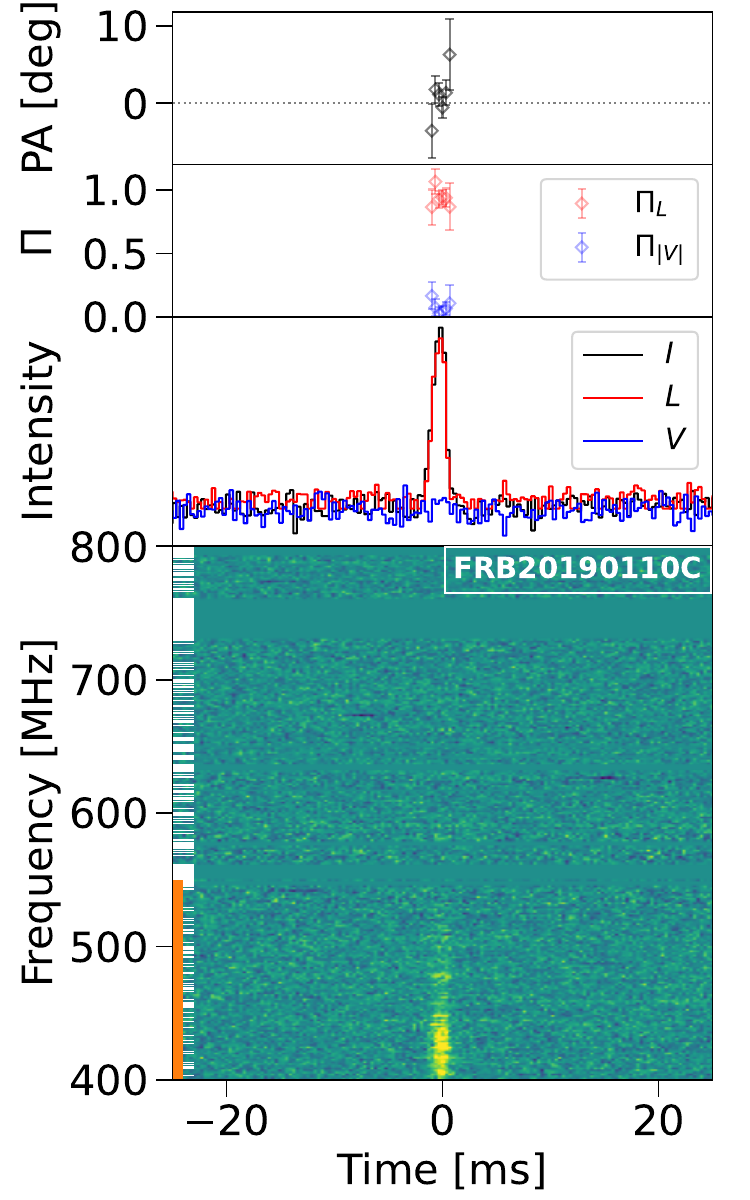}
    \includegraphics[width=0.188\textwidth]{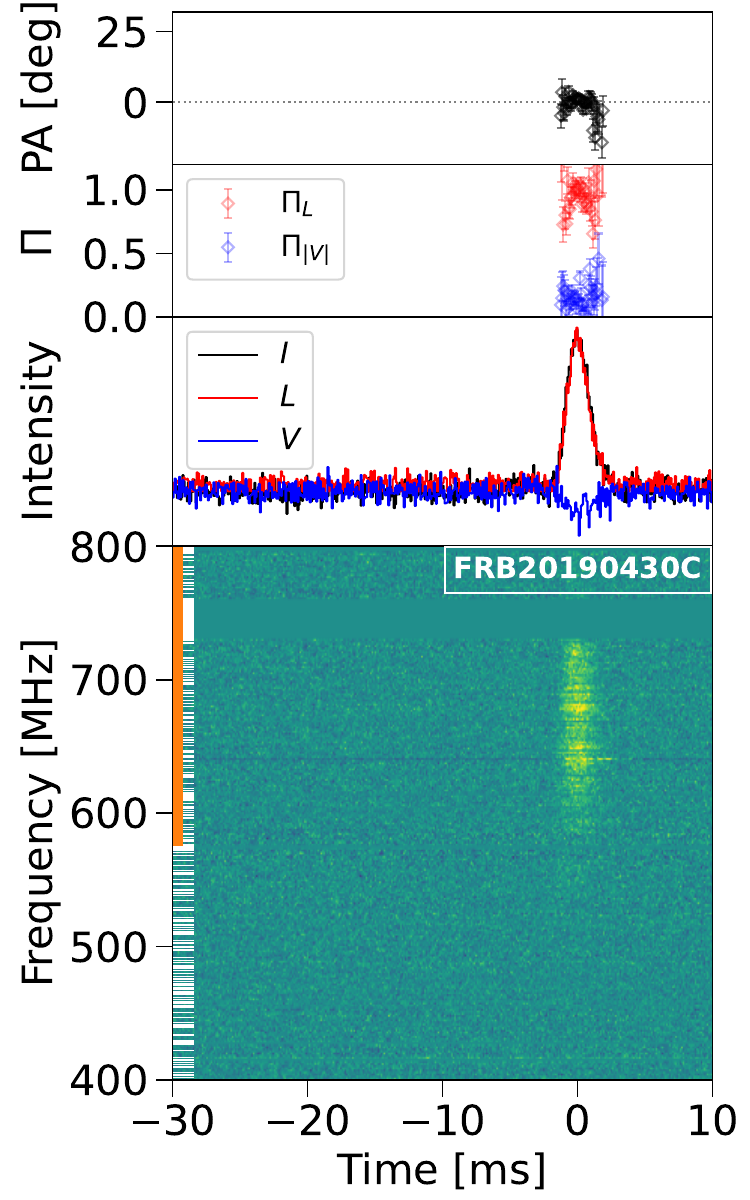}
    \includegraphics[width=0.188\textwidth]{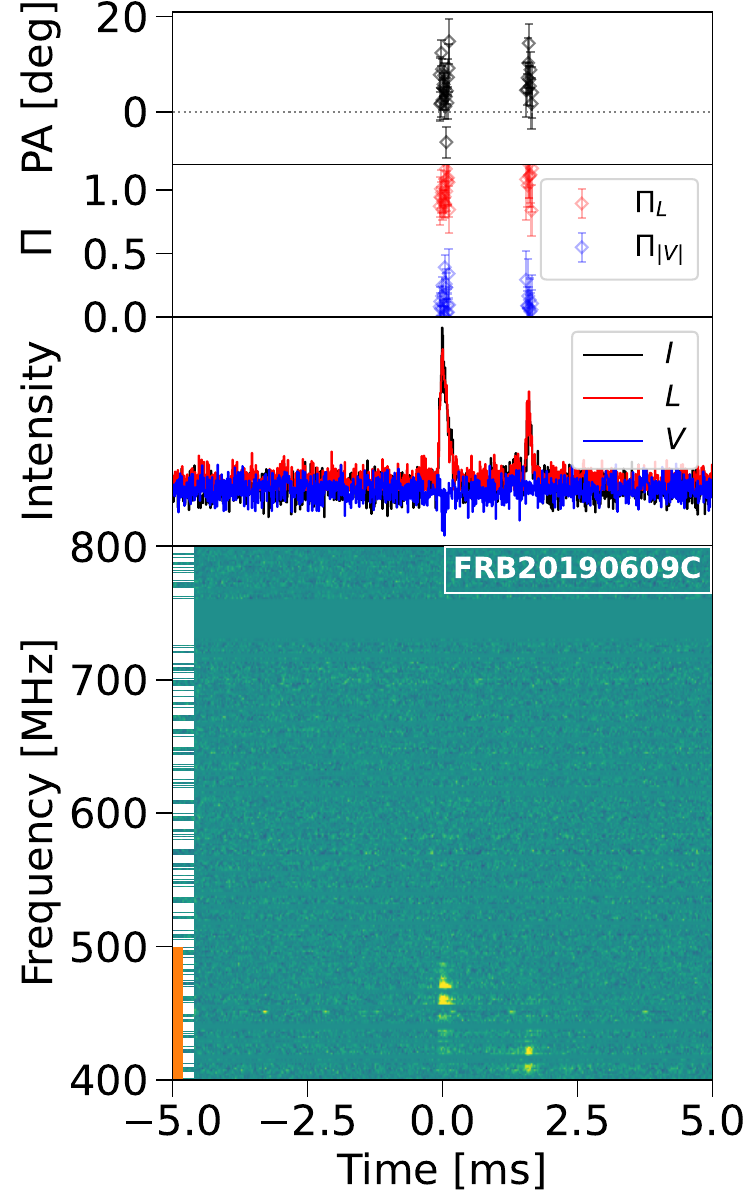}

    \includegraphics[width=0.188\textwidth]{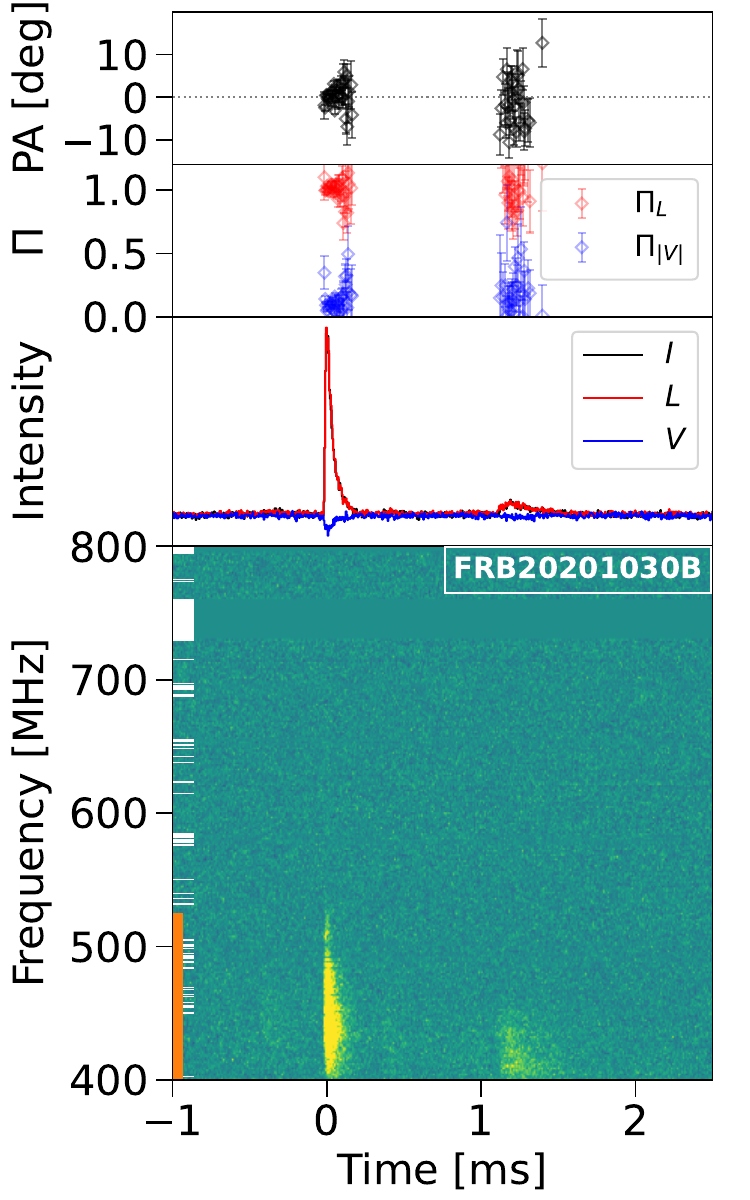}
    \includegraphics[width=0.188\textwidth]{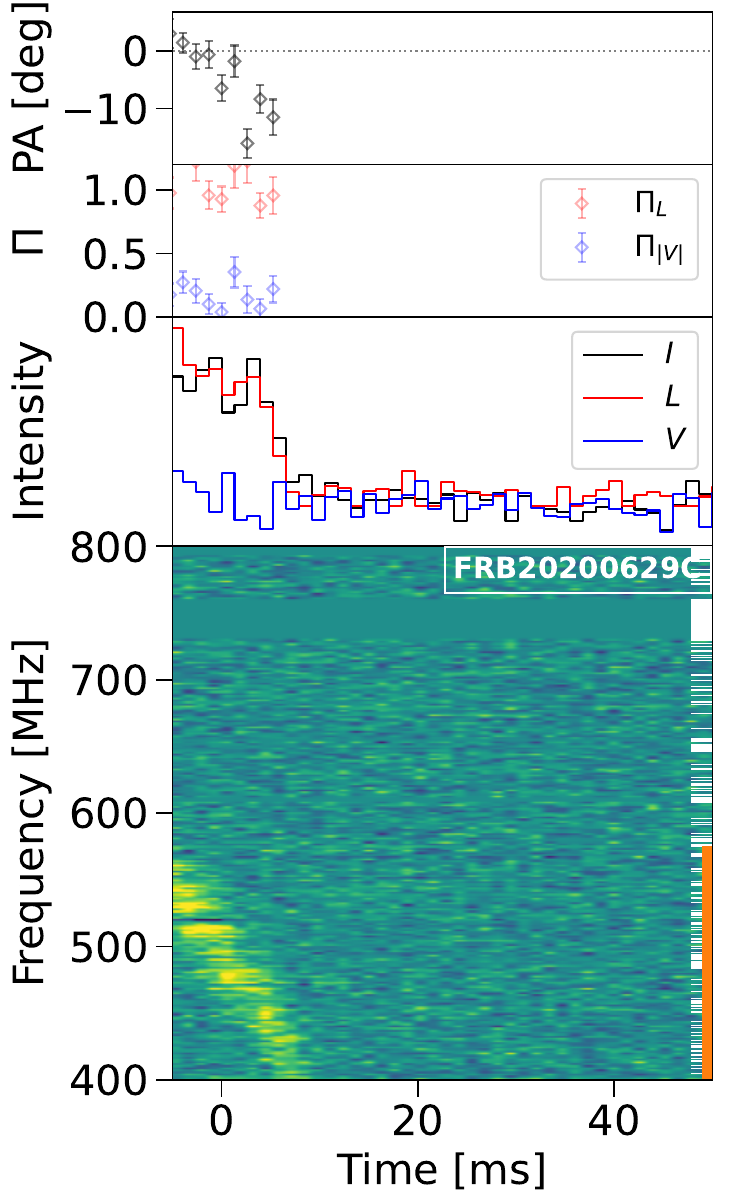}
    \includegraphics[width=0.188\textwidth]{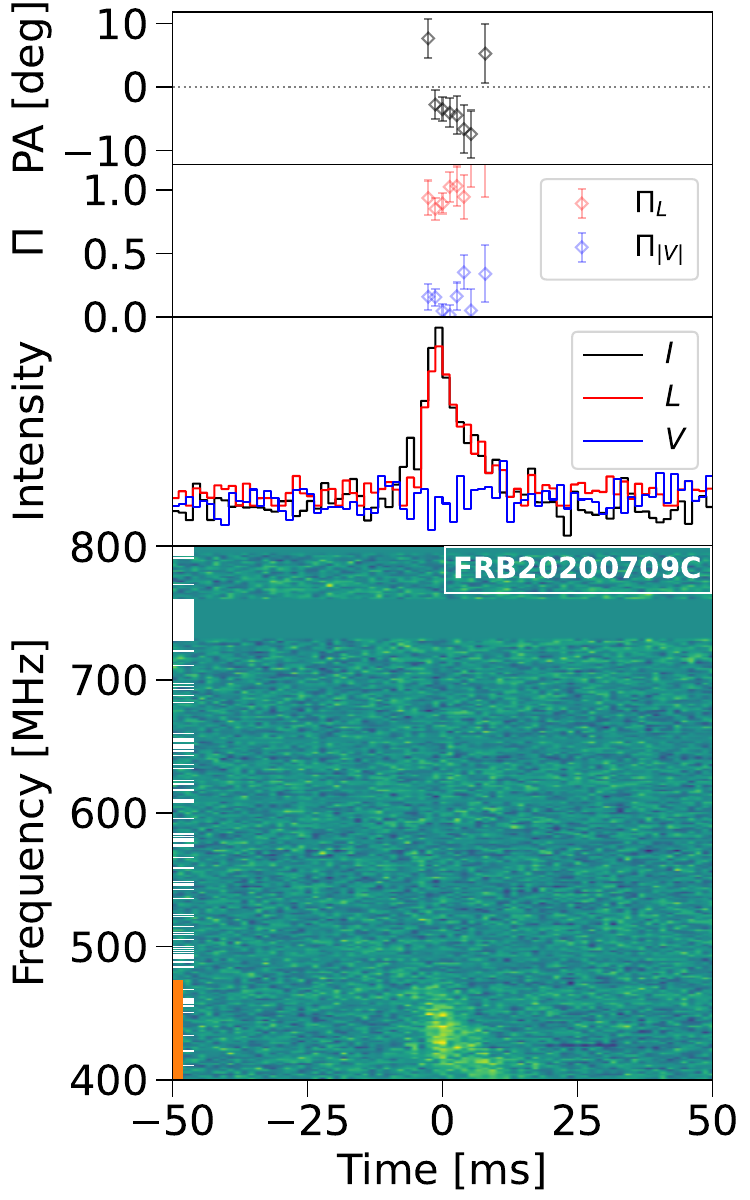}
    \includegraphics[width=0.188\textwidth]{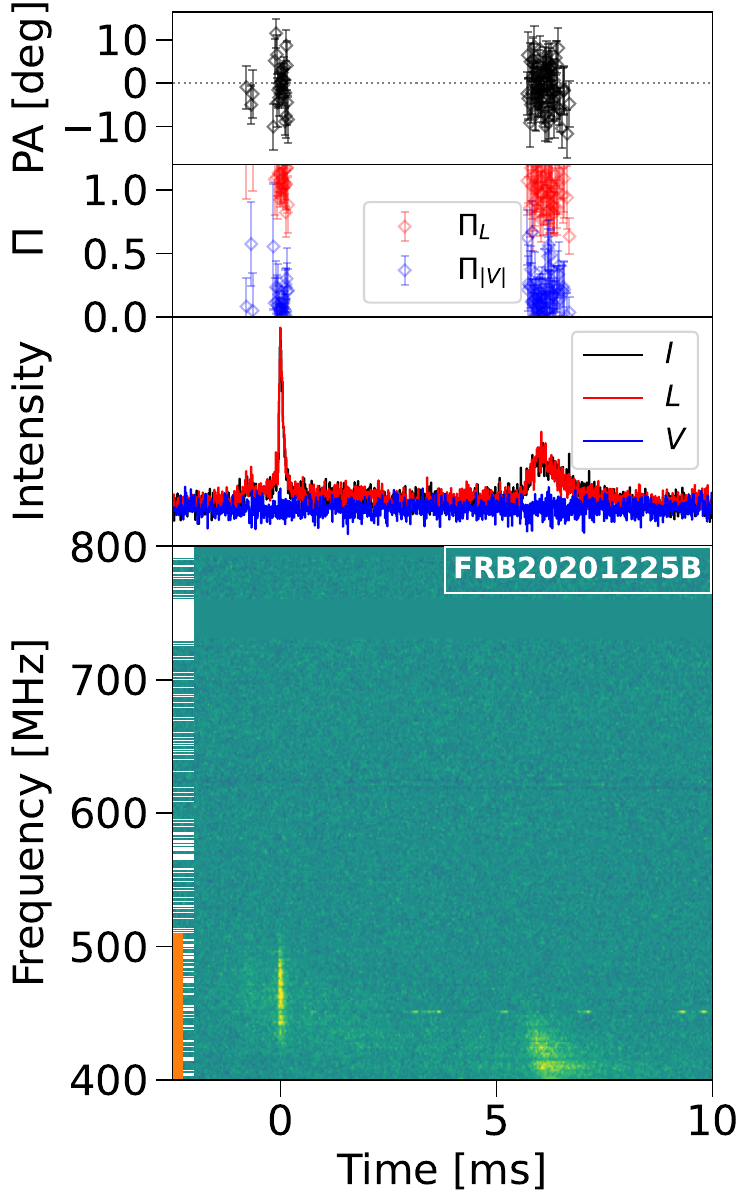}
    \includegraphics[width=0.188\textwidth]{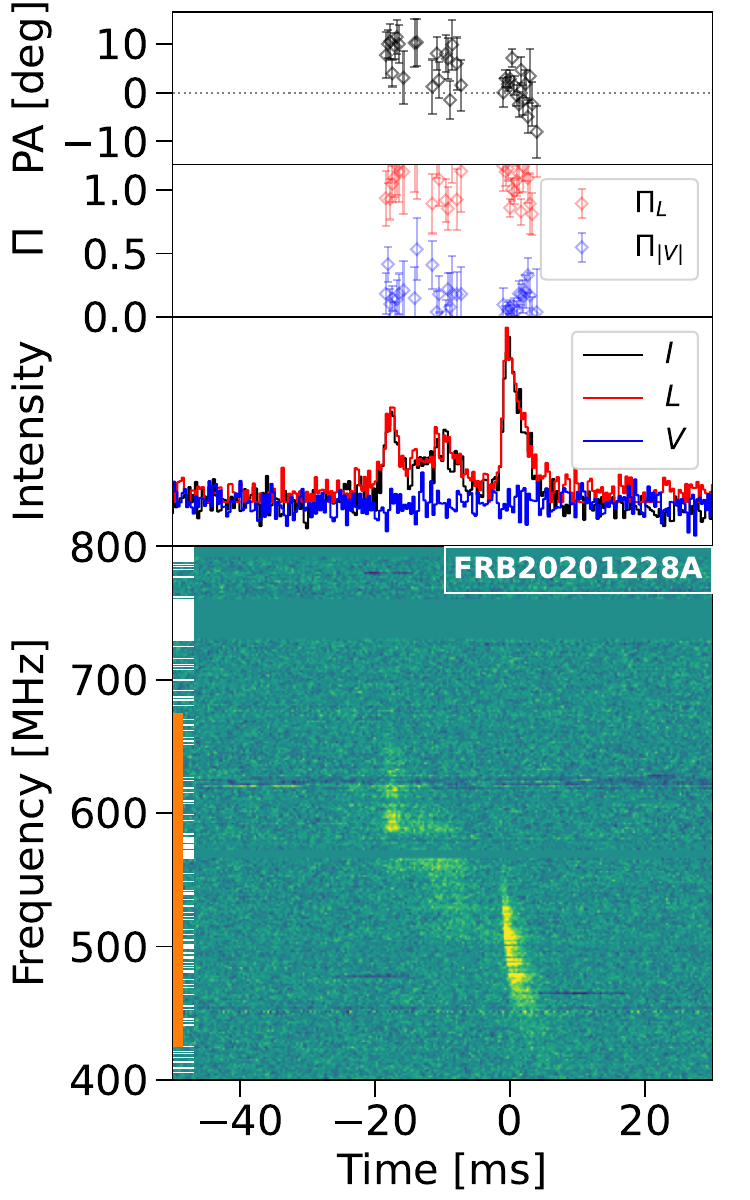}

    \includegraphics[width=0.188\textwidth]{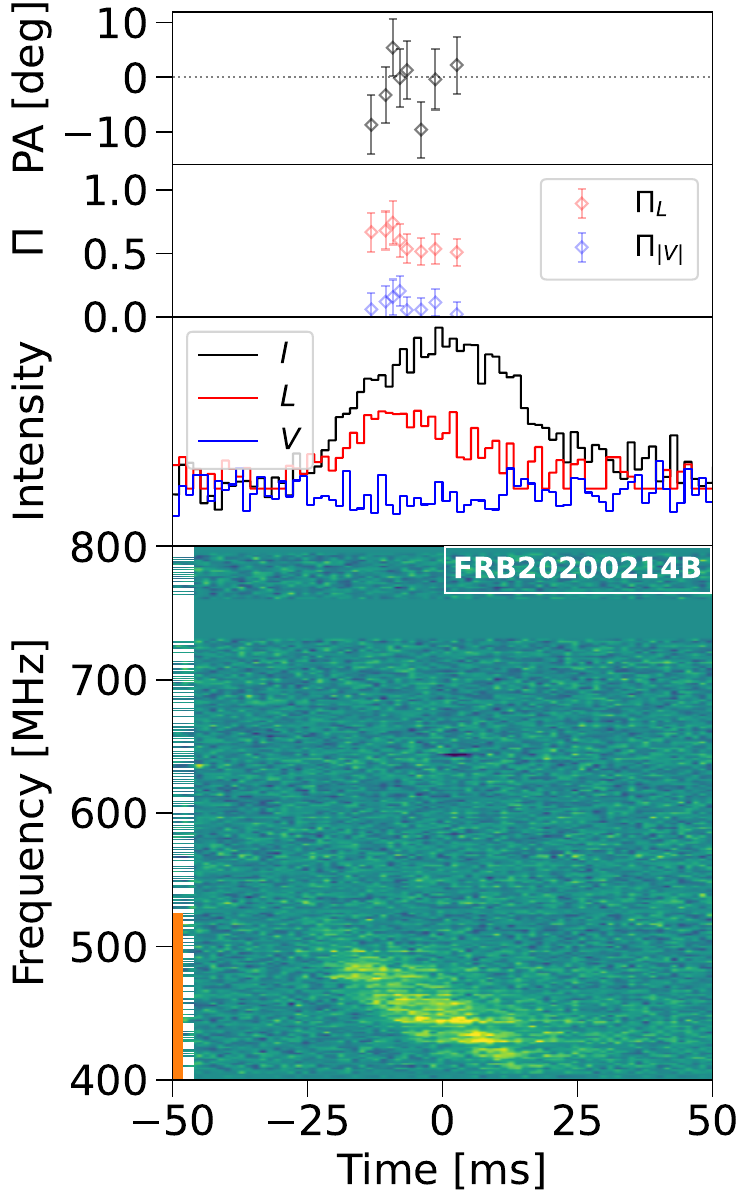}
    \includegraphics[width=0.188\textwidth]{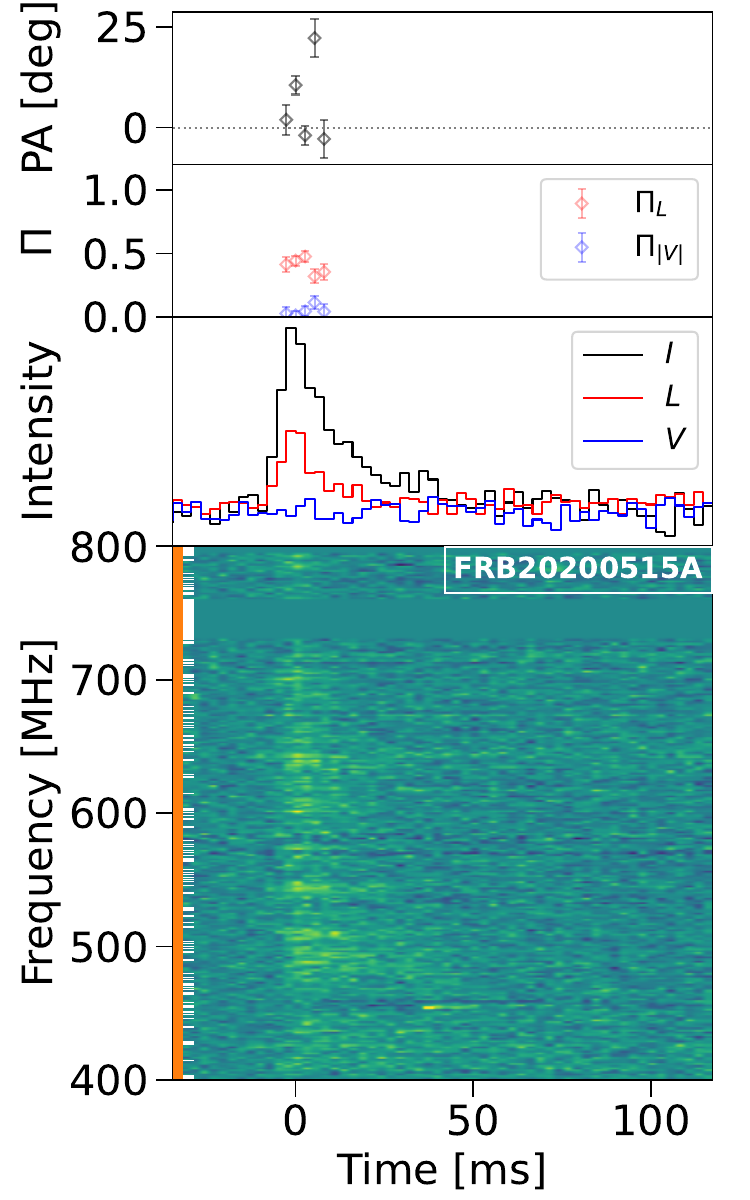}
    \includegraphics[width=0.188\textwidth]{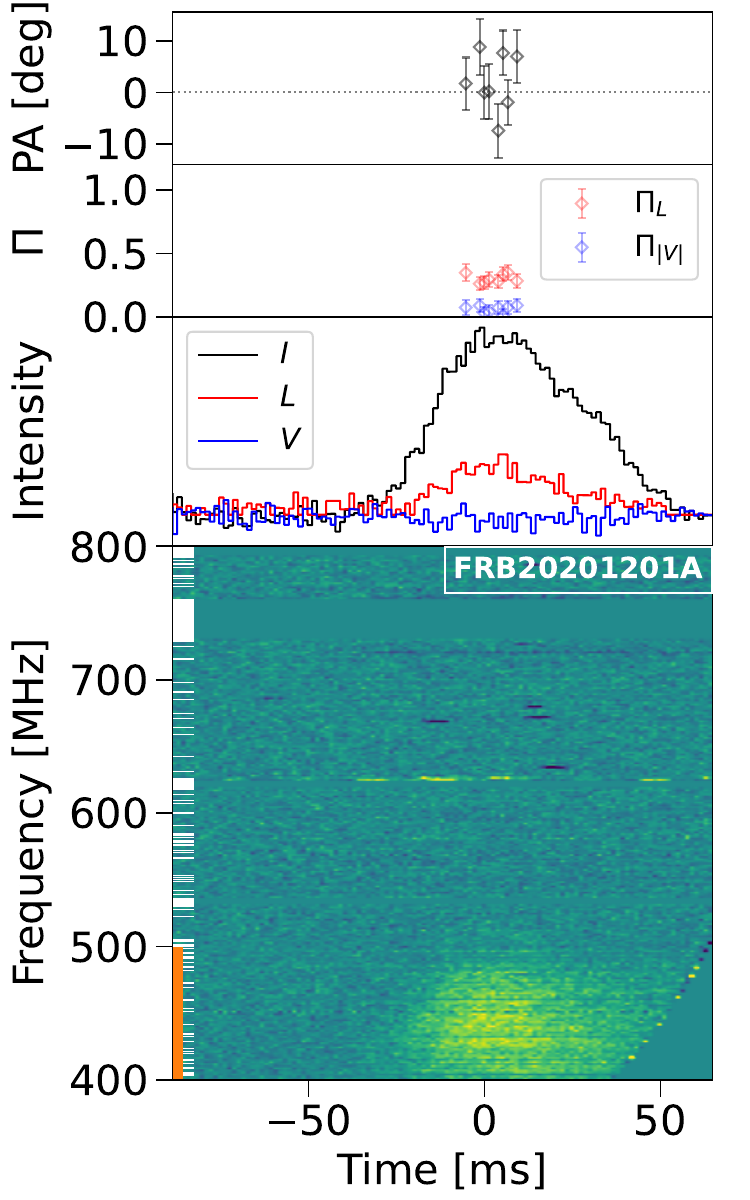}
    \includegraphics[width=0.188\textwidth]{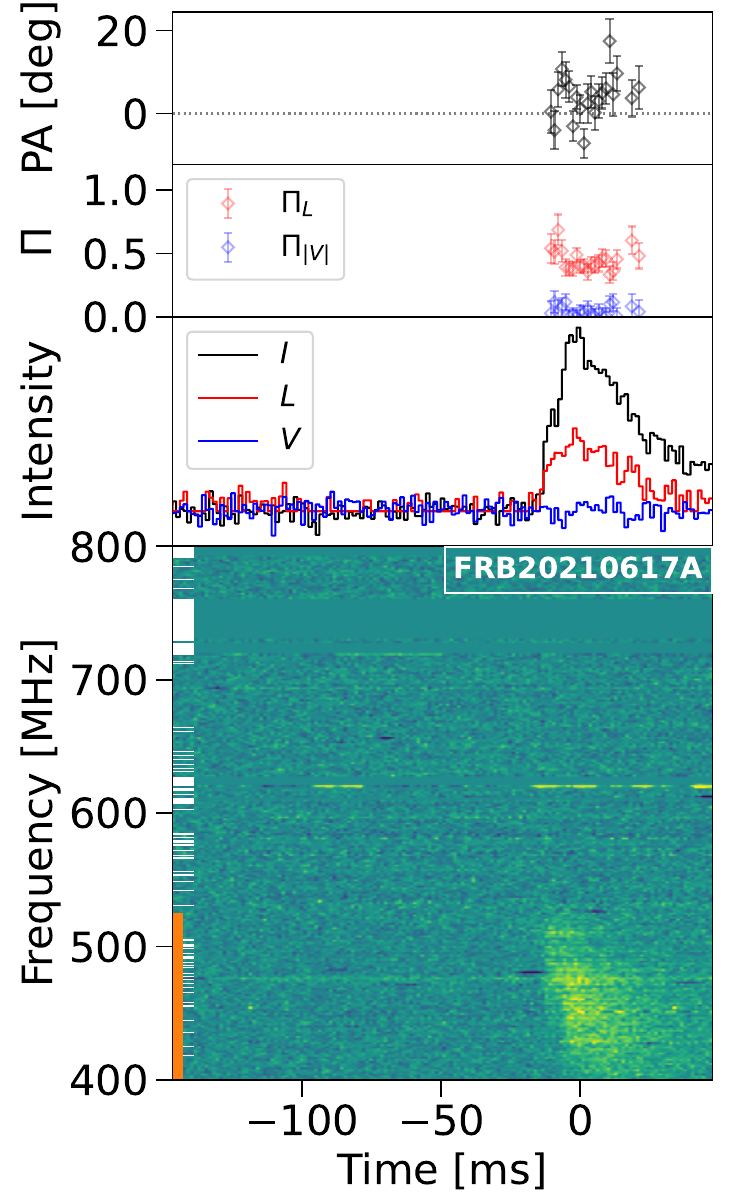}
    \includegraphics[width=0.188\textwidth]{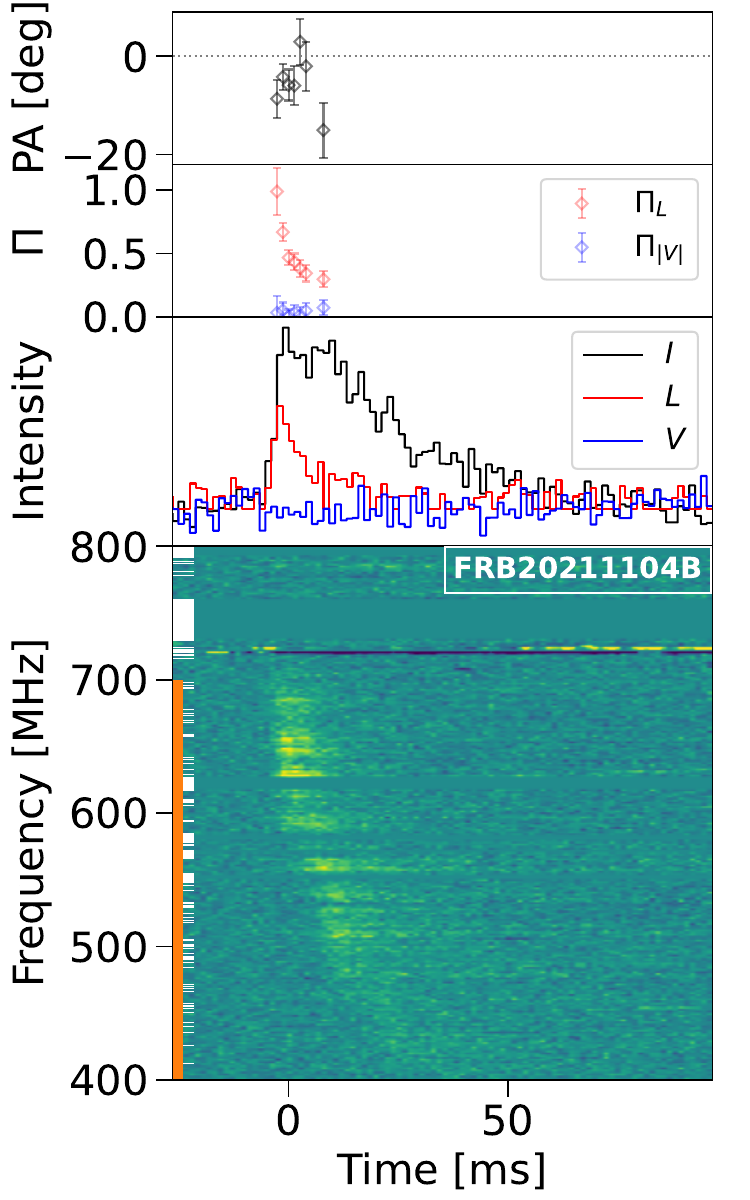}
    
    \includegraphics[width=0.188\textwidth]{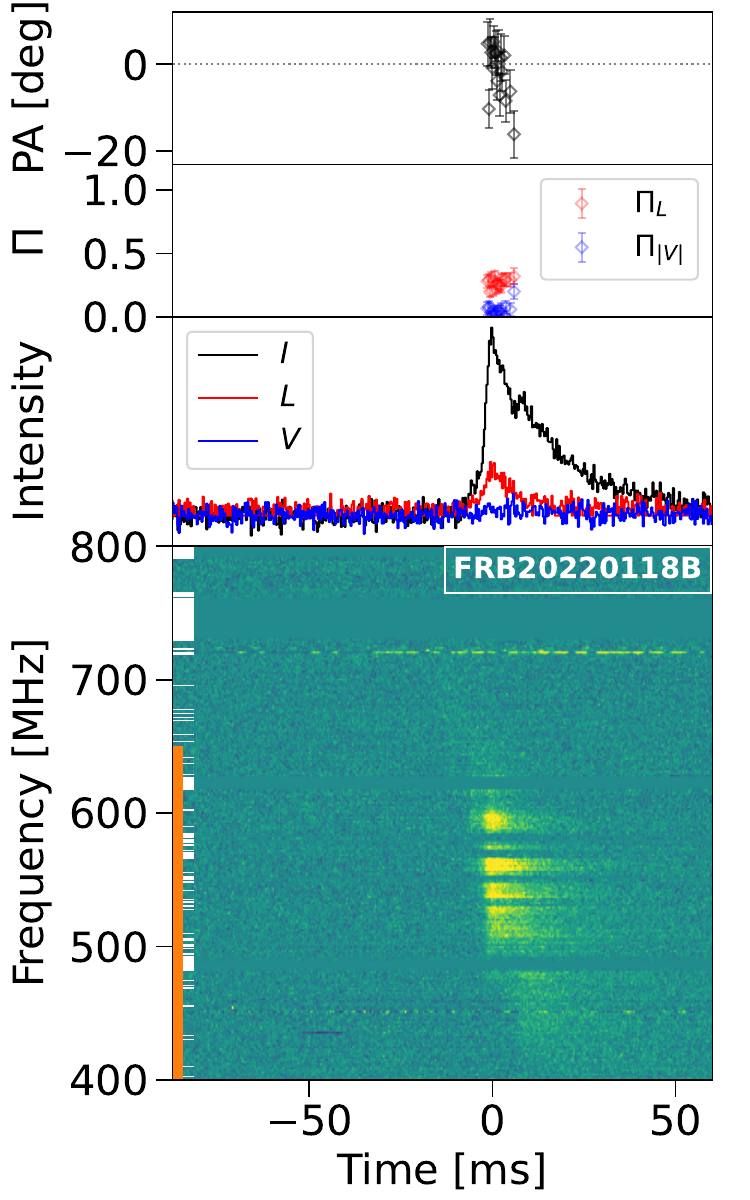}
    \includegraphics[width=0.188\textwidth]{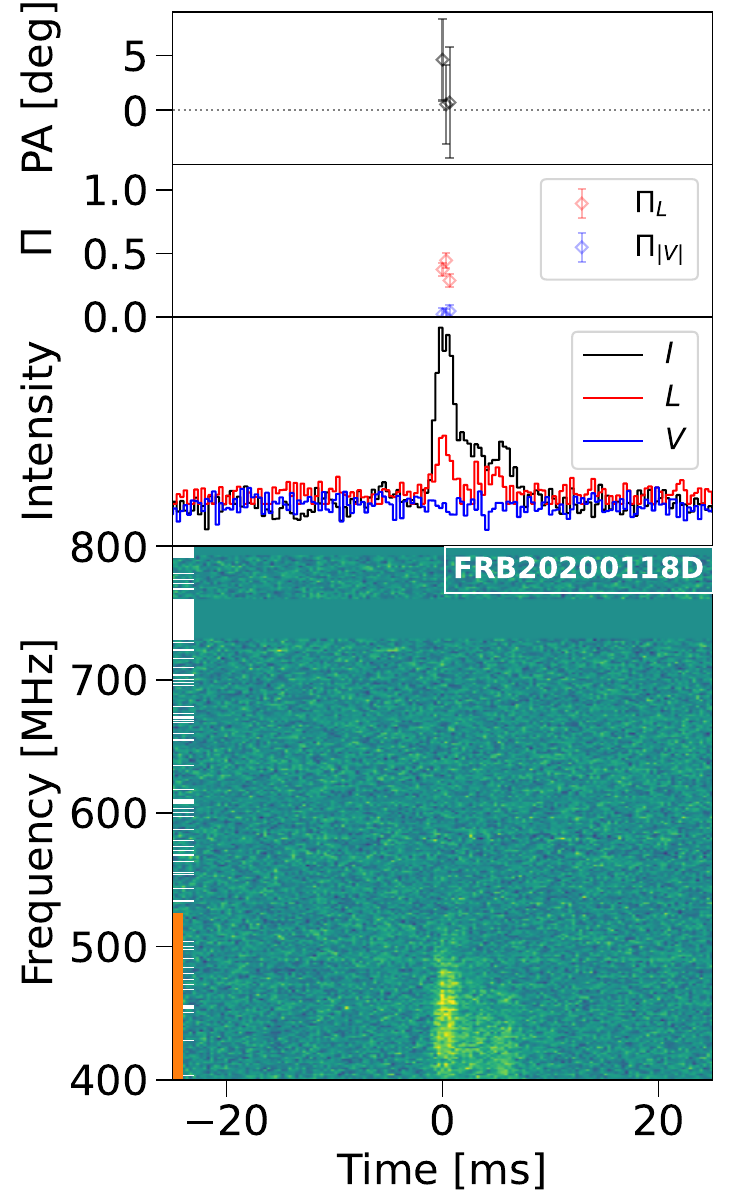}
    \includegraphics[width=0.188\textwidth]{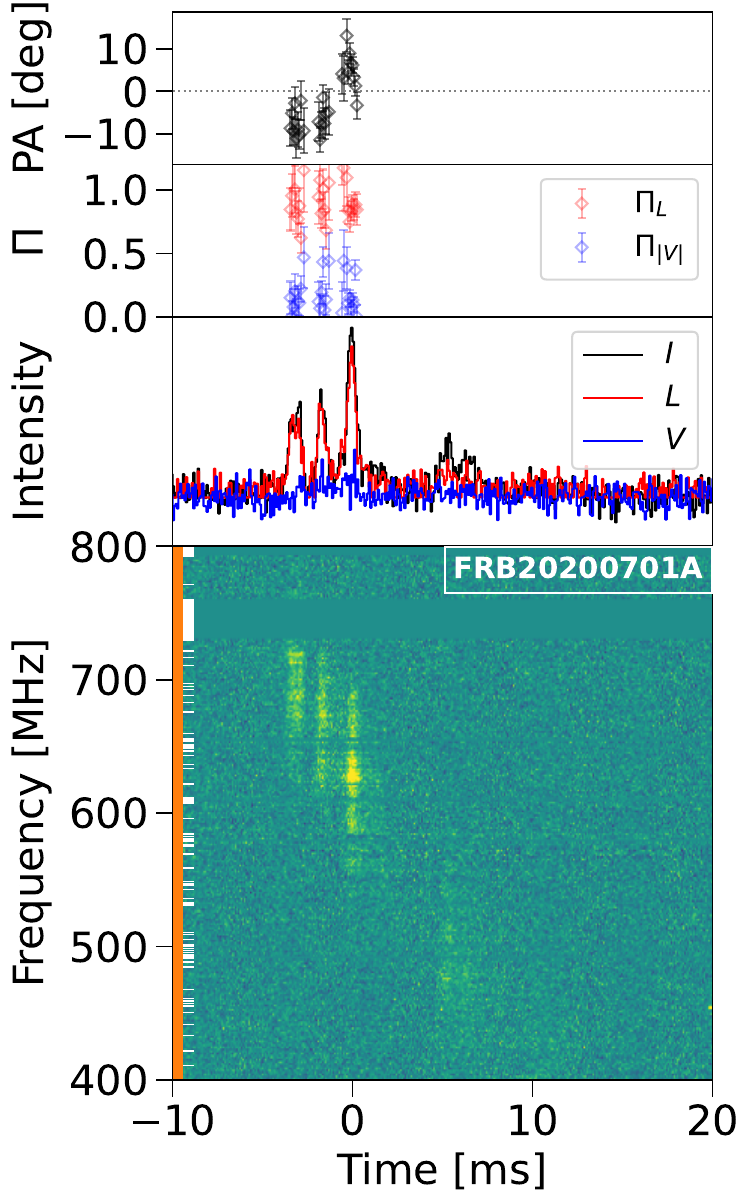}
    \includegraphics[width=0.188\textwidth]{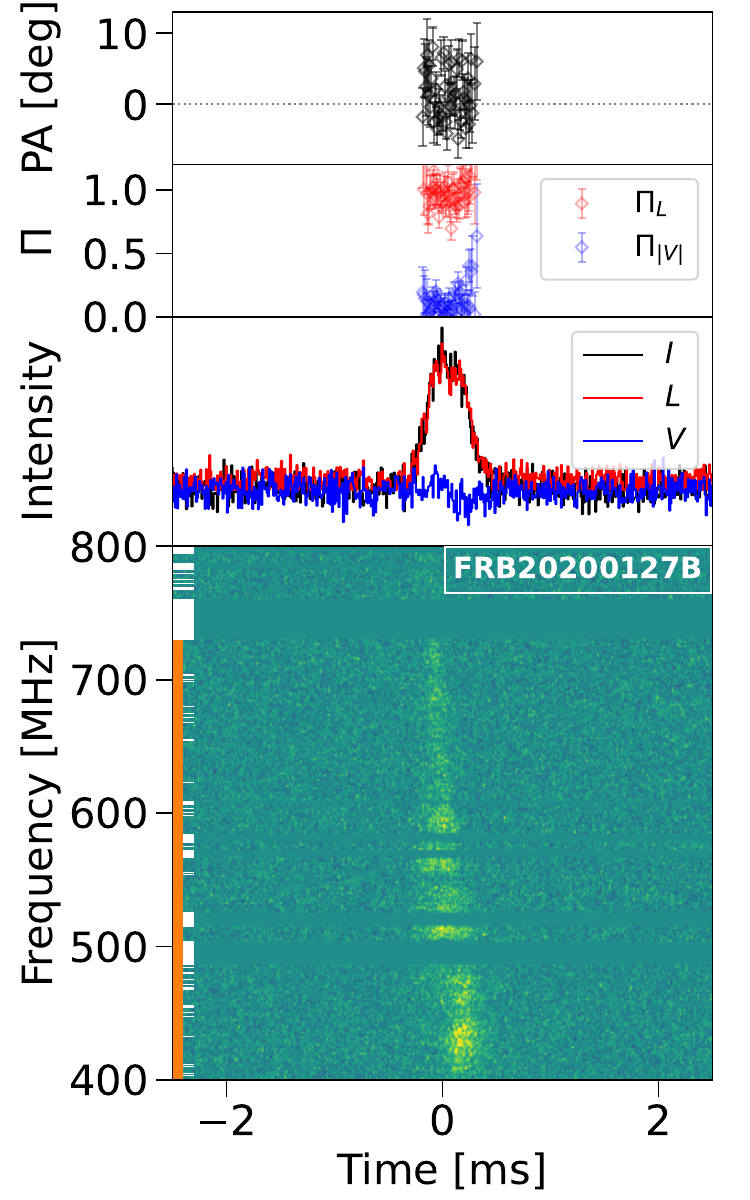}
    \includegraphics[width=0.188\textwidth]{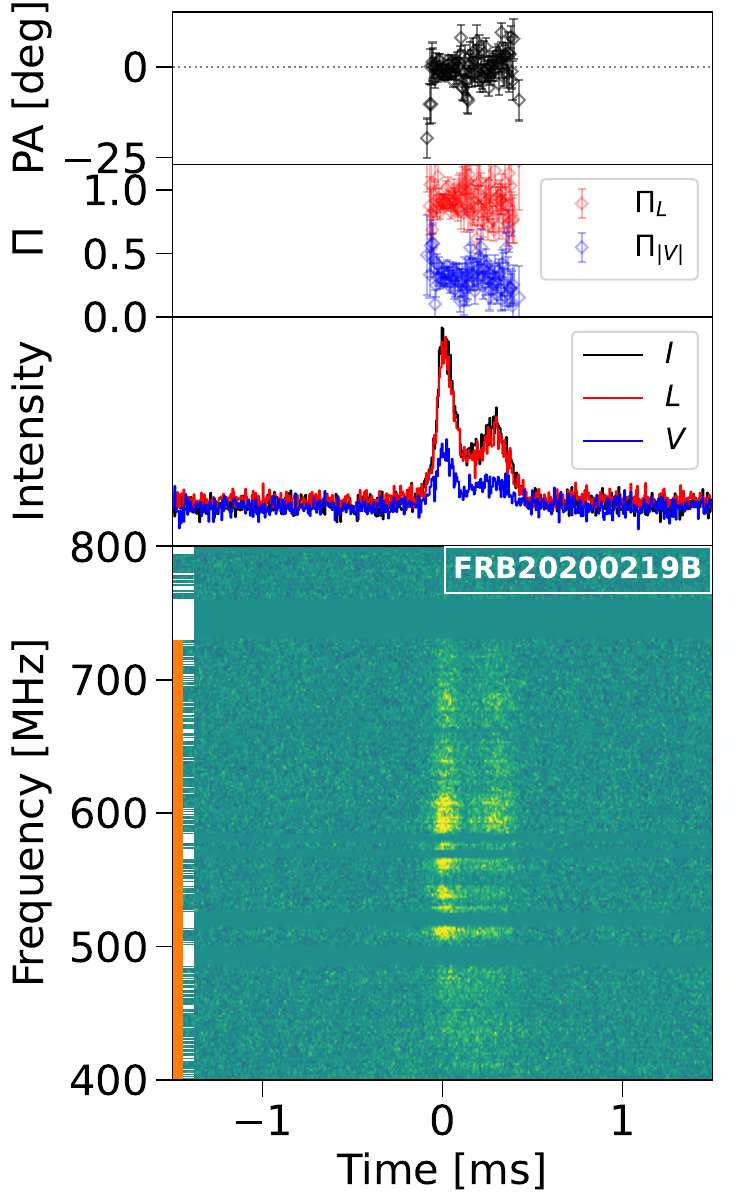}
    
    \caption{Total intensity waterfalls (de-dispersed to the associated $\mathrm{DM}_\mathrm{struct}$) and temporal profiles of Stokes $I$ (black line), $L = \sqrt{Q^2 + U^2}$ (red line), $V$ (blue line), linear polarization fraction ($L/I$; red circles), circular polarization fraction ($|V|/I$; blue circles), and linear polarization position angle (PA; black circles) for the FRBs from Tables~\ref{tab:RN3pol} and~\ref{tab:RN1-2pol}.  Channels masked due to radio frequency interference are highlighted by white streaks and the spectral limits are indicated by orange lines on the left-hand side of the total intensity waterfall plots. The panels are labeled with the TNS name of each event.}
    \label{fig:waterfalls}
\end{center}
\end{figure*}

\begin{figure*}[ht!]
\begin{center}

    \includegraphics[width=0.188\textwidth]{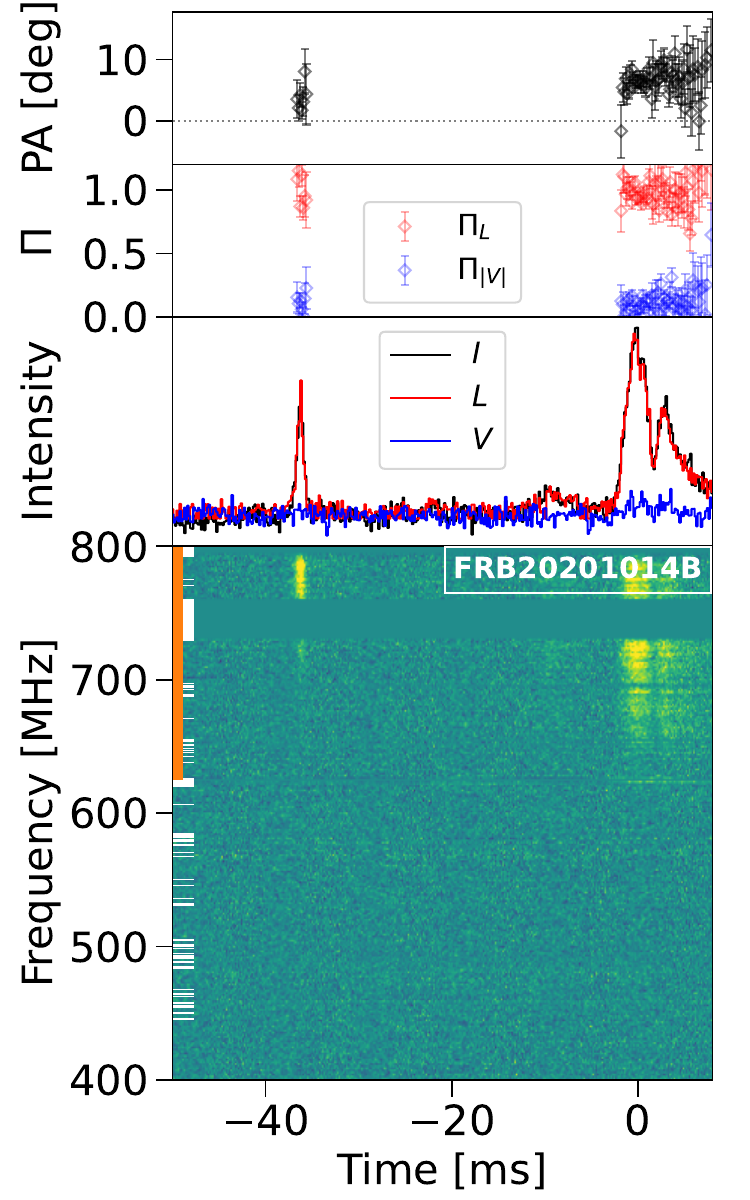}
    \includegraphics[width=0.188\textwidth]{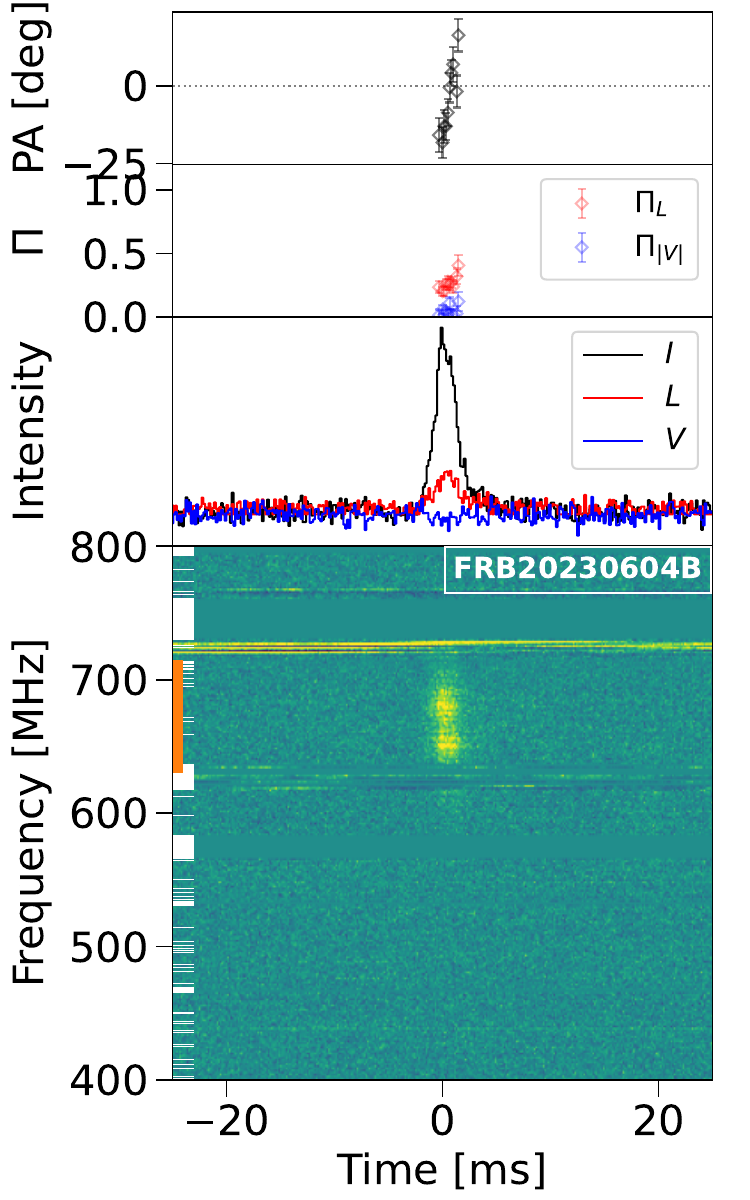}
    \includegraphics[width=0.188\textwidth]{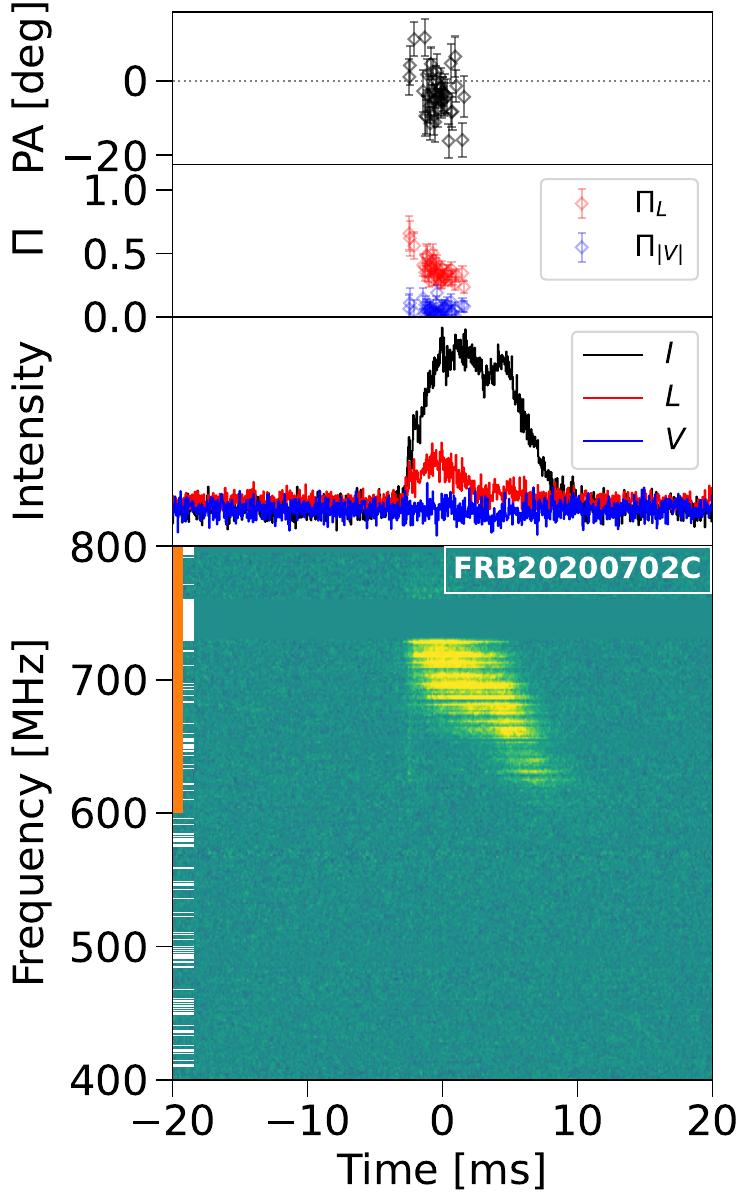}
    \includegraphics[width=0.188\textwidth]{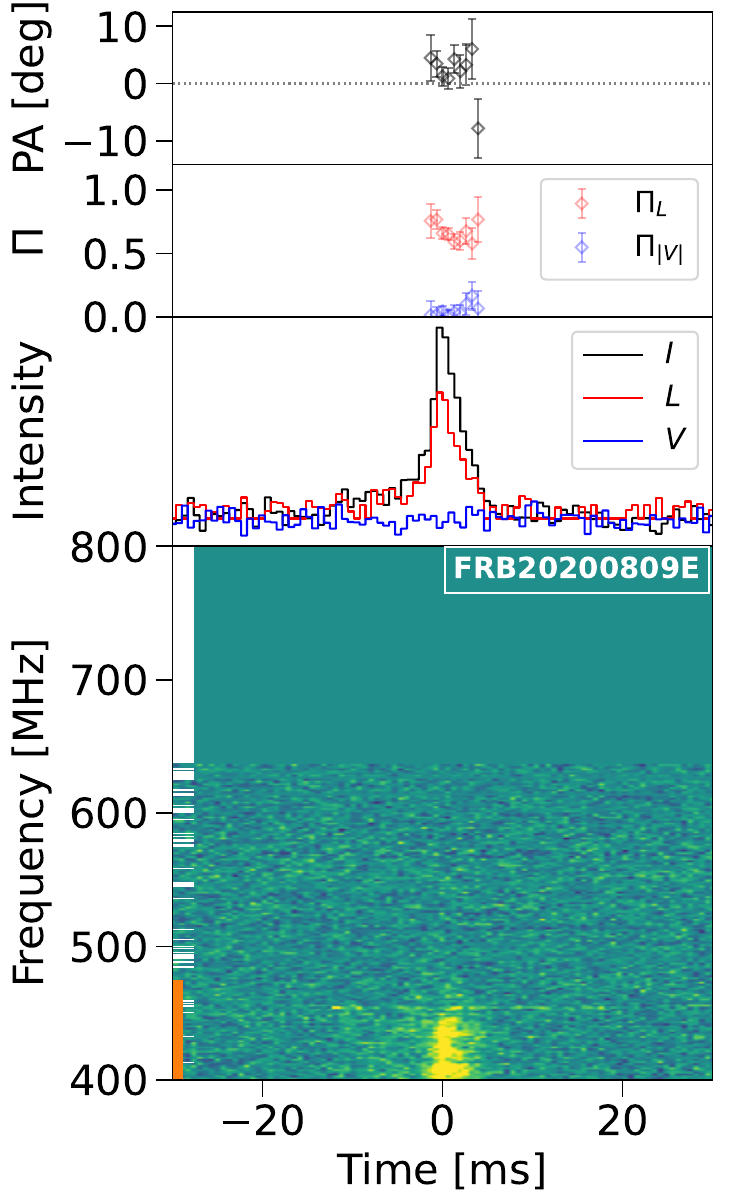}    
    \includegraphics[width=0.188\textwidth]{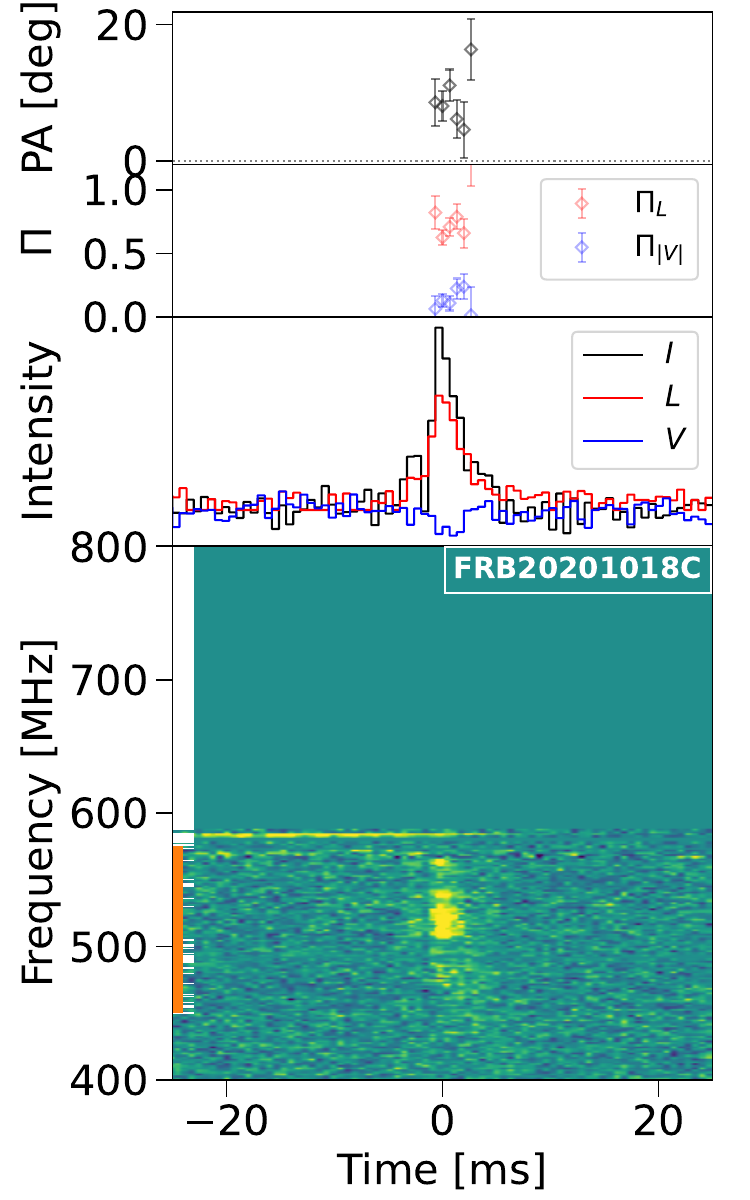}
    
    \includegraphics[width=0.188\textwidth]{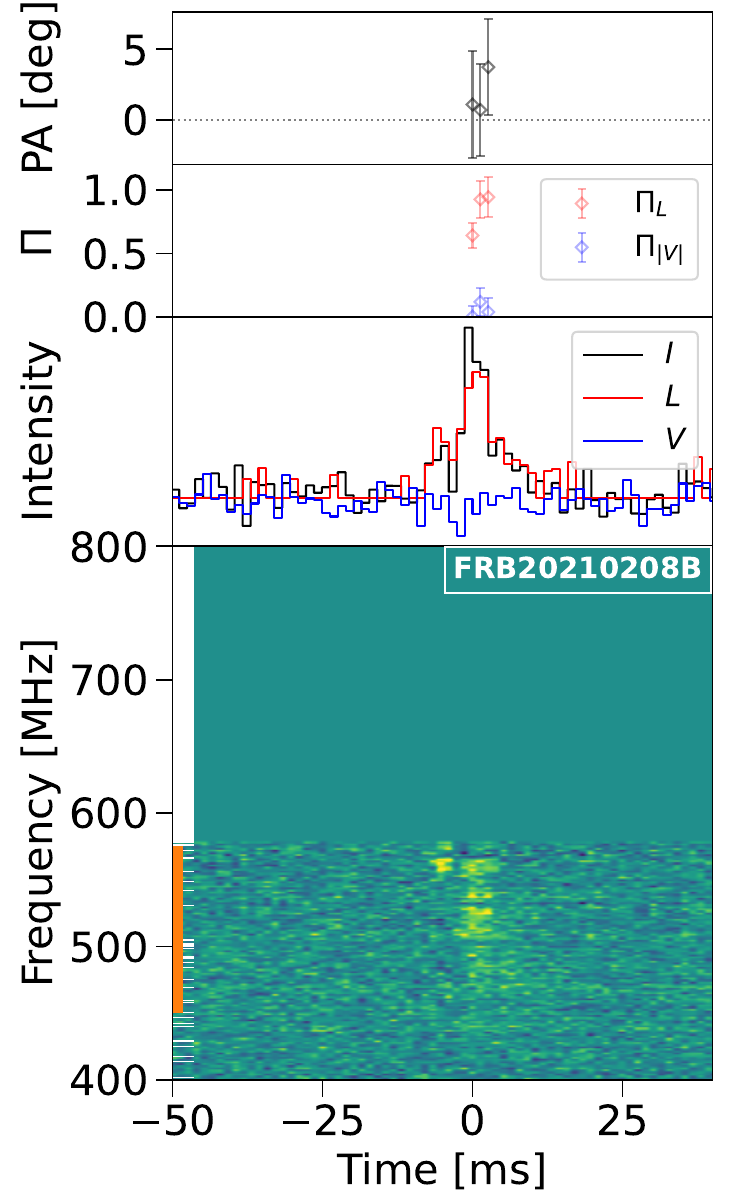}
    \includegraphics[width=0.188\textwidth]{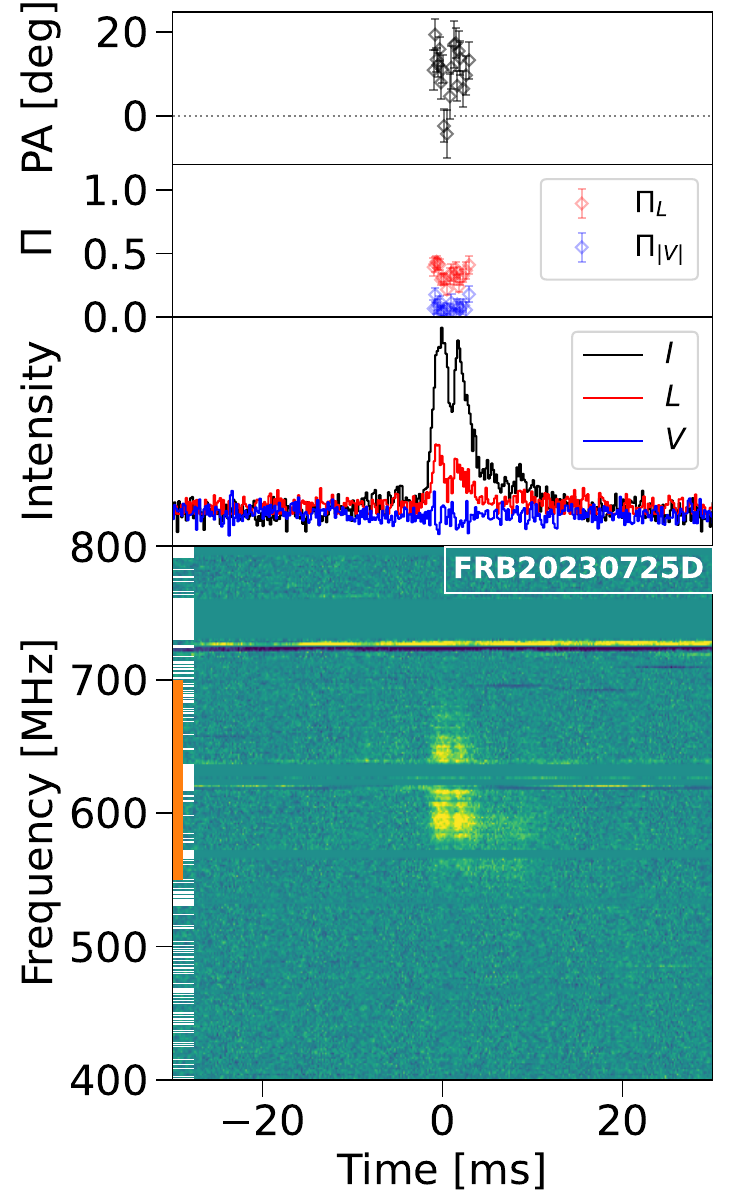}
    \includegraphics[width=0.188\textwidth]{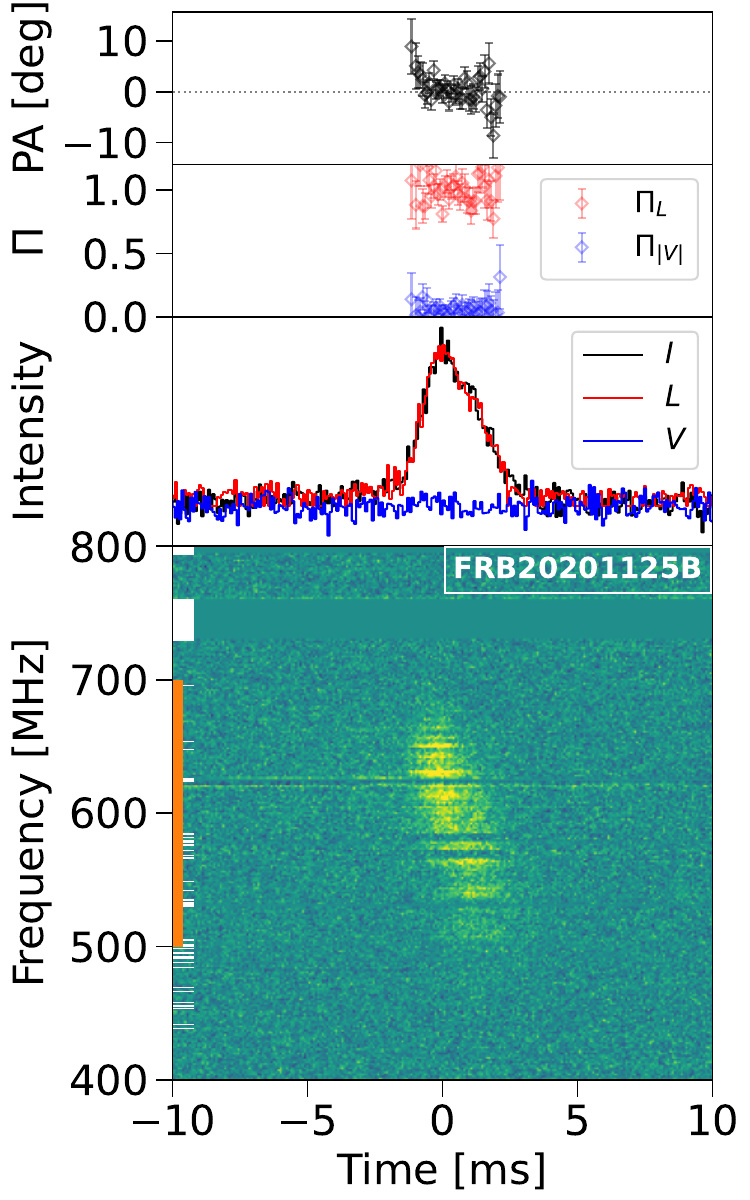}
    \includegraphics[width=0.188\textwidth]{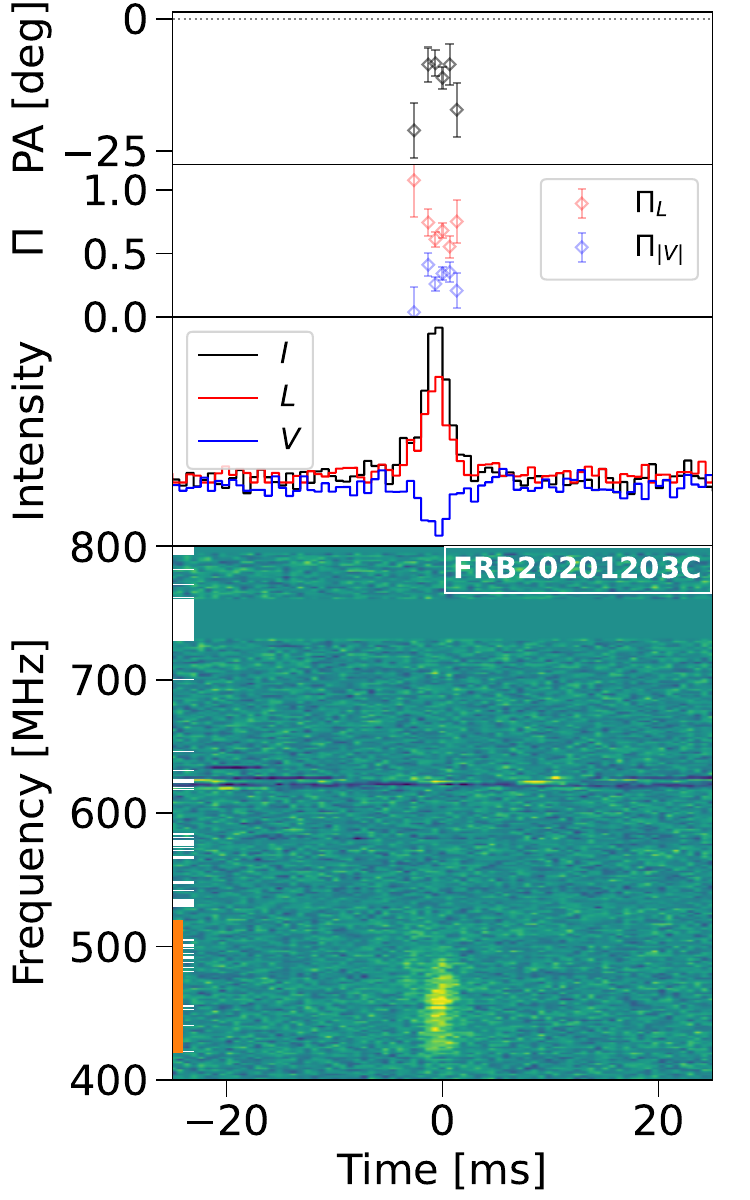}
    \includegraphics[width=0.188\textwidth]{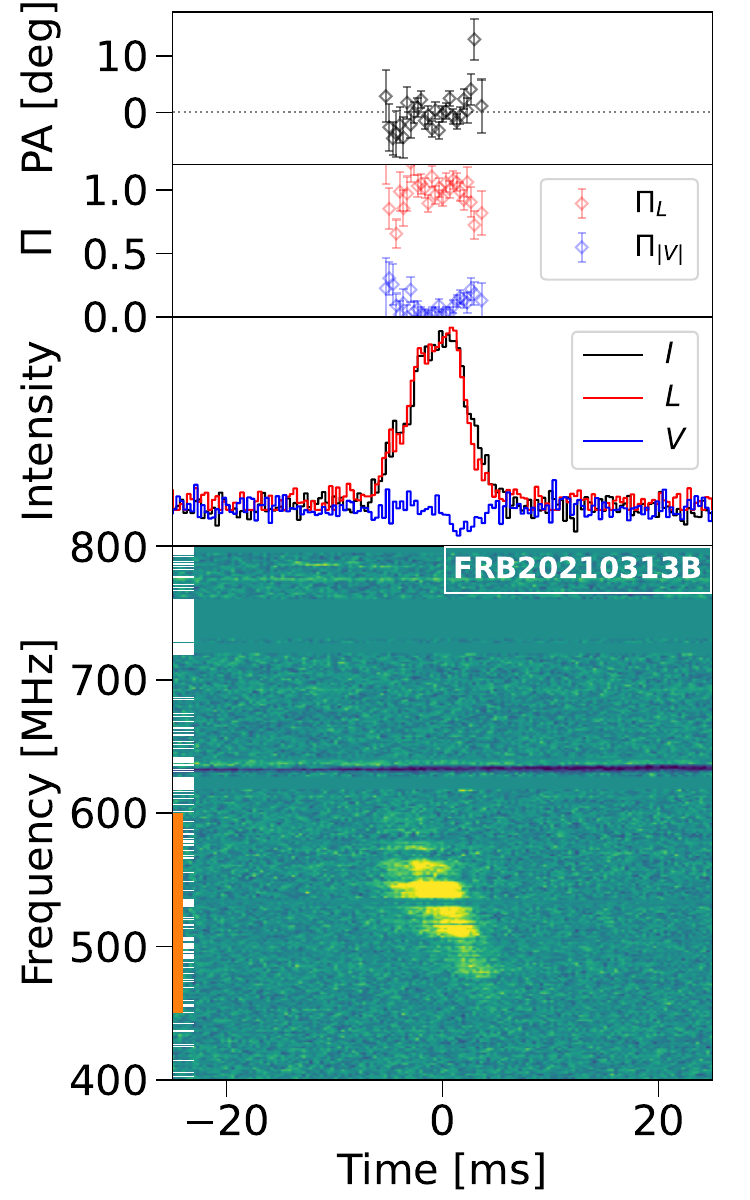}
    
    \includegraphics[width=0.188\textwidth]{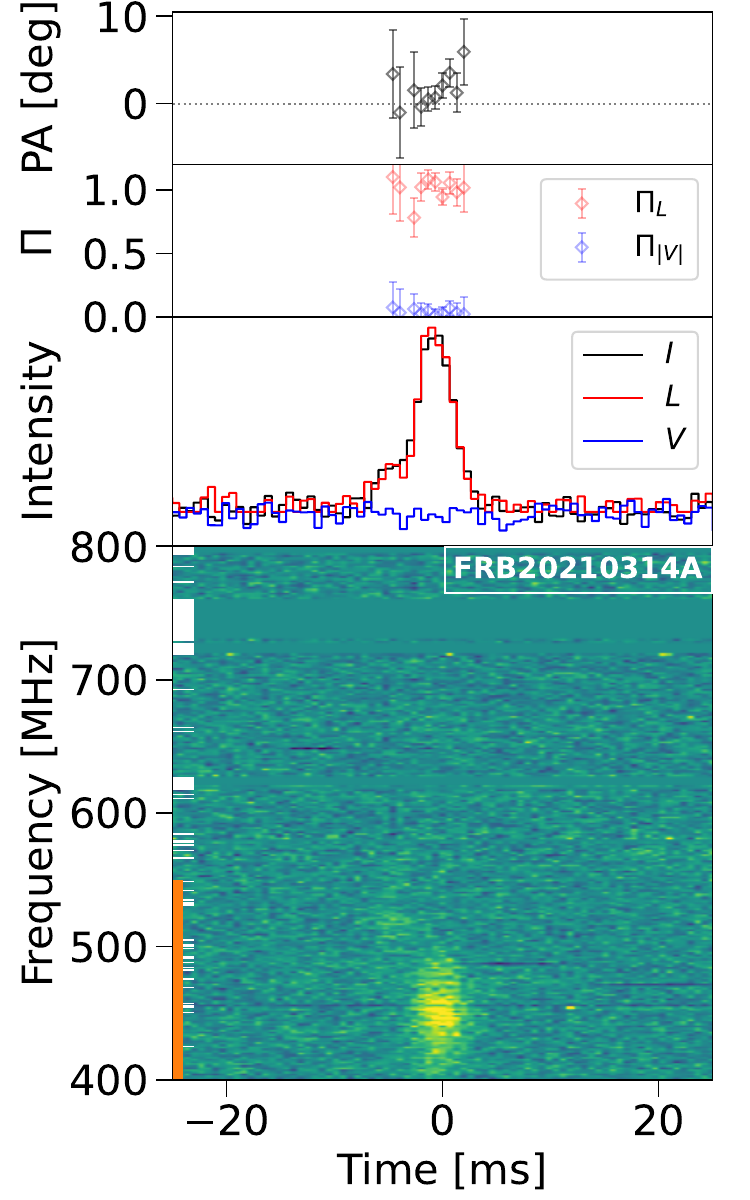}
    \includegraphics[width=0.188\textwidth]{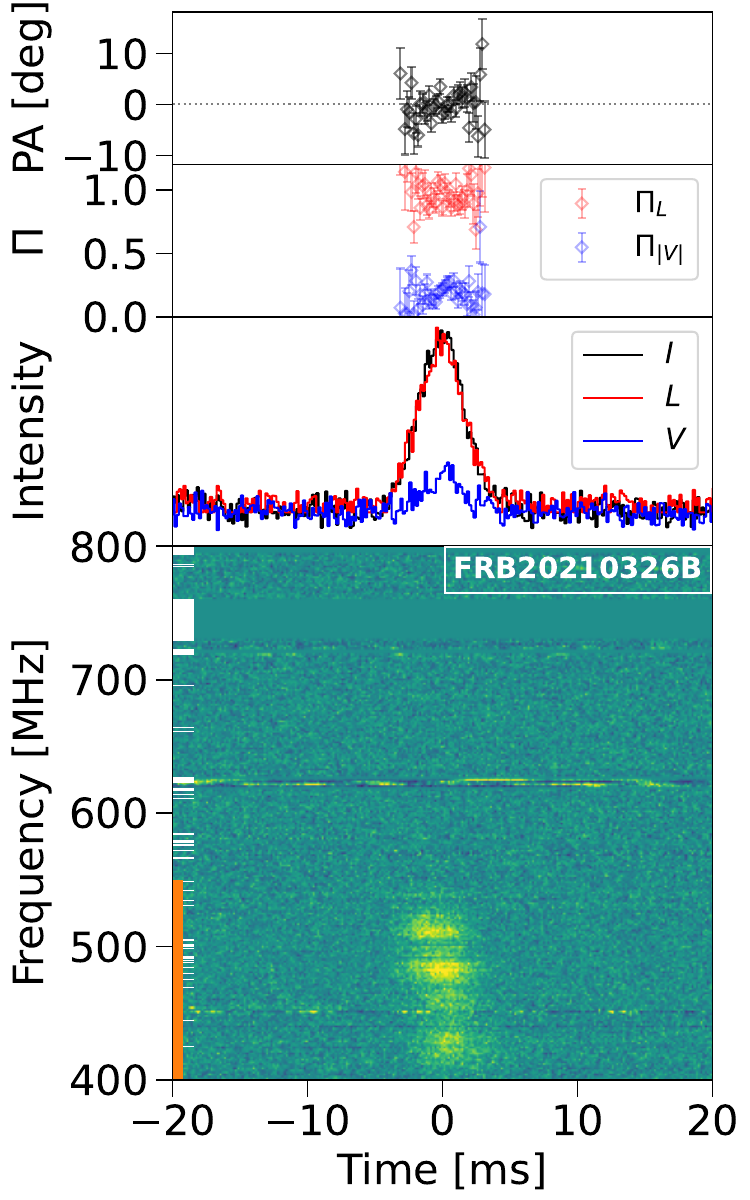}
    \includegraphics[width=0.188\textwidth]{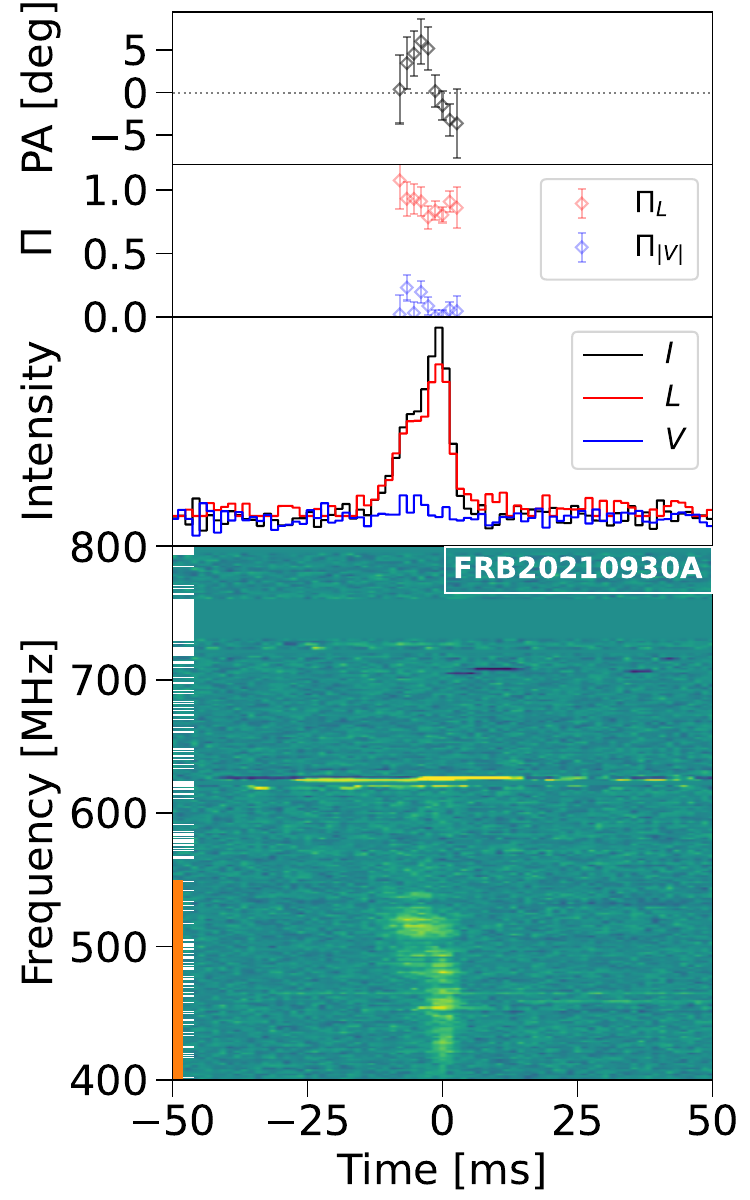}
    \includegraphics[width=0.188\textwidth]{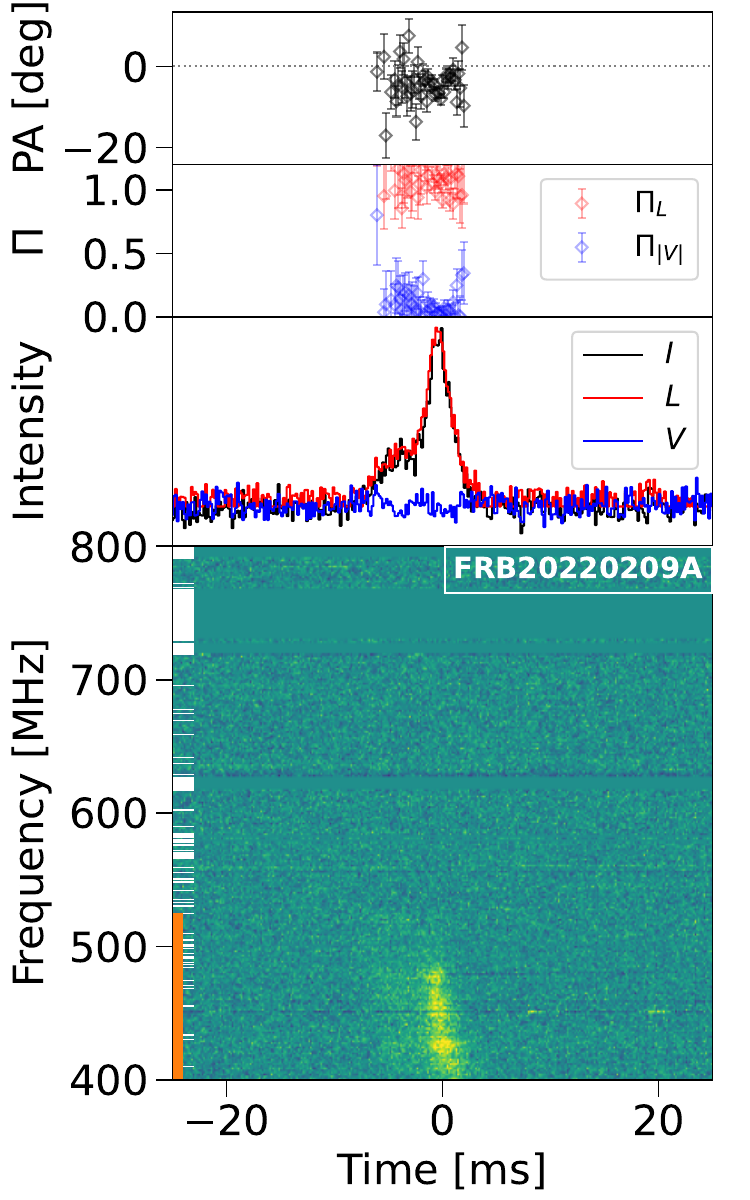}
    \includegraphics[width=0.188\textwidth]{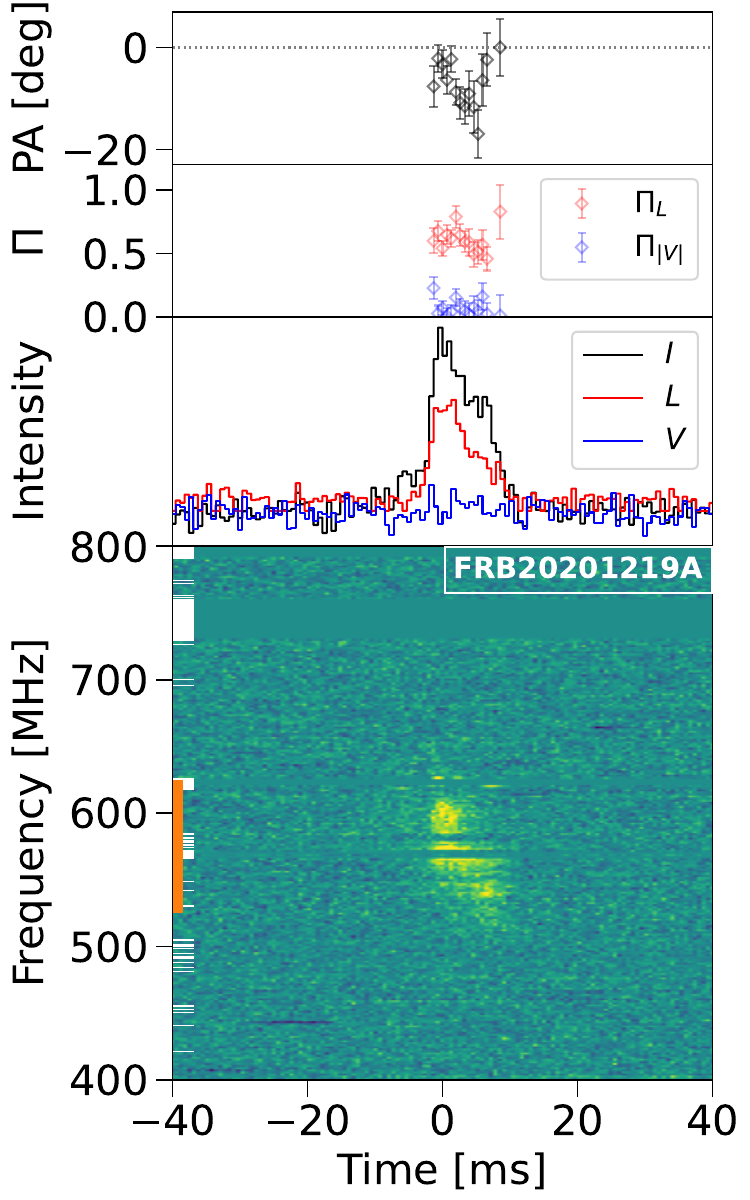}
    
    \includegraphics[width=0.188\textwidth]{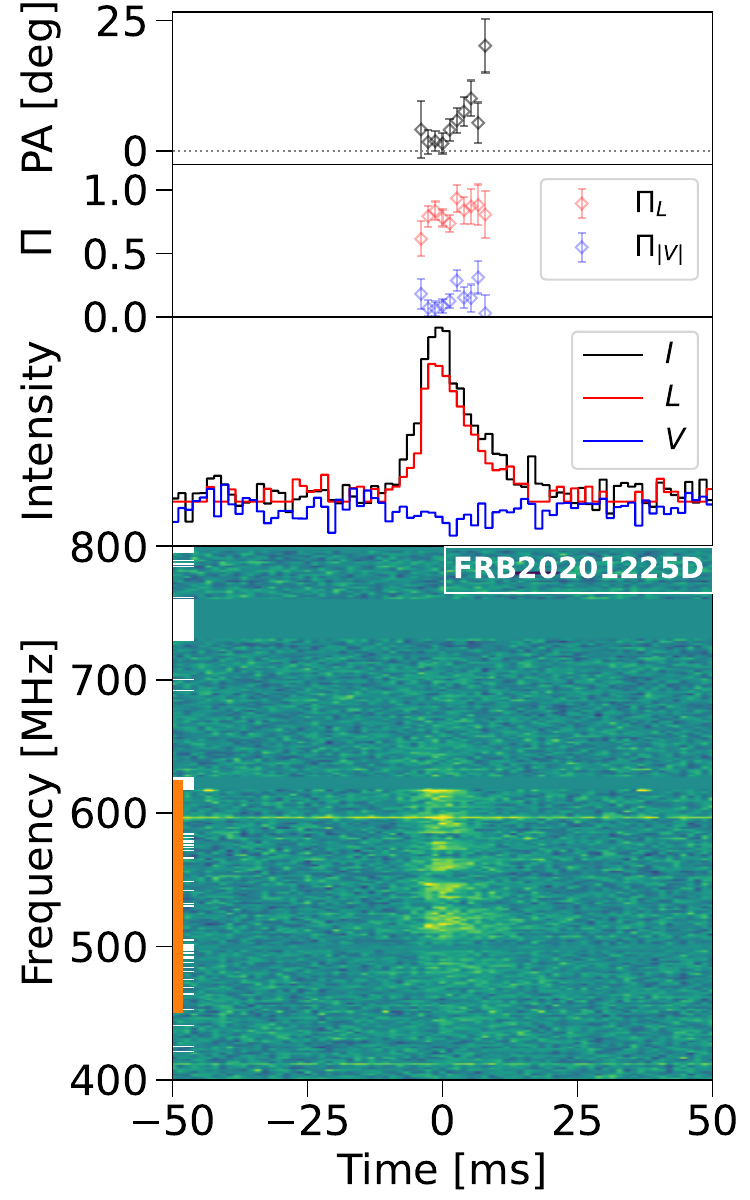}
    \includegraphics[width=0.188\textwidth]{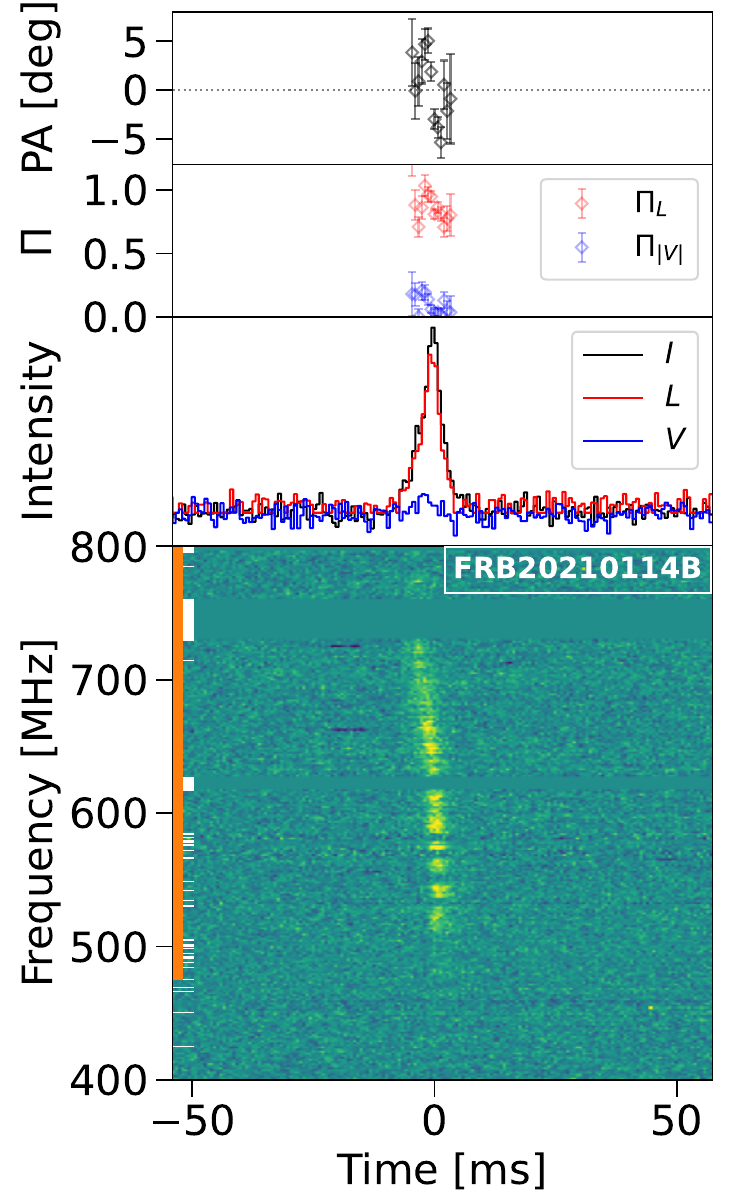}
    \includegraphics[width=0.188\textwidth]{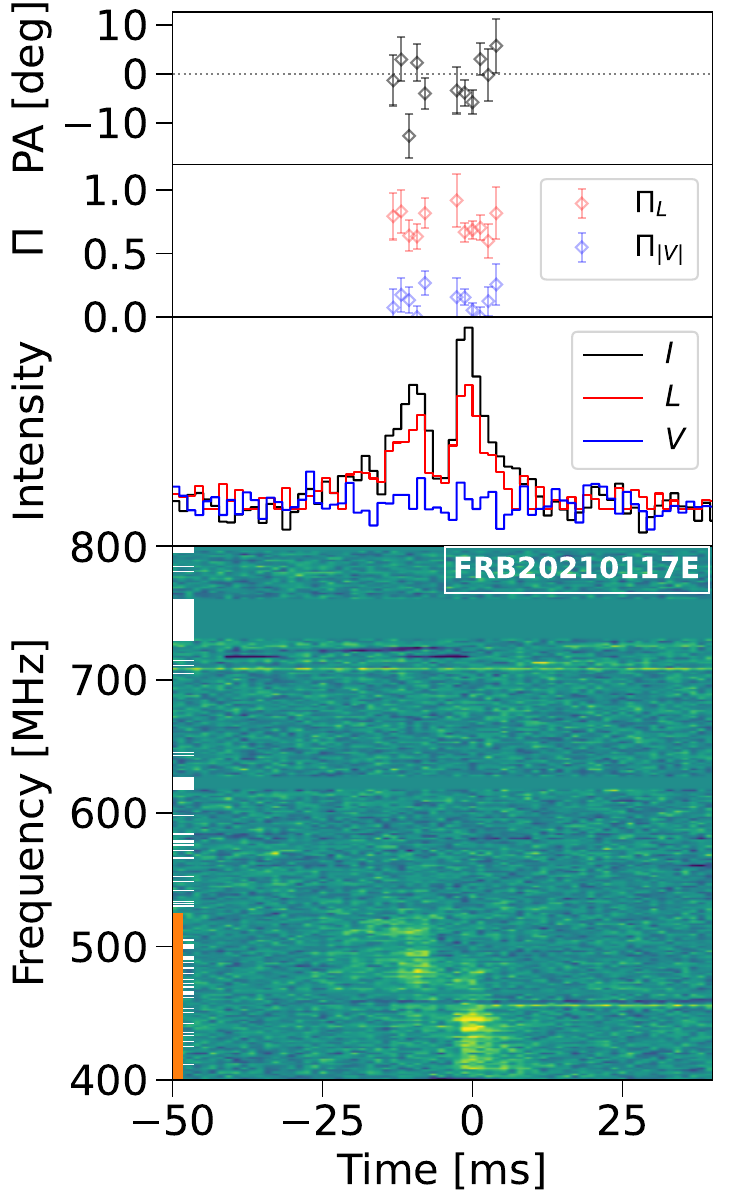}
    \includegraphics[width=0.188\textwidth]{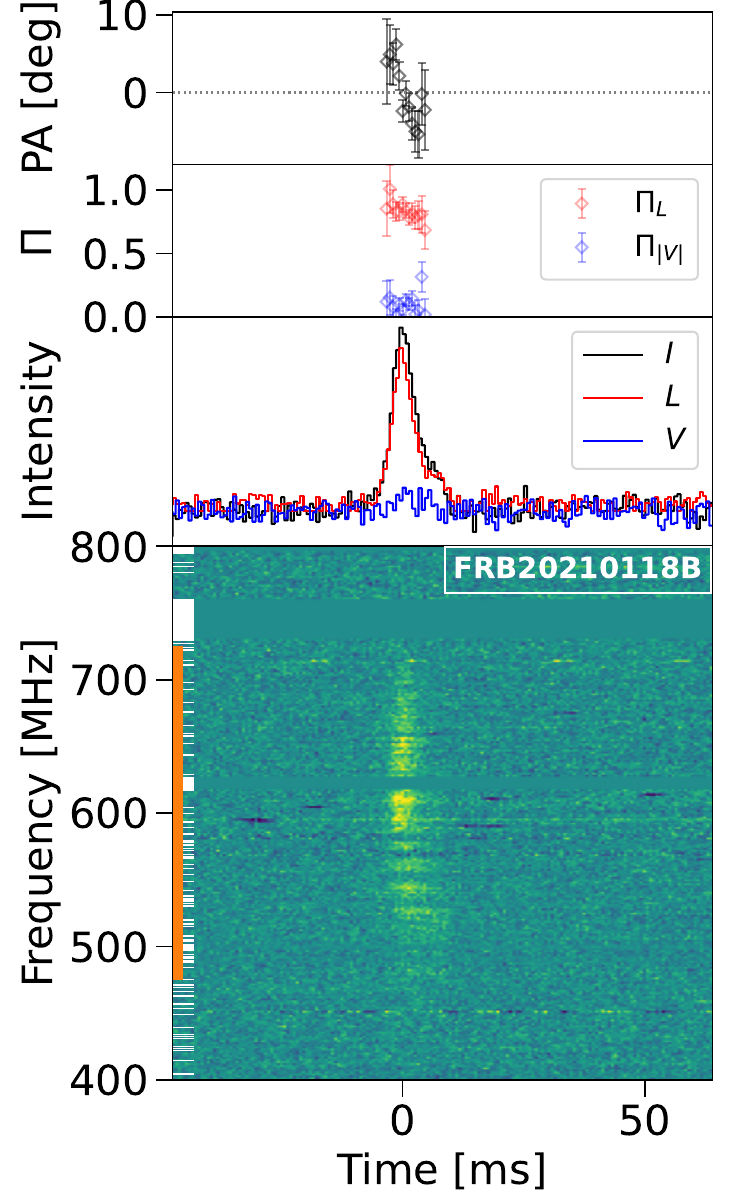}
    \includegraphics[width=0.188\textwidth]{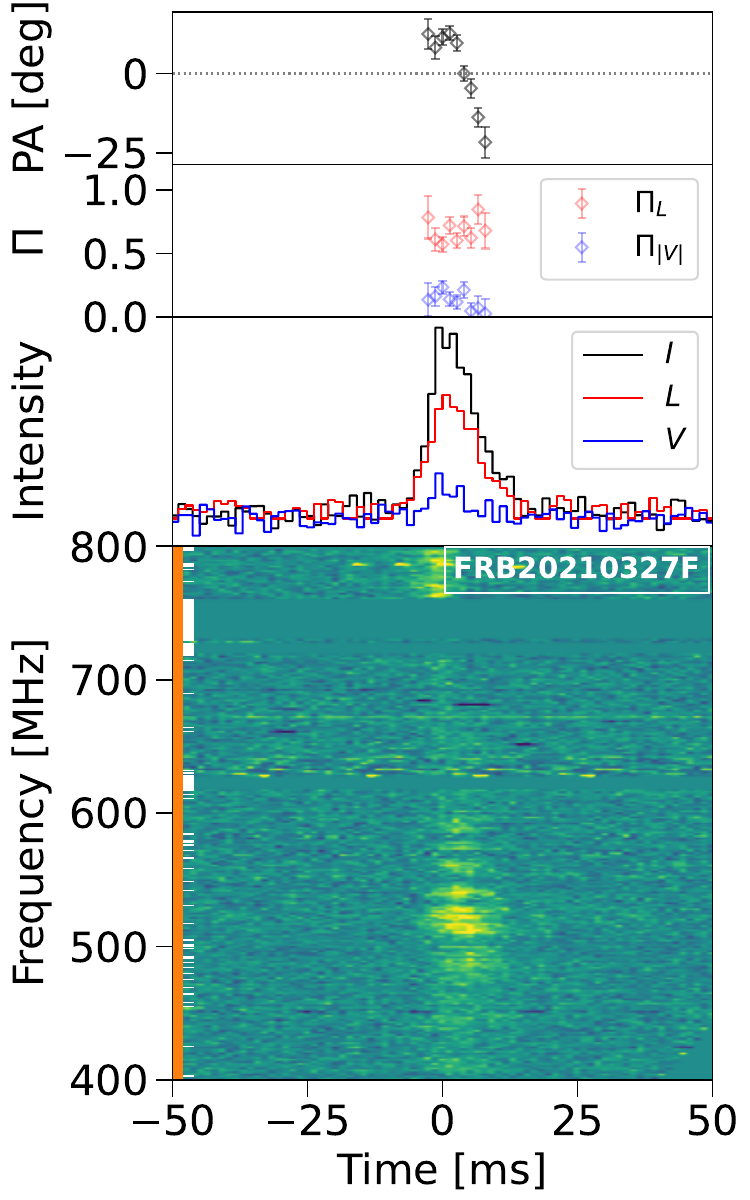}
    
    \caption{Further intensity waterfalls. Refer to the caption of Fig.~\ref{fig:waterfalls} for information on the legends and annotations. }
    \label{fig:waterfalls2}
\end{center}
\vspace{0.3cm}
\end{figure*}

\begin{figure*}[ht!]
\begin{center}

    \includegraphics[width=0.188\textwidth]{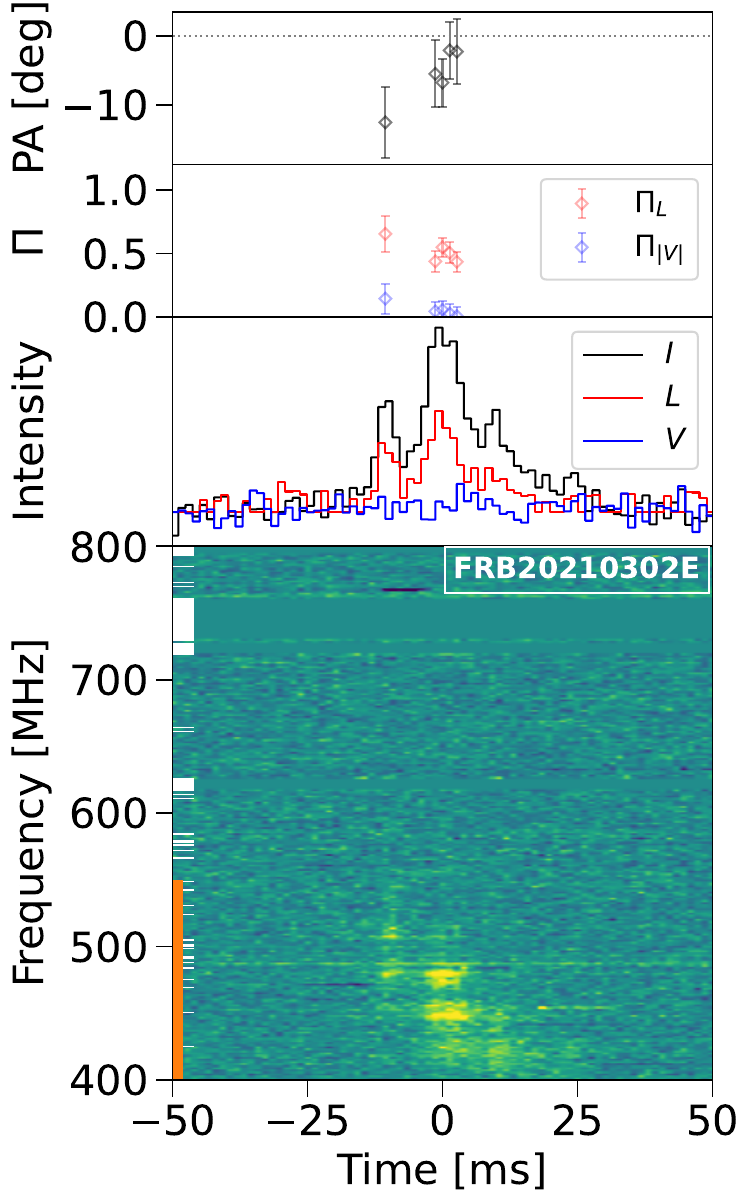}
    \includegraphics[width=0.188\textwidth]{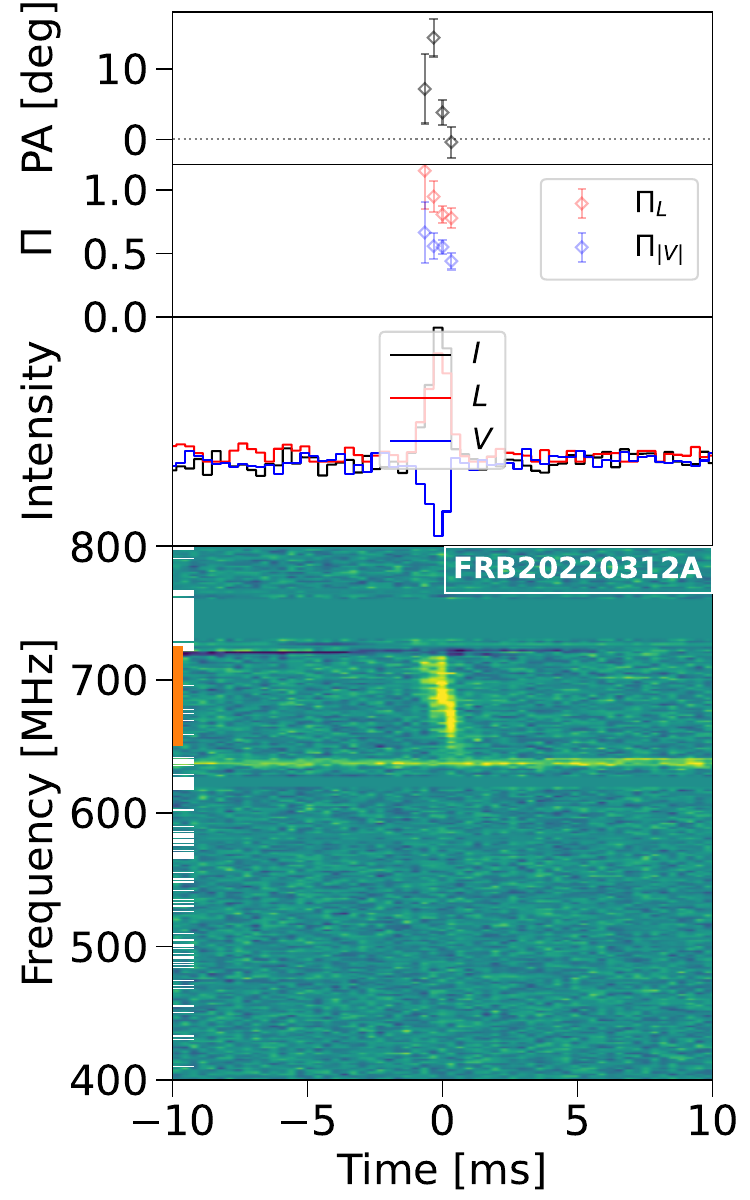}
    \includegraphics[width=0.188\textwidth]{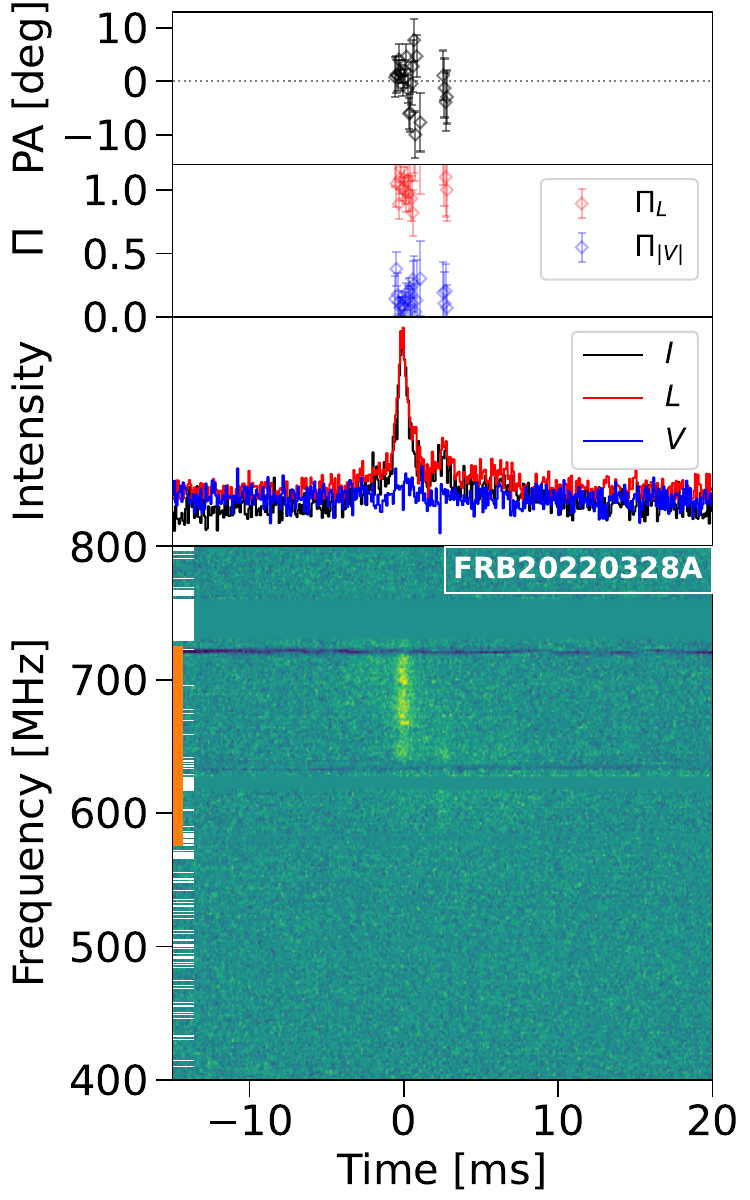}
    \includegraphics[width=0.188\textwidth]{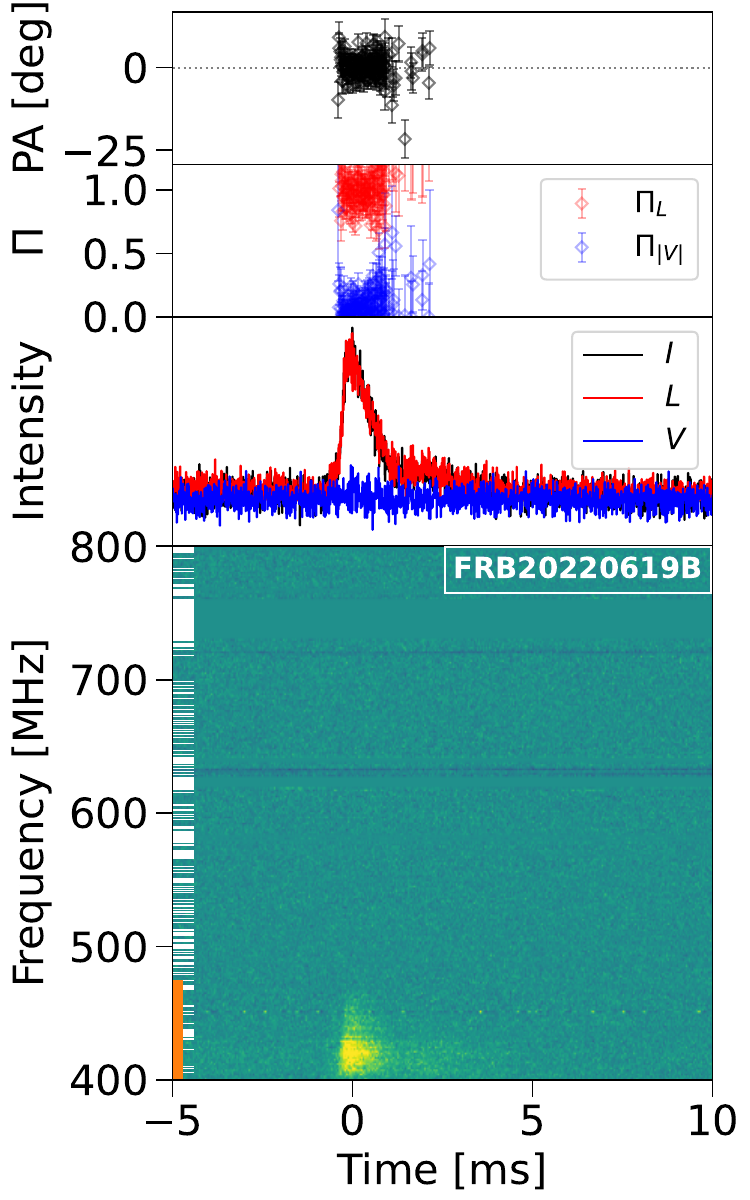}
    \includegraphics[width=0.188\textwidth]{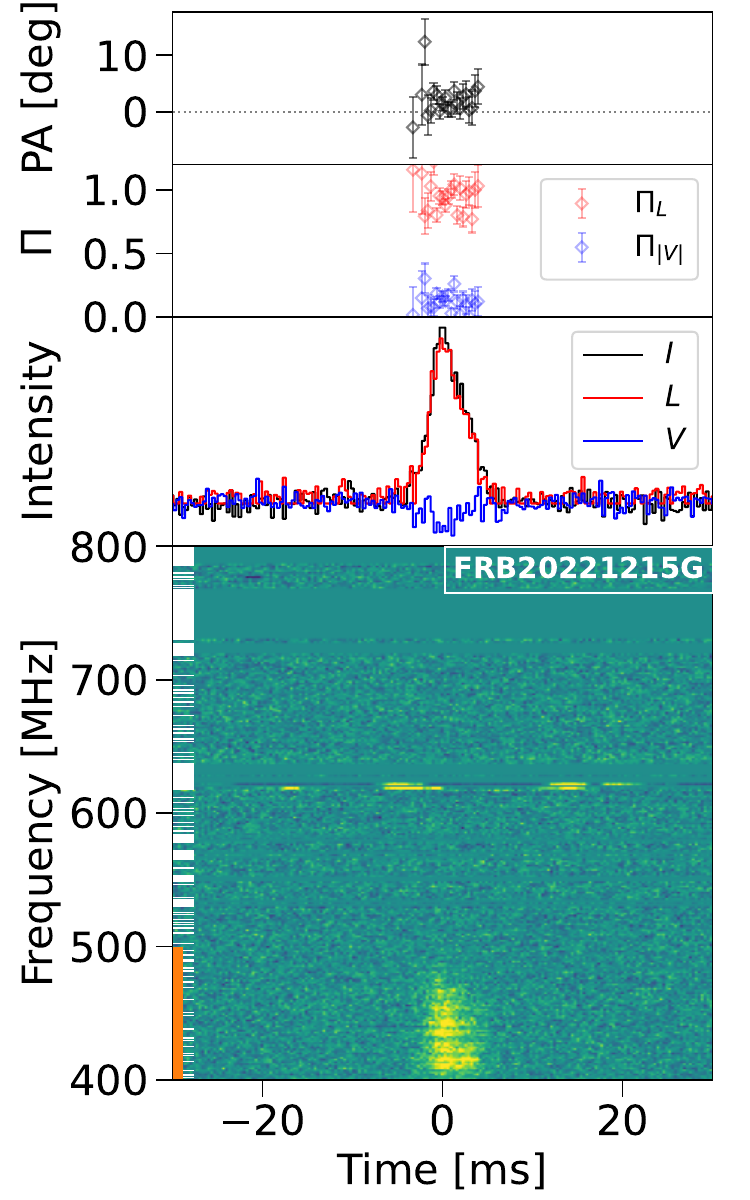}
 
    \includegraphics[width=0.188\textwidth]{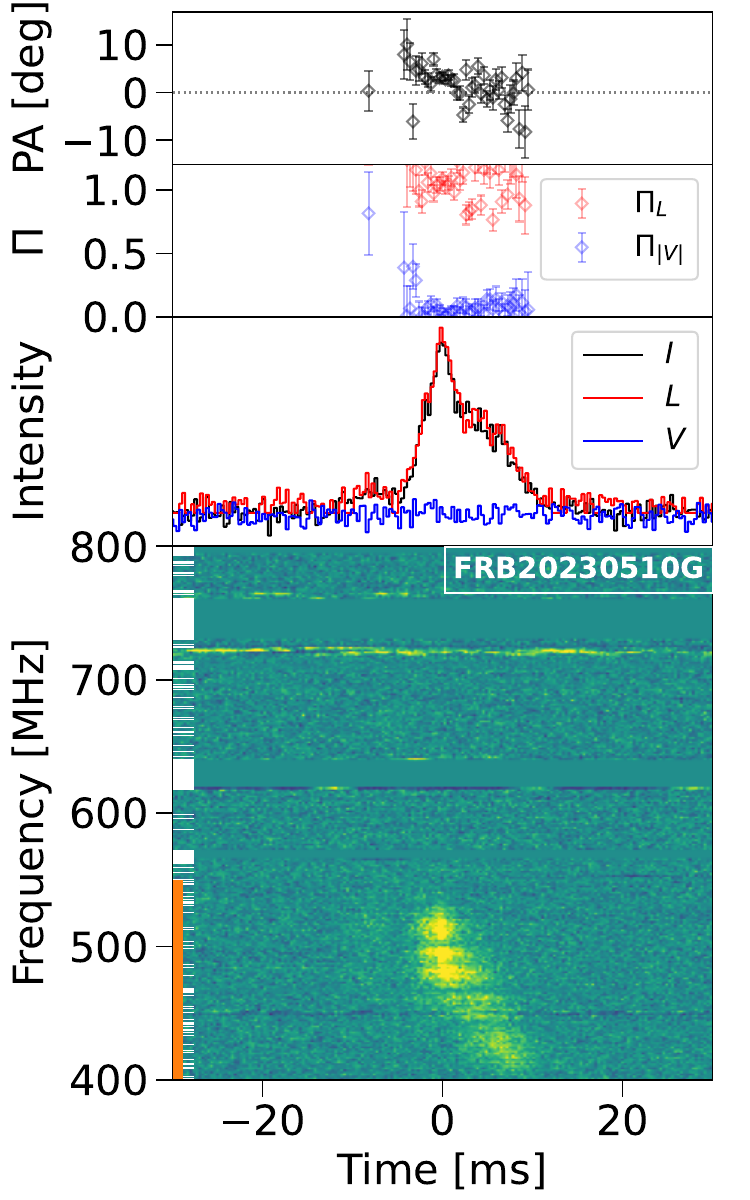}
    \includegraphics[width=0.188\textwidth]{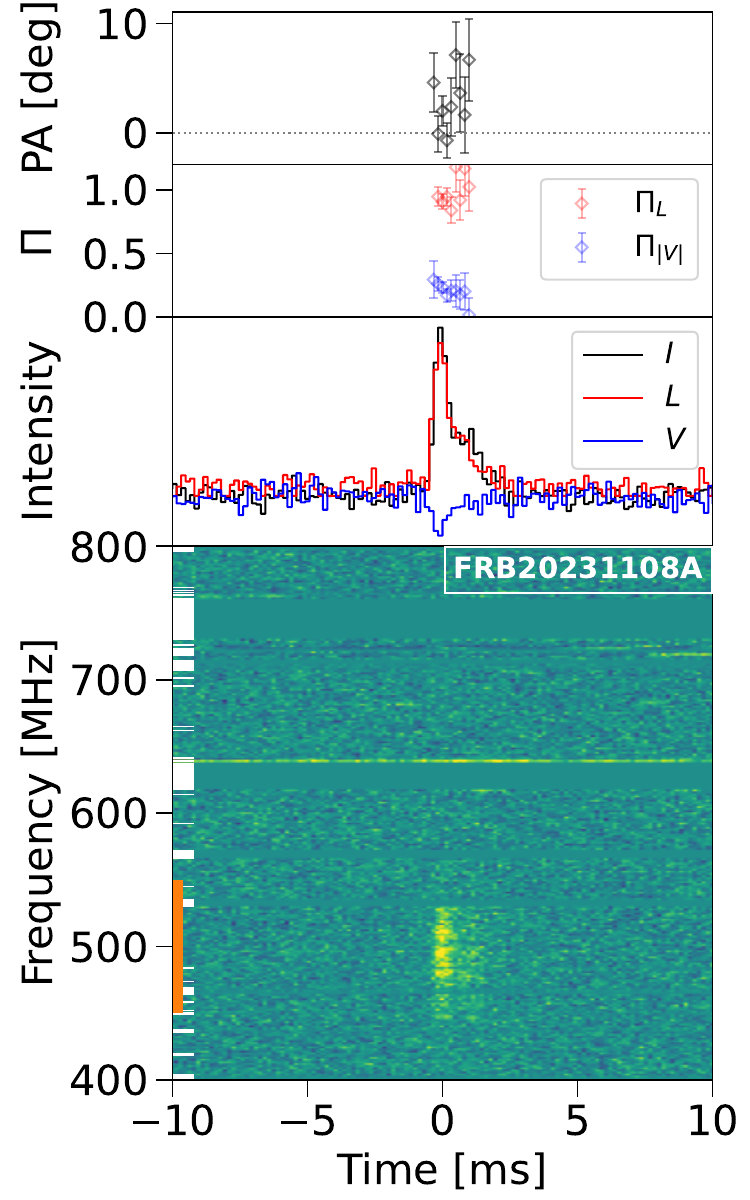}
    \includegraphics[width=0.188\textwidth]{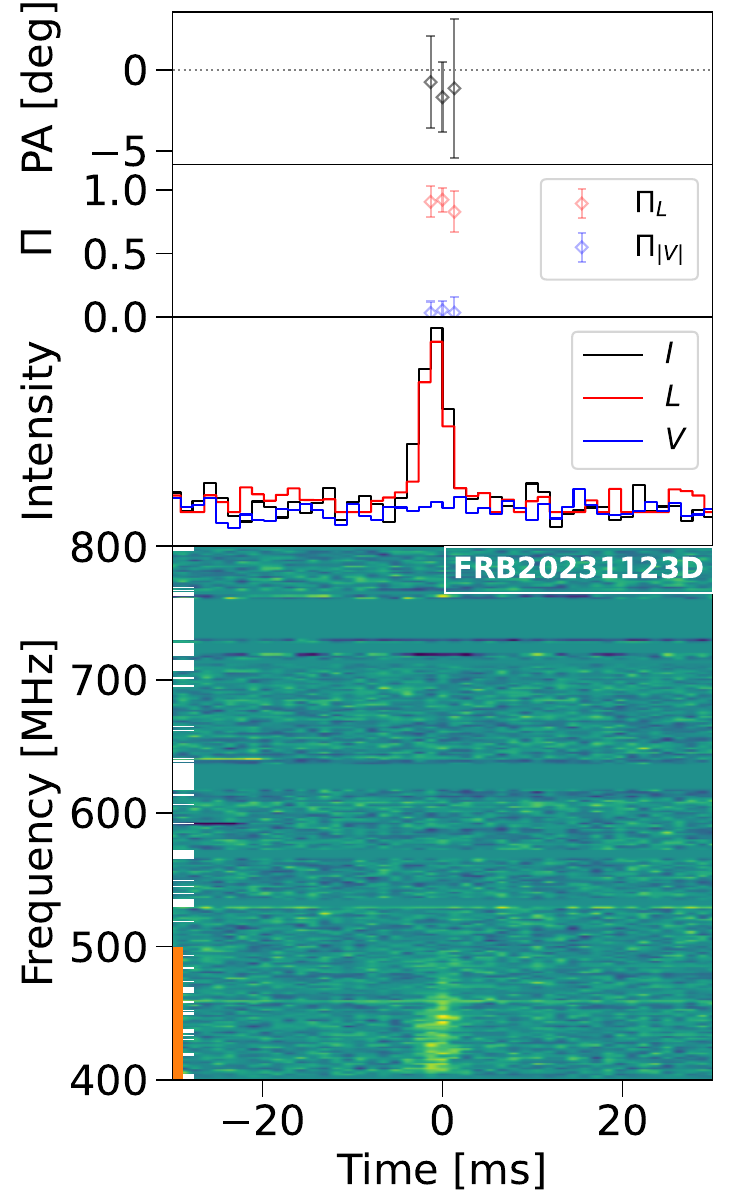} 
    \includegraphics[width=0.188\textwidth]{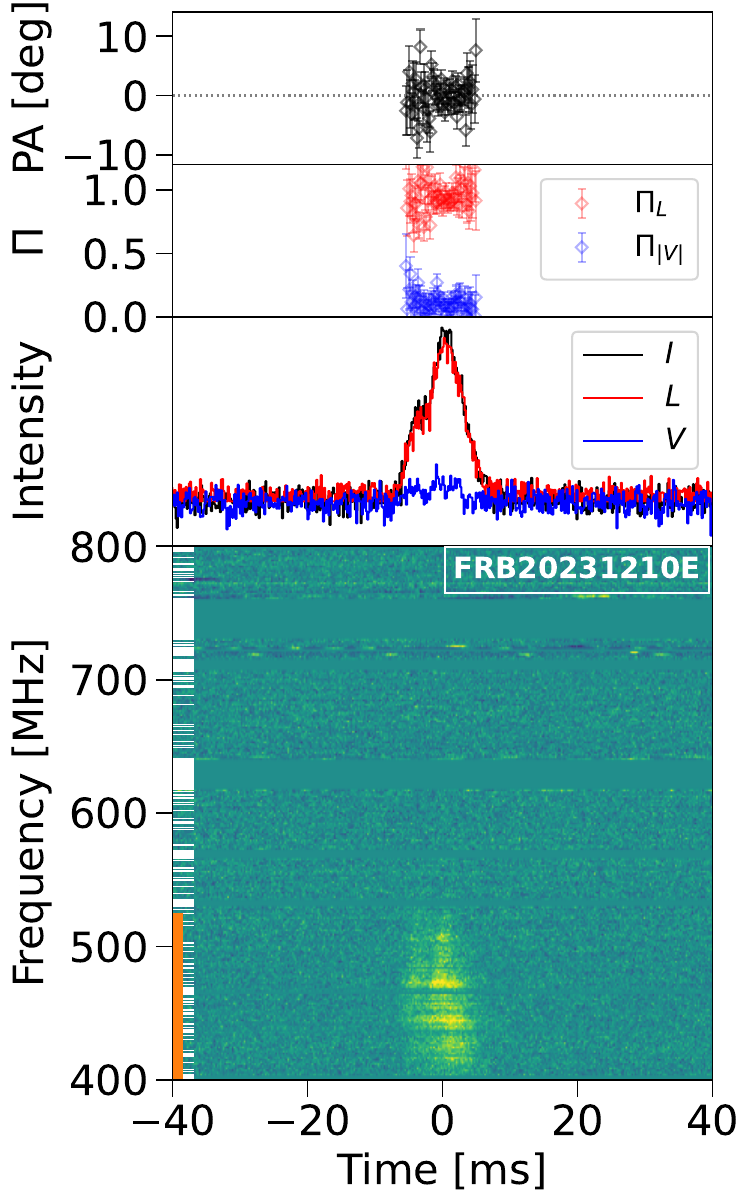}
   \includegraphics[width=0.188\textwidth]{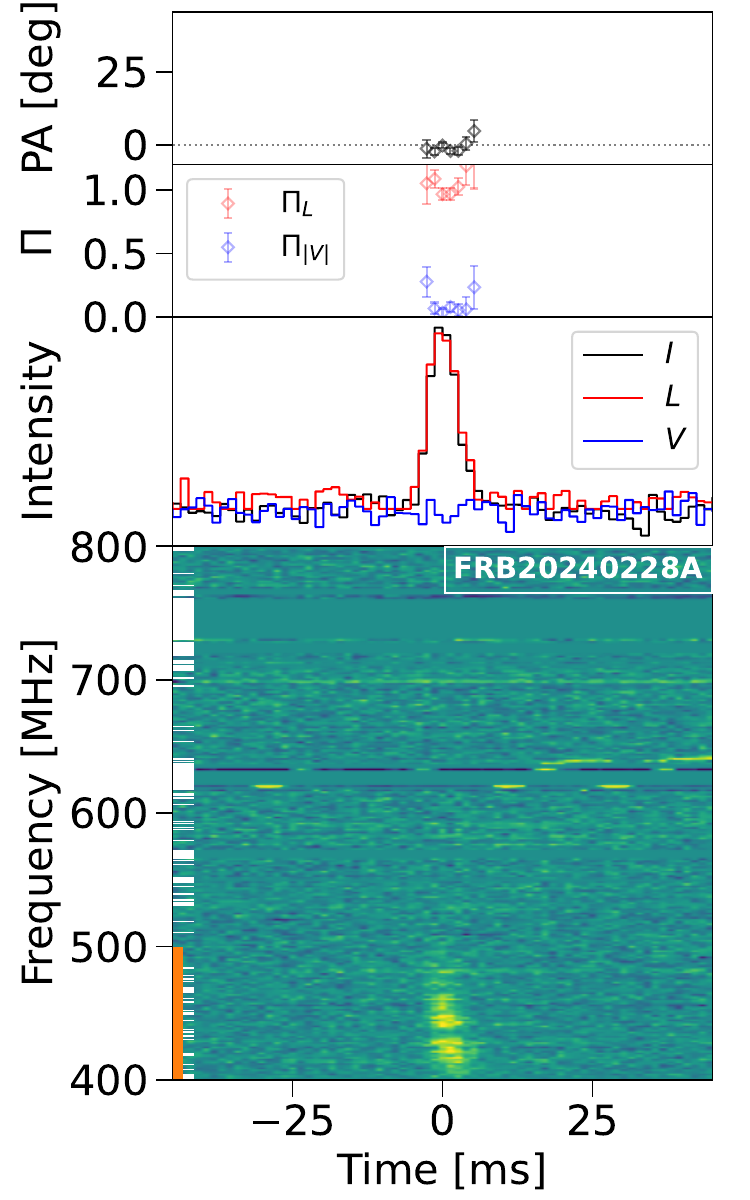}
   
    \includegraphics[width=0.188\textwidth]{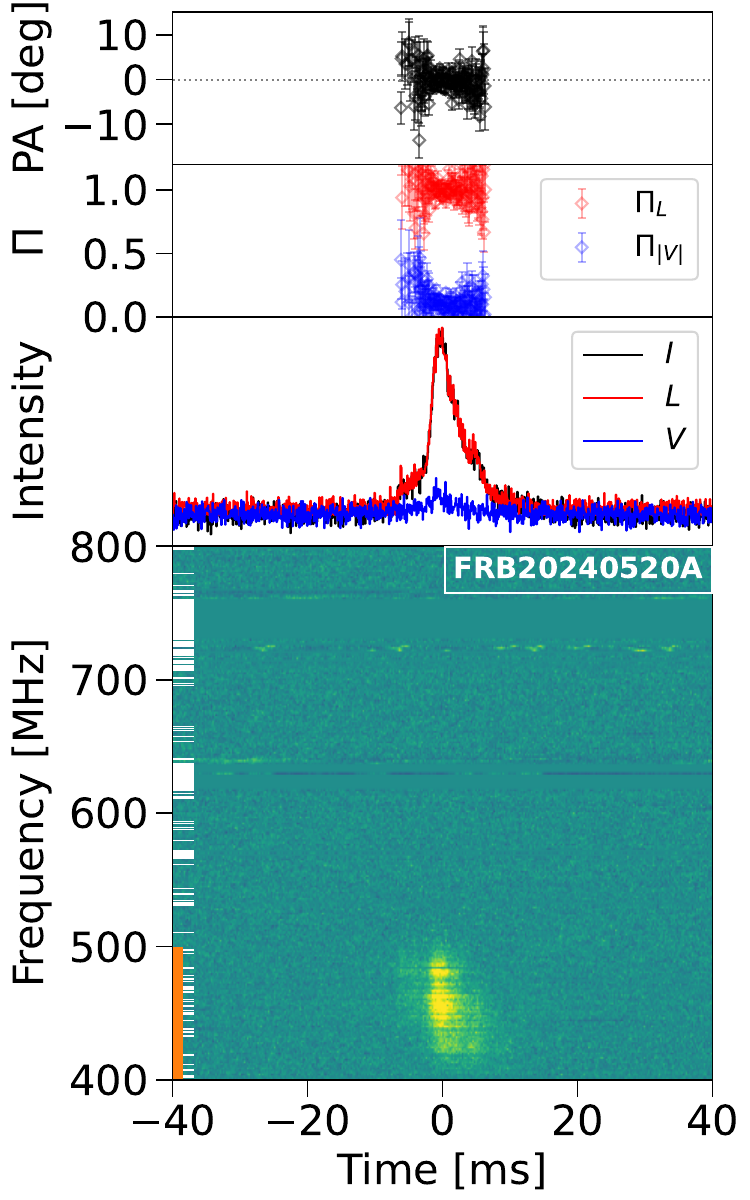}
    \includegraphics[width=0.188\textwidth]{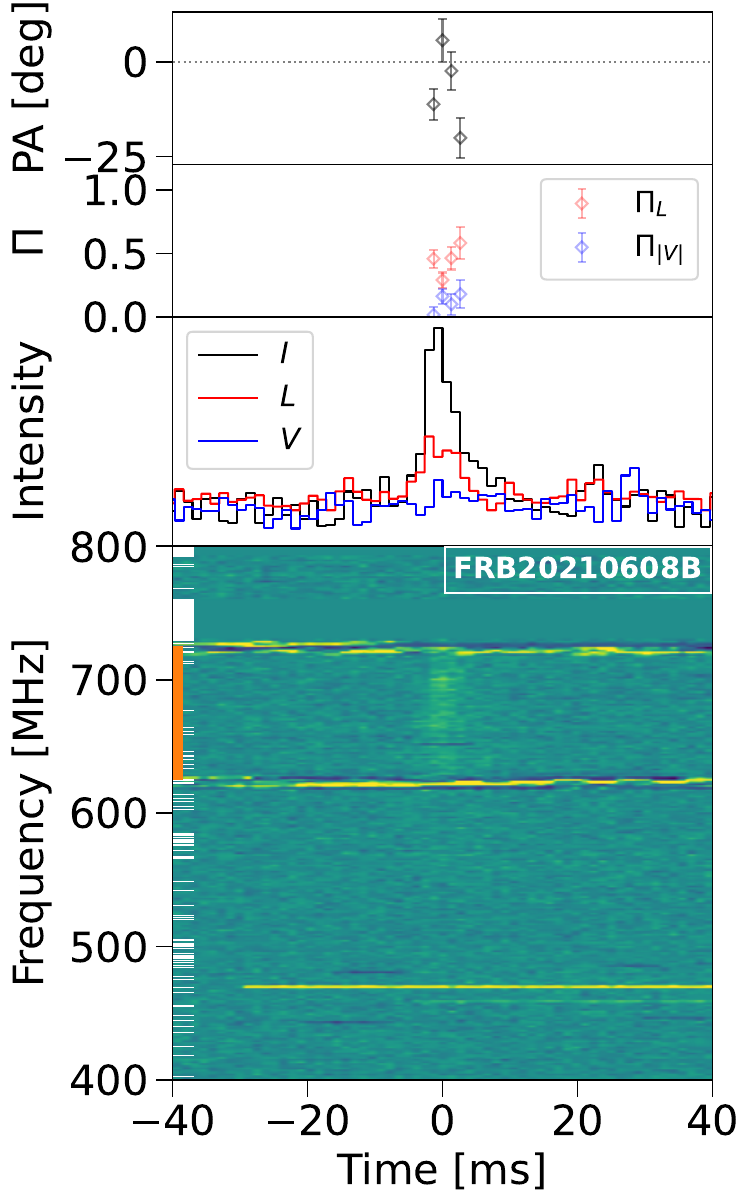}
    \includegraphics[width=0.188\textwidth]{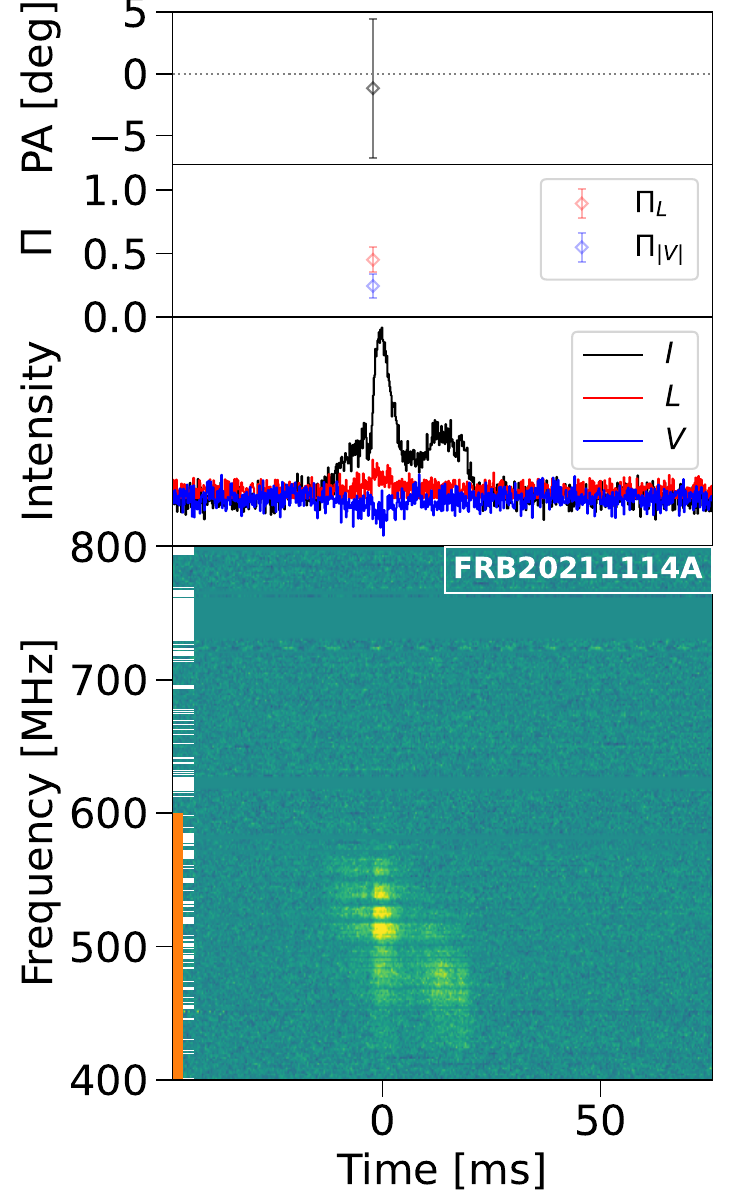}
    \includegraphics[width=0.188\textwidth]{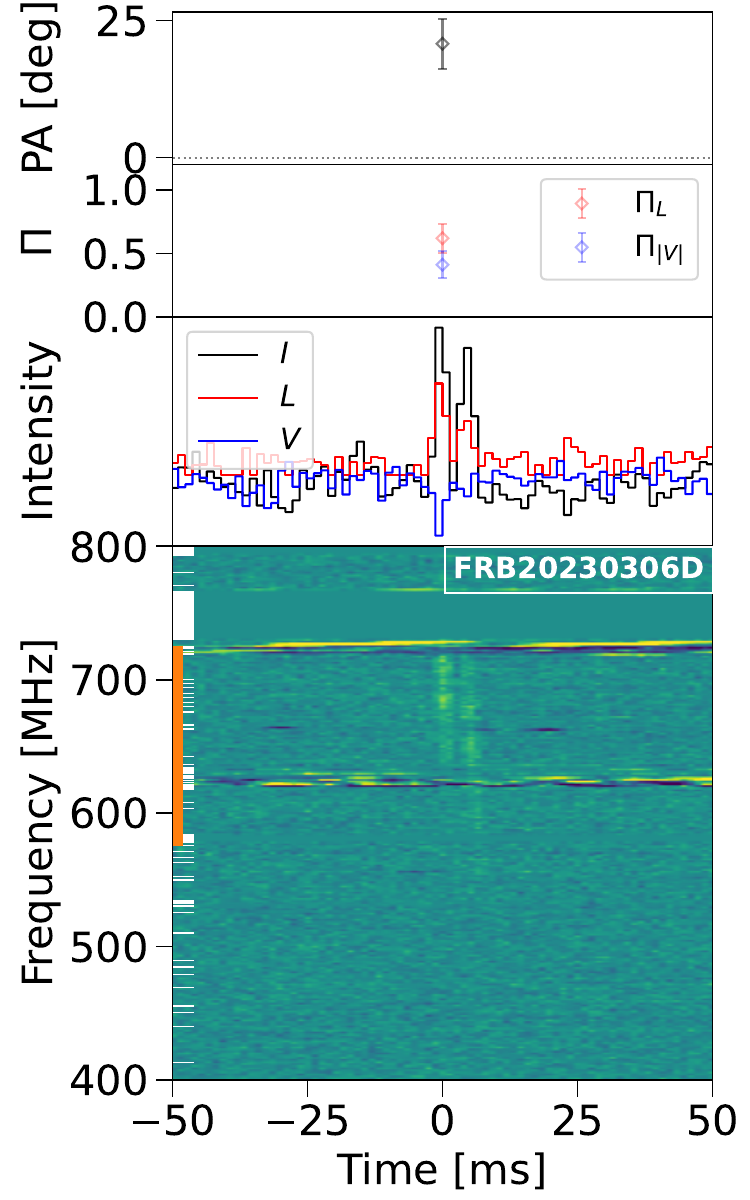}  
    \includegraphics[width=0.188\textwidth]{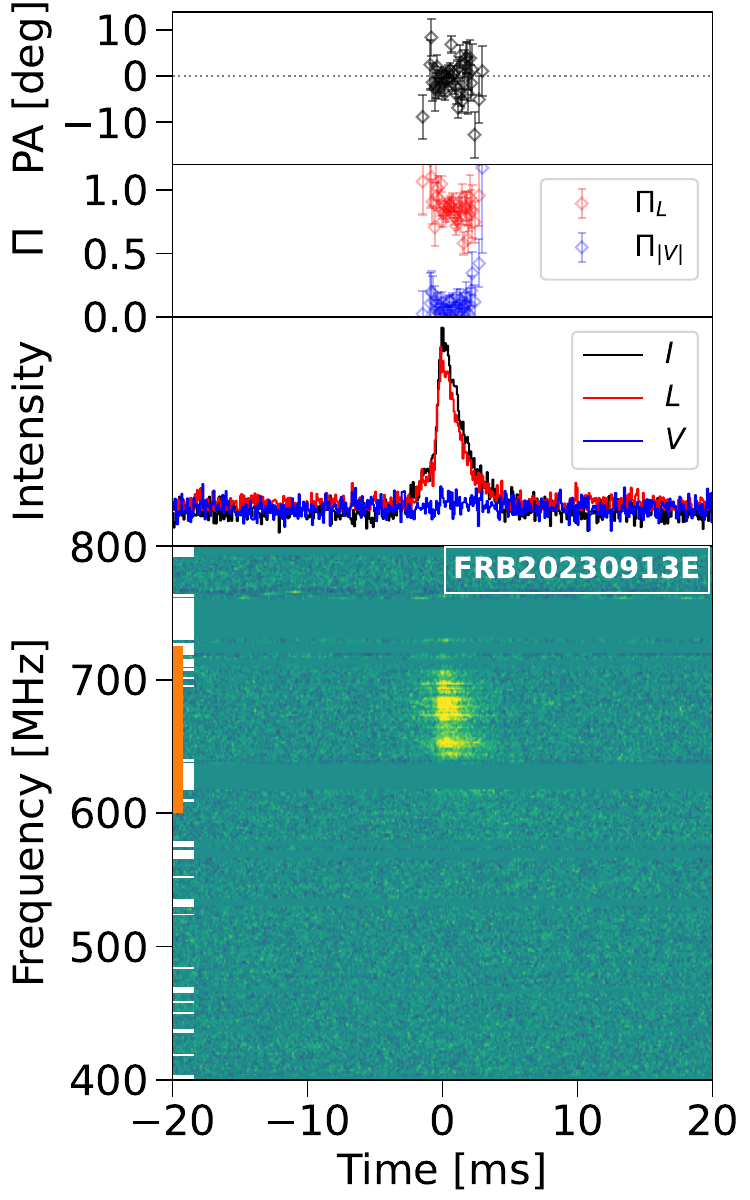}
    
    \includegraphics[width=0.188\textwidth]{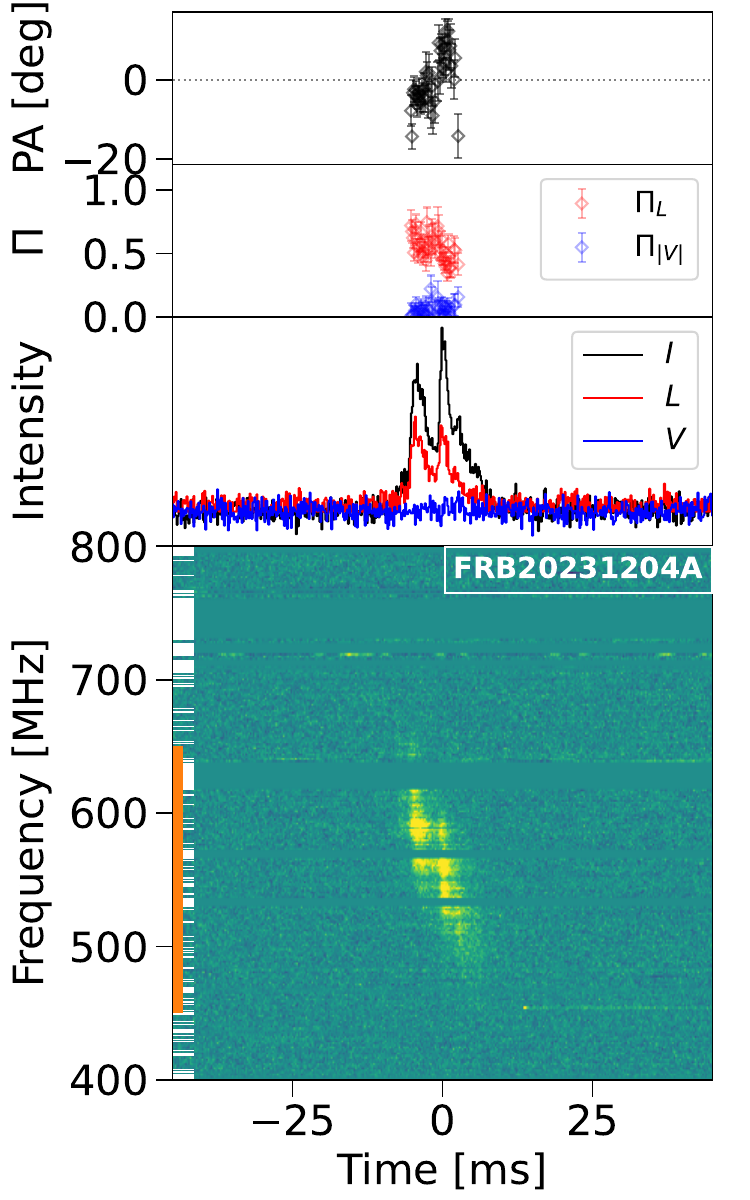}
    \includegraphics[width=0.188\textwidth]{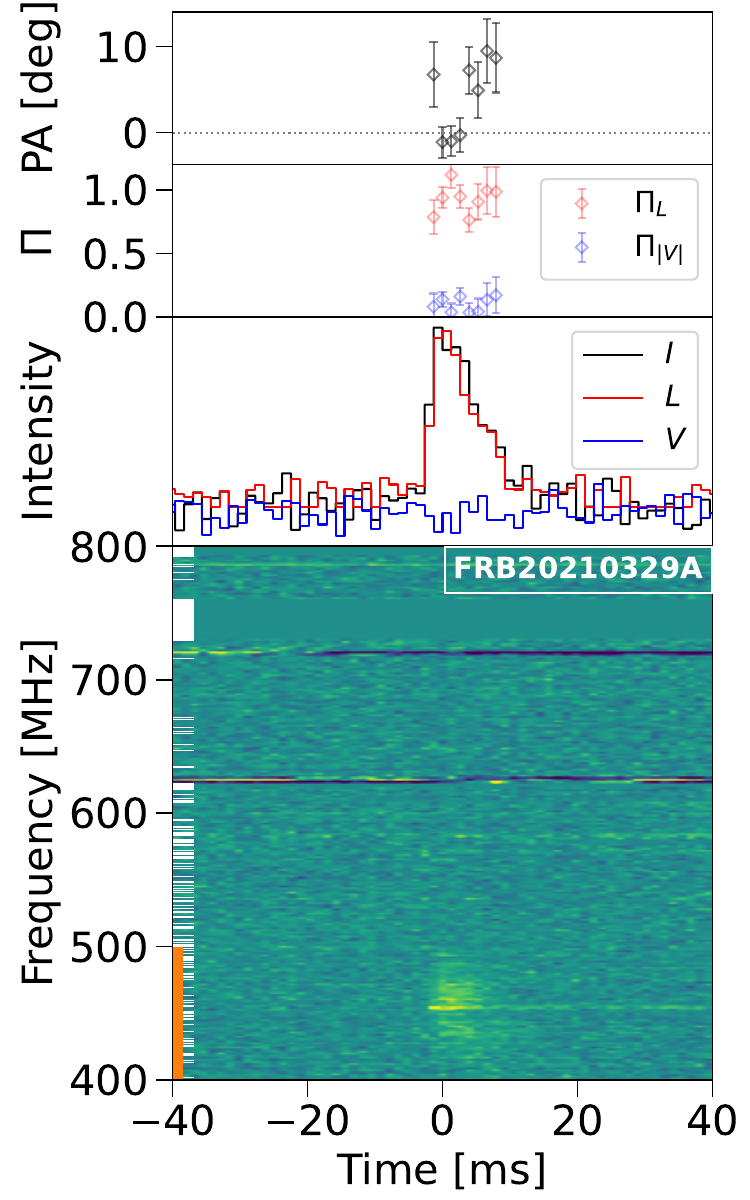}   
    \includegraphics[width=0.188\textwidth]{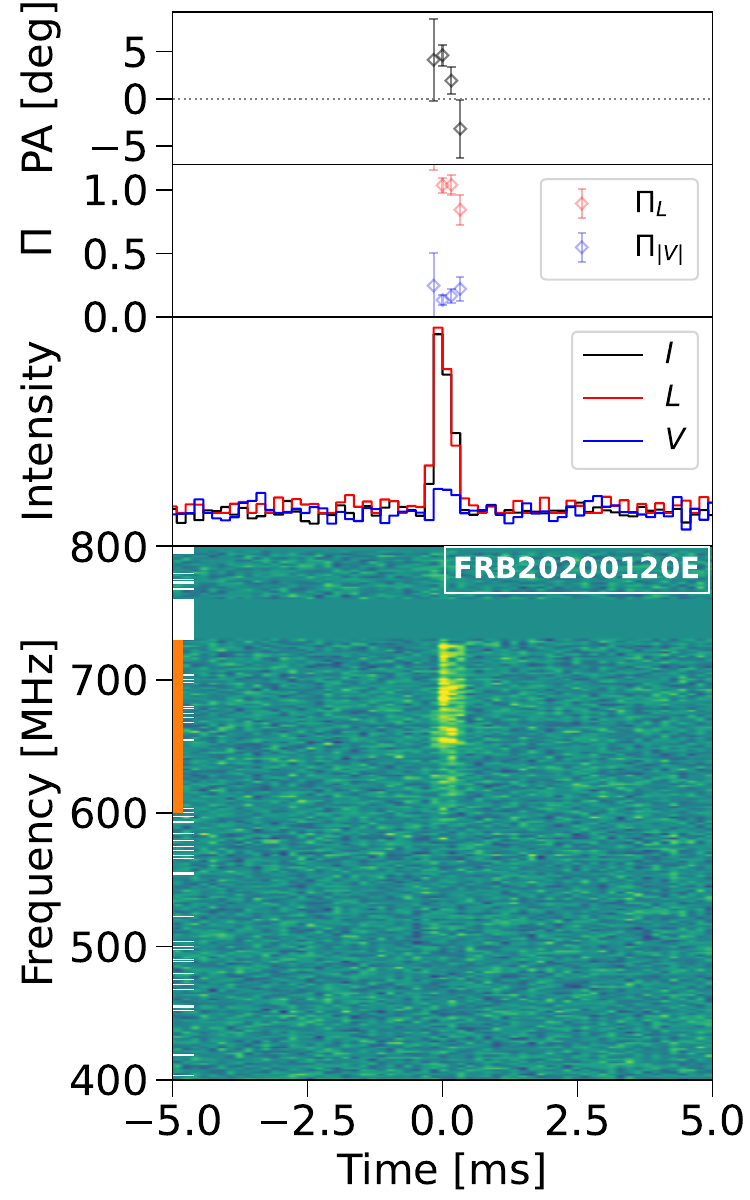}
    \includegraphics[width=0.188\textwidth]{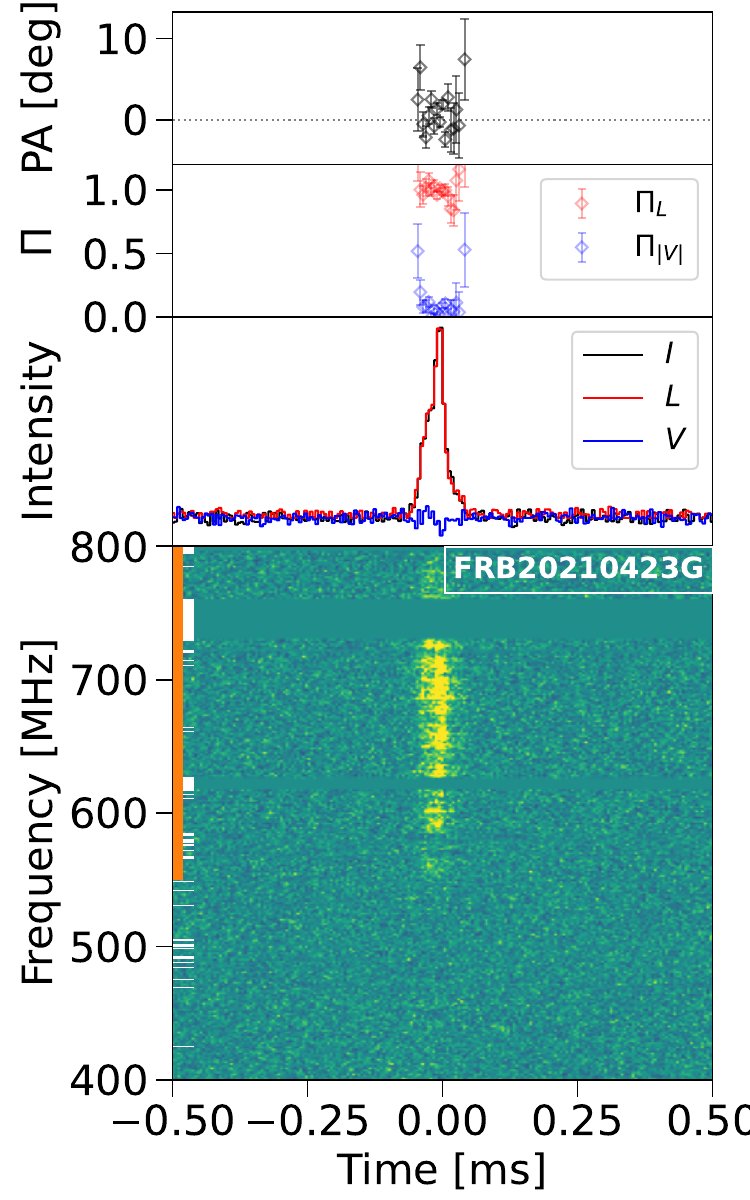}
    \includegraphics[width=0.188\textwidth]{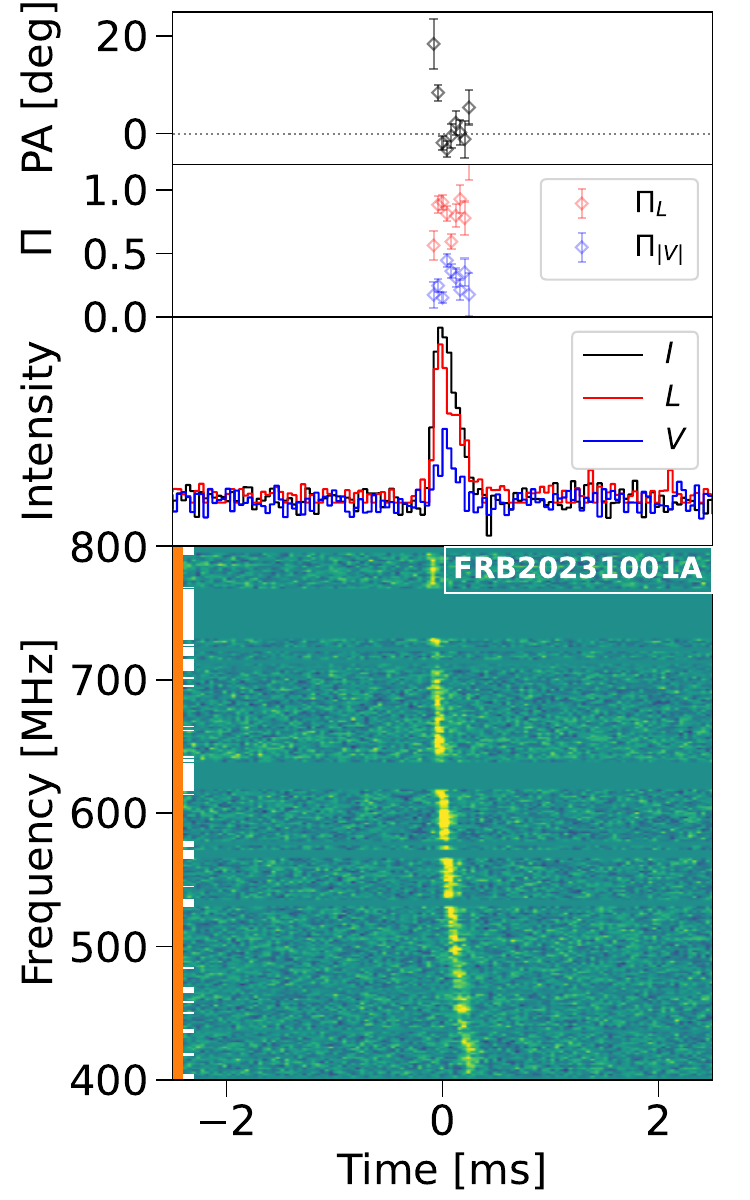}
         
\caption{Further intensity waterfalls. Refer to the caption of Fig.~\ref{fig:waterfalls} for information on the legends and annotations. }
    \label{fig:waterfalls3}
\end{center}
\vspace{0.3cm}
\end{figure*}

\begin{figure*}[ht!]
\begin{center}

    \includegraphics[width=0.188\textwidth]{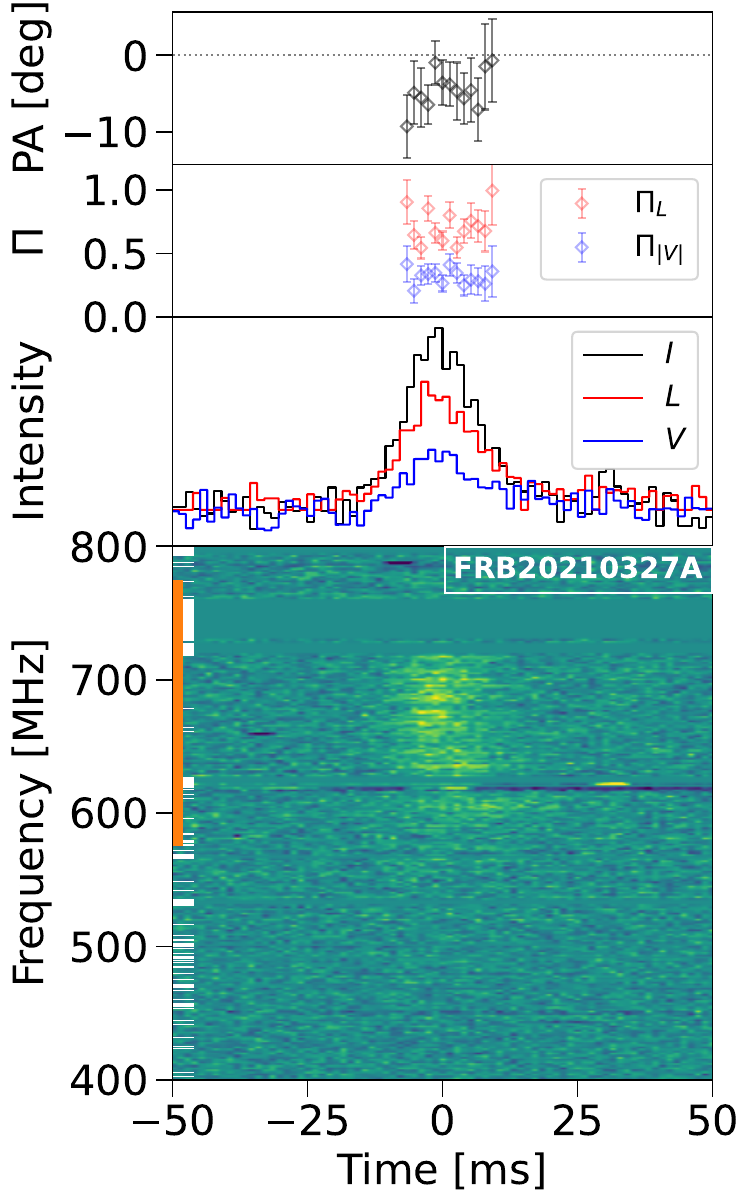}
    \includegraphics[width=0.188\textwidth]{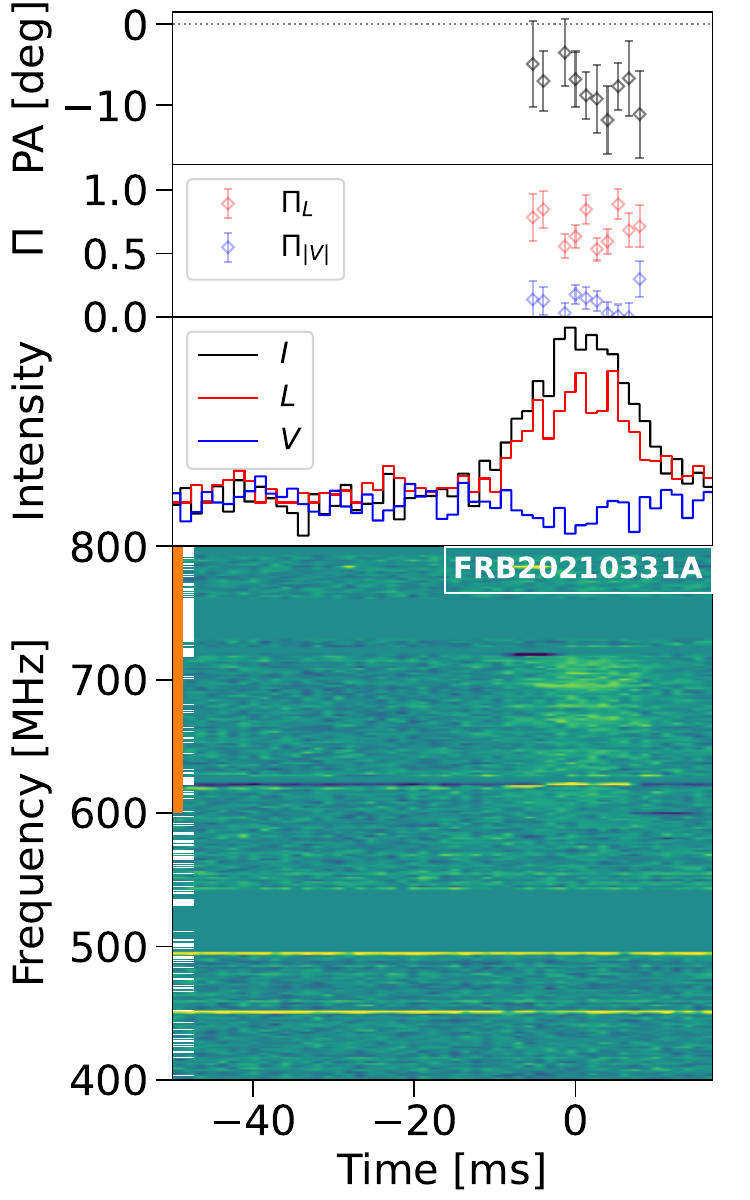} 
    \includegraphics[width=0.188\textwidth]{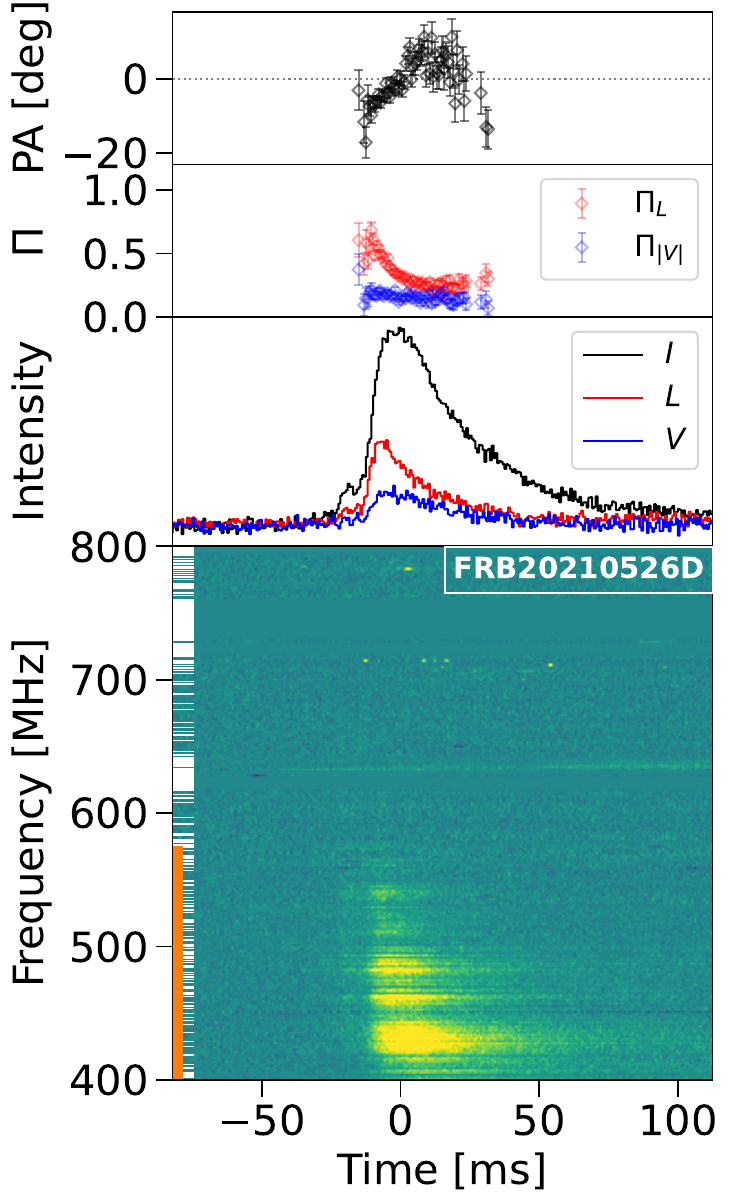}
\caption{Further intensity waterfalls. Refer to the caption of Fig.~\ref{fig:waterfalls} for information on the legends and annotations. }
    \label{fig:waterfalls4}
\end{center}
\end{figure*}

Table~\ref{tab:RN3pol} lists the measured and fitted burst properties from the baseband data for the repeaters initially discovered in CHIME23, while the same information for the newer bursts of the previously published repeaters can be found in Table~\ref{tab:RN1-2pol}.
Unconstrained parameters are shown as ``$-$". Uncertainties are reported at the 1-$\sigma$ confidence level.
The topocentric times-of-arrival (TOAs) provided in the third column are in Modified Julian Date (MJD) format, referenced at 400\,MHz with $\sim$1~second precision.
The DMs reported in the fifth column are obtained through structure-optimization fitting \citep[refer to][]{Curtin2024}. 
Certain bursts do not have significant S/N at our highest downsampling factor, and hence we are unable to determine a structure maximized DM. In these cases, we visually determine the DMs and do not report any uncertainties for these measurements in Tables~\ref{tab:RN3pol} and~\ref{tab:RN1-2pol}.
These cases are discussed in further detail in \citet{Curtin2024}. 
The downsampling factor in the sixth column provides the time resolution (i.e. $n_{\mathrm{down}}\times2.56\,\mu$s) of the dynamic spectrum (waterfall) plots displayed in Figs.~\ref{fig:waterfalls}-~\ref{fig:waterfalls4}.
The linear polarization fraction ($L/I$) in the seventh column is calculated by integrating $L$ over the burst profile. 
The upper limits on $L/I$ for unpolarized bursts are indicated by ``$<$" symbols.
It is possible that RM measurements with small magnitudes could be a result of instrumental effects, 
for example, leakage from Stokes I to Q induced by differential sensitivity of the X, Y polarized feeds \citep[see e.g., ][]{Mckinven2021}.
However, if the associated $L/I$ fraction is high and a burst signal is seen in both the Stokes Q and U waterfalls, then the RM value is unlikely to be purely instrumental (e.g., the case of burst \#1 (TNS name 20190609C) from FRB~20190609C). We are thus reporting small RM values only if the associated $L/I$ is high. 

Some bursts appear unpolarized despite semi-coherent RM searches up to $|$10$^{6}|$\,rad\,m$^{-2}$. As concluded by \citet{Mckinven2023b}, for repeaters with multiple bursts, it is likely that some events (typically also the ones with lower S/N) appear unpolarized because the polarized signal from those bursts is below the limits of our sensitivity.  
For these unpolarized bursts, we use the total intensity burst S/N to calculate the $L/I$ that would be required to produce a polarization detection at S/N = 5. This value is used as a conservative upper limit on $L/I$ in Table~\ref{tab:RN3pol}. 

Figs.~\ref{fig:waterfalls}--~\ref{fig:waterfalls4} are the frequency vs time waterfall plots for the 63 bursts that have well-constrained RM values. In the top two panels ($L/I$, $|V|/I$, and PA profiles) for each burst, only data points with $L$ S/N limits of $>$5 are shown.  
Some faint, low linear polarization FRBs thus can have very few (or no) points on their profiles.
For the circular polarization measurements, even though a first-order correction has been included to account for the instrumental effect of non-zero cable delay, there remain secondary effects (e.g., due to the frequency-dependent beam phase) that still result in residual instrumental circular polarization which could be of the order of $|V|/I\sim20\%$. 
Hence, the circular polarization information shown in Figs.~\ref{fig:waterfalls}--~\ref{fig:waterfalls4} represents only the relative changes and is not absolutely calibrated.

\setlength{\LTcapwidth}{0.9\textwidth}
\begin{longtable*}[l]{cccccccc}
\caption{Individual burst properties of repeating CHIME/FRB sources from the CHIME23 sample. We include the TNS name, the topocentric times-of-arrival referenced at 400\,MHz, the structure-optimized DM, the downsample factor in time for the waterfall plots in Figs.~\ref{fig:waterfalls}--~\ref{fig:waterfalls4}, the linear polarization fraction ($L/I$) integrated over the burst profile, as well as the RM from the FDF method. Uncertainties are reported at the 1-$\sigma$ confidence level.
The $L/I$ upper limits for unpolarized bursts are indicated by ``$<$" symbols and the corresponding unconstrained RMs are shown as ``$-$".
} 
\endfirsthead
\caption{\textit{continued}} \endhead
\hline
TNS names \hspace{0.5pt} & Arrival Date & Arrival Time & S/N & $\rm{DM_{struct}}$ & $\rm{n_{down}}$ & $\rm{\langle L/I \rangle}$ & $\rm{RM_{FDF}}$  \\
  & (YYYY/MM/DD) & (MJD) & & (pc~cm$^{-3}$) & &  & (rad m$^{-2}$)  \\
\hline
 \multicolumn{8}{c}{FRB~20171019A} \\  
 \hline
20200801C & 2020/08/01 & 59062.40088 & 18.2 & 455.0058(14) & 256 & 0.851(15) &  $-$2.0(3)  \\
\hline 
\multicolumn{8}{c}{FRB~20180910A} \\  
 \hline
20200621D & 2020/06/21 & 59021.04801 & 65.9 & 696.36(16) & 32 & 0.343(8) &  $-$340.37(16)  \\ 
 \hline
\multicolumn{8}{c}{FRB~20190110C} \\  
 \hline
20190110C &2019/01/10 & 58493.71640 & 23.9 & 222.03(3) & 64 & 0.93(2) &  $+$118.5(2)  \\ 
\hline
 \multicolumn{8}{c}{FRB~20190430C} \\  
 \hline
20190430C & 2019/04/30 & 58603.49572 & 66.5 & 400.35(16) & 16 & 0.913(19) &  $-$71.25(17)   \\ 
\hline
 \multicolumn{8}{c}{FRB~20190609C} \\  
 \hline
20190609C & 2019/06/09 & 58643.81885 & 22.4 & 479.8612(5) &  2  & 0.9337(15)&  $-$3.7(3) \\ 
20201030B & 2020/10/30 & 59152.42714 & 58.0 & 479.7960(7) &  1  & 0.966(2)  &  $-$39.46(2)  \\ 
20210113D & 2021/01/13 & 59227.22623 & 6.3 & 479.6  & N/A & $<$0.55(18)  &  $-$  \\ 
\hline
 \multicolumn{8}{c}{FRB~20190804E} \\  
  \hline
20200629C & 2020/06/29 & 59029.28288 & 23.5 & 362.77(4)  & 256 & 0.435(2) &  $-$200.75(14)  \\ 
20200709C & 2020/07/09 & 59039.25722 & 11.5 & 362.3      & 256 & 0.55(3)  &  $-$206.4(6)   \\ 
20201225B & 2020/12/25 & 59208.79707 & 27.3 & 362.724(5) & 1   & 0.991(12)&  $-$202.53(15)    \\ 
20201228A & 2020/12/28 & 59211.78653 & 21.2 & 362.96(2)  & 64  & 0.809(15)&  $-$196.0(2) \\ 
20220203A & 2022/02/03 & 59613.68909 & 8.7  & 362(13)    & N/A & $<$0.51(12) &  $-$        \\ 
\hline
 \multicolumn{8}{c}{FRB~20190915D} \\  
  \hline
20200214B & 2020/02/14 & 58893.96581 & 22.1 & 487.6(3) & 256 & 0.274(12) &  $-$170.3(4) \\ 
\hline
 \multicolumn{8}{c}{FRB~20191013D} \\  
 \hline
20200515A & 2020/05/15 & 58984.79805 & 24.3 & 522.6(3)  & 256 & 0.357(13) &  $-$35.7(2)  \\ 
\hline
\multicolumn{8}{c}{FRB~20191106C} \\  
 \hline
20201201A & 2020/12/01 & 59184.69025 & 41.9 & 330.6   & 256 & 0.172(5)    &  $-$263.3(2) \\ 
20210212C & 2021/02/12 & 59257.48754 & 7.0  & 330.78  & N/A & $<$0.33(15) &  $-$ \\ 
20210617A & 2021/06/17 & 59382.14621 & 38.5 & 330.6   & 256 & 0.209(2)    &  $-$468.5(2)  \\ 
20210822A & 2021/08/22 & 59448.96952 & 22.1 & 331.2   & 256 & $<$0.525(7)    &  $-$  \\
20211104B & 2021/11/04 & 59522.76580 & 28.0 & 331.60(16) & 256 & 0.186(9) &  $-$921.8(3)   \\ 
20220118B & 2022/01/18 & 59597.56025 & 59.5 & 330.56(9)  & 64  & 0.219(4)    &  $-$1044.43(17)  \\
20220514C & 2022/05/14 & 59713.24328 & 21.9 & 331.38(11) & N/A & $<$0.27(4)  &  $-$            \\
\hline
 \multicolumn{8}{c}{FRB~20200118D} \\  
 \hline
20200118D & 2020/01/18& 58866.30273 & 21.7 & 625.22(2) & 64 & 0.258(2)  &  $+$125.4(8) \\ 
20200701A & 2020/07/01 &59031.84983 & 40.8 & 625.28(2) & 16 & 0.696(13) &  $+$132.6(3)\\ 
\hline
 \multicolumn{8}{c}{FRB~20200127B} \\  
 \hline
20200127B &2020/01/27 & 58875.83372 & 70.3  & 351.3 & 2 & 0.703(10) & $+$39.32(5)\\ 
20200219B &2020/02/19 & 58898.76908 & 117.1 & 351.3 & 1 & 0.909(7) & $+$40.58(4) \\ 
\hline
 \multicolumn{8}{c}{FRB~20200202A} \\  
 \hline
20201014B & 2020/10/14 & 59136.34382 & 30.8 & 726.44(7) & 32 & 0.833(9) &  $+$52.6(2) \\ 
20230604B & 2023/06/04 & 60099.70443 & 40.5 & 734.3(2) & 16 & 0.230(9) & $+$61.5(10) \\ 
\hline
 \multicolumn{8}{c}{FRB~20200223B} \\  
 \hline
20200702C & 2020/07/02 & 59032.57511 & 163.9 & 200.354(10)  & 8  & 0.182(2) &  $+$70.4(4)  \\ 
20210115C & 2021/01/15 & 59229.04023 & 45.9  & 200.521(10)  & N/A & $<$0.328(8) &  $-$  \\ 
\hline
 \multicolumn{8}{c}{FRB~20200619A} \\  
  \hline
 20201022D & 2020/10/22 & 59144.00771 & 6.9    & 439(23) & N/A & $<$0.31(15) &  $-$ \\ 
 20210130E & 2021/01/30 & 59244.72817 & 10.0   & 440.2 & N/A & $<$0.21(10) &  $-$ \\ 
\hline
 \multicolumn{8}{c}{FRB~20200809E} \\  
 \hline
20200809E & 2020/08/09 & 59070.49237 & 23.7 & 1702.89(9) & 128 & 0.70(2) &  $-$40.6(6)  \\ 
20201018C & 2020/10/18 & 59140.32924 & 21.2 & 1702.85(3) & 128 & 0.80(2) &  $-$36.5(3)   \\ 
20210208B & 2021/02/08 & 59253.00241 & 14.6 & 1703.35(2) & 256 & 0.58(3) &  $-$36.3(5)  \\ 
\hline
 \multicolumn{8}{c}{FRB~20200926A} \\  
 \hline
20201223A & 2020/12/23 & 59206.86524 & 6.8  & 759.3 & N/A & $<$0.51(16) &  $-$ \\ 
20211206A & 2021/12/06 & 59554.91389 & 4.0  & 759.4 & N/A & $<$0.86(5)  &  $-$  \\
20230725D & 2023/07/25 & 60150.27956 & 52.4 & 758.62(3) & 32  & 0.266(7) &  $+$274.1(4)\\ 
 \hline
 \multicolumn{8}{c}{FRB~20200929C} \\  
  \hline
20201125B & 2020/11/25 & 59178.20015 & 63.9 & 413.66(6) & 16 & 0.879(8)  &  $-$4.85(9) \\ 
20201203C & 2020/12/03 & 59186.17777 & 20.0 & 413.54(3) & 128 &0.65(2)   &  $-$8.4(4)   \\ 
20210313B & 2021/03/13 & 59286.90499 & 49.3 & 413.67(10) & 64  & 0.906(7) &  $+$9.23(11)  \\ 
20210314A & 2021/03/14 & 59287.90333 & 26.1 & 413.54(17) & 128 & 0.897(16)&  $-$1.55(19)  \\ 
20210326B & 2021/03/26 & 59299.86586 & 50.0 & 413.74(12) & 32  & 0.874(7) &  $+$11.34(7)  \\ 
20210930A & 2021/09/30 & 59487.35525 & 30.0 & 413.43(19) & 256 & 0.531(8) &  $-$42.37(13) \\ 
20220209A & 2022/02/09 & 59619.99073 & 36.9 & 413.20(6)  & 32  & 0.908(11)&  $-$12.25(11)  \\ 
\hline
 \multicolumn{8}{c}{FRB~20201114A} \\  
 \hline
20201219A & 2020/12/19 & 59202.70239 & 31.1 & 321.31(13) & 128 & 0.541(12) &  $+$1348.7(3)  \\ 
\hline
\multicolumn{8}{c}{FRB~20201130A} \\  
 \hline
20201225D  & 2020/12/25 & 59208.24932 & 26.7 & 287.6(5)   & 256 &  0.475(13) &  $+$183.62(16)  \\ 
20210114B  & 2021/01/14 & 59228.19620 & 38.6 & 288.2(2)   & 128 &  0.618(13) &  $+$183.53(15)  \\ 
20210117E  & 2021/01/17 & 59231.18612 & 7.4  & 287.82(10) & 256 &  0.46(2)   &  $+$188.9(6)   \\ 
20210118B  & 2021/01/18 & 59232.18424 & 41.0 & 287.9(5)   & 128 &  0.633(11) &  $+$184.01(13)  \\ 
20210327F  & 2021/03/27 & 59300.00032 & 39.7 & 287.67(11) & 256 &  0.430(11) &  $+$182.94(16)  \\ 
\hline
 \multicolumn{8}{c}{FRB~20201221B} \\  
\hline
20210224A & 2021/02/24 & 59269.24936 & 4.1 & 510.0  & N/A & $<$0.6(2)     &  $-$ \\ 
20210302E & 2021/03/02 & 59275.24211 & 19.6 & 510.5 & 256 & 0.455(13) &  $-$1.7(2) \\ 
20210303F & 2021/03/03 & 59276.23157 & 38.2 & 509.71(14) & N/A & $<$0.155(8)  &  $-$  \\ 
\hline
\label{tab:RN3pol}
\end{longtable*}

\begin{longtable*}[l]{cccccccc}
\endlastfoot
\caption{Updated burst properties of repeating FRB sources with previously published polarization information, with the same columns as described in the caption of Table~\ref{tab:RN3pol}. } 
\endfirsthead
\caption{\textit{continued}} \endhead
\hline
TNS names \# \hspace{0.5pt} & Arrival Date & Arrival Time & S/N & $\rm{DM_{struct}}$ & $\rm{n_{down}}$ & $\rm{\langle L/I \rangle}$ &  $\rm{RM_{FDF}}$ \\
  & (YYYY/MM/DD) & (MJD) & & (pc~cm$^{-3}$) & & & (rad m$^{-2}$)  \\
\hline
\hline \multicolumn{8}{c}{FRB~20180916B} \\   
 \hline
20220312A & 2022/03/12 & 59650.93438 & 15.0 & 348.51(10)  & 64 & 0.64(2)   &  $-$54.1(23)  \\ 
20220328A & 2022/03/28 & 59666.89952 & 27.1 & 348.96(7)   & 16 & 0.868(18) &  $-$56.6(5)    \\
20220619B & 2022/06/19 & 59749.66674 & 56.3 & 348.986(11) & 2  & 0.945(6)  &  $-$53.30(12)  \\
20221215G & 2022/12/15 & 59928.17600 & 32.1 & 349.25(2)   & 64 & 0.888(12) &  $-$59.80(17)\\
20230510G & 2023/05/10 & 60074.77908 & 45.4 & 348.95(12)  & 64 & 0.834(9)  &  $-$54.49(7)\\
20231108A & 2023/11/08 & 60256.29047 & 21.9 & 348.89(2)   & 32 & 0.716(19) &  $-$53.8(3)   \\
20231123D & 2023/11/23 & 60271.24684 & 9.4  & 349.44(6)   &256 & 0.74(3)   &  $-$51.8(4)  \\
20231210E & 2023/12/10 & 60288.18768 & 62.1 & 349.4       &32  & 0.866(7)  &  $-$57.94(11)\\ 
20240228A & 2024/02/28 & 60368.98383 & 15.1 & 349.11(18)  &256 & 0.81(2)   &  $-$52.2(2) \\
20240520A & 2024/05/20 & 60450.74919 & 69.7 & 349.02(8)  & 16 & 0.968(6)  &  $-$56.14(6) \\ 

\hline \multicolumn{8}{c}{FRB~20181119A} \\   
\hline
20210608B & 2021/06/08 & 59373.14569301 &  13.8 & 364.939(15) & 256 & 0.274(14) & $+$355.2(11) \\ 
\hline
 \multicolumn{8}{c}{FRB~20190117A} \\  
 \hline
20211114A & 2021/11/14 & 59532.10691 & 9.9  & 395.89(4) & 256 & 0.476(19) &  $+$28.9(2)  \\
\hline
 \multicolumn{8}{c}{FRB~20190208A} \\  
 \hline
20230306D & 2023/03/06 & 60009.66716 & 8.5  & 579.81(18) & 256 & 0.460(18) &  $+$26.3(10) \\
 \hline
 \multicolumn{8}{c}{FRB~20190303A} \\  
 \hline
20230913E & 2023/09/13 & 60200.93406 & 47.8 &  221.48(6) & 16  & 0.805(7) & $+$371.7(3)  \\
20231204A & 2023/12/04 & 60282.70886 & 67.0 & 221.24(3) & 32  & 0.490(14)&  $+$287.8(2)\\ 
 \hline
 \multicolumn{8}{c}{FRB~20190604A} \\  
 \hline
20210329A & 2021/03/29 & 59302.42051 & 11.0 & 552.4(5) & 256 & 0.37(2) &  $-$14.6(4)  \\
 \hline
 \multicolumn{8}{c}{FRB~20200120E} \\  
 \hline
20200120E & 2020/01/20 & 58868.41500  & 17.0 & 87.85(10) & 64  & 0.95(2)   &  $-$29.8(2)    \\
20210423G & 2021/04/23 & 59327.15873  & 70.1 & 87.760(10) & 1   & 0.993(7)  &  $-$28.78(14)\\
20231001A & 2023/10/01 & 60218.72630  & 42.0 & 87.7 & 8   & 0.75(2)   &  $-$24.45(13) \\
 \hline
 \multicolumn{8}{c}{FRB~20201124A} \\  
 \hline
20210327A & 2021/03/27 & 59300.03220  & 30.8 & 416.5 & 256 & 0.263(9) &  $-$543.6(3) \\
20210331A & 2021/03/31 & 59304.02299  & 25.6 & 416.2 & 256 & 0.439(9) &  $-$576.3(3) \\
20210526D & 2021/05/26 & 59360.86503  & 87.2 & 410.8 & 64  & 0.266(2) &  $-$602.01(5) \\
 \hline
\label{tab:RN1-2pol}
\end{longtable*}

\subsection{Individual repeaters} \label{sec:IndivdualRepeaters}
\begin{figure}[ht]
    \centering
    \includegraphics[width=3.3in]{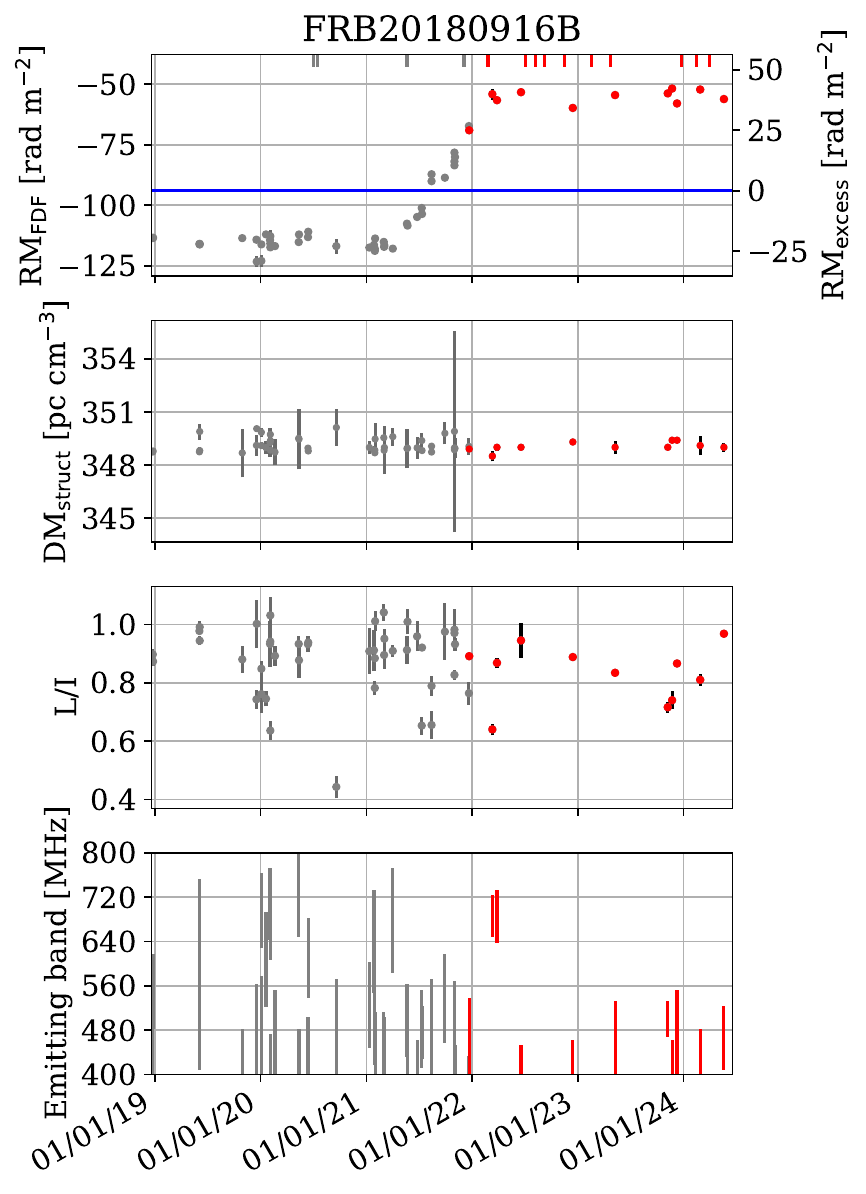}
    \caption{Burst properties ($\rm{RM_{FDF}}$, $\rm{DM_{struct}}$, $L/I$ and emission bandwidth) as a function of time for FRB~20180916B. The gray points correspond to data previously published by \citet{Mckinven2023b}, whereas the red ones are new results from this work.The small ticks on the top of the top panel show epochs when baseband data were recorded but no significant polarization measurements were detected. Error bars correspond to 1-$\sigma$ for the RM and $L/I$ panels and 3-$\sigma$ for the DM panel.}
        \label{fig:temporalR3}
\end{figure}

\textbf{FRB~20180916B}: A periodic ($\sim$16.35\,days) activity window was found for this FRB \citep{chime/frb2020b}. \citet{Mckinven2023b} have reported temporal RM variations that are unrelated to the 16.35-day cycle, showing stochastic behavior between January 2019 and January 2021, and then a secular increase until early 2022. In our extended data set of FRB~20180916B taken between January 2022 and May 2024, 
CHIME recorded 26 baseband events, of which significant polarization measurements can be obtained for 10 events that are bright enough, as listed in Table~\ref{tab:RN1-2pol}. 
From these, we report that the RM of FRB~20180916B once again shows a stochastic trend through to mid 2024 (Fig.~\ref{fig:temporalR3}). A similar trend was reported by \citet{Bethapudi2024} during the preparation of this work.

This temporal variation pattern does not support the long-orbit binary models proposed by \citet{Wang2022}, \citet{Zhao2023} and \citet{Lan2024} being the reason for this observed polarization fluctuation, 
particularly since no clear RM sign change nor turning point is seen during the five years of observations, which argues against a potential magnetic field reversal caused by the orbital motion of the binary star along the line of sight. 
Although this does not preclude FRB~20180916B to be in a binary system.
We note that the expected Galactic RM contribution at the position of FRB~20180916B is quite uncertain, with RM$_{\mathrm{MW}}$=$-$94$\pm$45\,rad\,m$^{-2}$.
If the true Galactic contribution is $<-$59\,rad\,m$^{-2}$, then our latest observations would indicate a RM sign change happened in 2022. This may or may not indicate a magnetic field reversal. The ambiguity is due to the fact that we cannot distinguish the fractional RM contribution from the host ISM.

\begin{figure}[ht]
    \centering
    \includegraphics[width=3.3in]{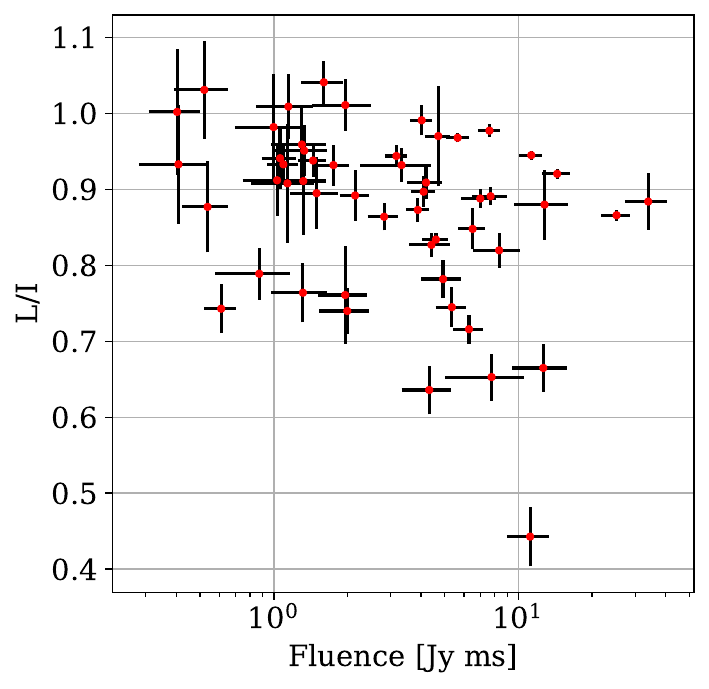}
    \caption{The $L/I$ vs fluence for FRB~20180916B. Statistically these two parameters are uncorrelated, with a Spearman coefficient $p$-value of 0.07.}
        \label{fig:R3-ML-Fl}
\end{figure}

We see no correlation between the RM variation and other burst parameters, namely emission bandwidth, $L/I$ and DM. 
Note that the emitting bandwidth shown in Fig.~\ref{fig:temporalR3} represents a by-eye estimation of the portion of the 400$-$800\,MHz CHIME band over which emission is observed, uncorrected for the non-uniform bandpass of the instrument.
The emission bandwidth appears to drift to lower frequencies in more recent observations, as was already noted by \citet{Mckinven2023a}, although correlated changes are not observed in the other parameters.
Fig.~\ref{fig:R3-ML-Fl} shows the $L/I$ vs fluence for the bursts we have detected from FRB~20180916B. We do not see any obvious correlation between the two parameters, with a Spearman correlation coefficient $p$-value of 0.07. There is perhaps a wider spread of $L/I$ values at the higher fluence end, which is likely due to the fact that it is easier to detect weakly polarized signals in brighter events.

\textbf{FRB~20181119A}: 
Three baseband polarization measurements (RMs of $+$606.2(1.2), $+$1343(2), $+$480.9(5)\,rad\,m$^{-2}$) from FRB~20181119A have already been published in \citet{Mckinven2023a}. In this paper, we include a fourth event (RM of $+$355.2(11)\,rad\,m$^{-2}$) during the time period of June 2020 to June 2021. The RMs vary by hundreds of units during the time. The cable delay for some of these events are quite large ($\sim$19\,ns), which introduced some uncertainty in the signs of the RMs, although the positive RM fits showed slightly higher likelihood. 

\textbf{FRB~20190117A}: Eight baseband events were recorded between June 2019 and May 2023.  Only one burst with significant RM detection is shown in Table~\ref{tab:RN1-2pol}, while the remaining bursts have no constraining measurements likely due to low S/N.

\begin{figure*}[ht]
    \centering
    \includegraphics[width=0.98\textwidth]{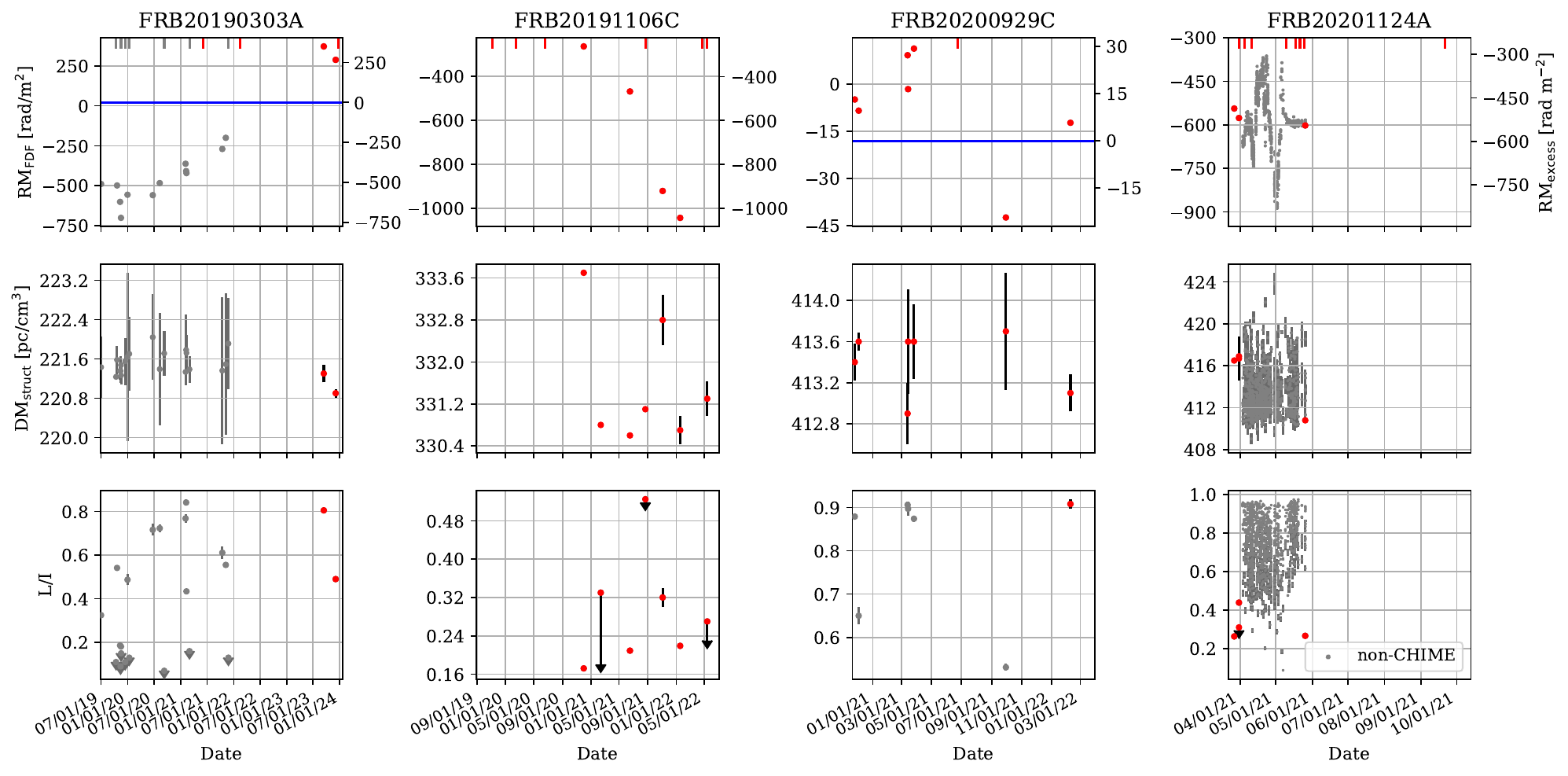}
    \caption{Burst properties ($\rm{RM_{FDF}}$, $\rm{DM_{struct}}$, and $L/I$) as a function of time for (far left) FRB~20190303A, (left)  FRB~20191106C, (right) FRB~20200929C and (far right) FRB~20201124A. 
    The gray points are data from the literature, including measurements from  \citet{Mckinven2023b,Kumar2022,Hilmarsson2021b,Xu2022}.
    Red points are new results from this paper.
    The error bars in the RM and $L/I$ panels represent the 1-$\sigma$ uncertainty range, and 3-$\sigma$ uncertainty range in the DM panel.
    Downwards arrows in the $L/I$ panel indicate upper limits. 
    The times of arrival (TOA) of the bursts with unconstrained RM are indicated as small ticks on the top of the top panel. The blue horizontal lines in the top panel indicate the level of RM=0\,rad\,m$^{-2}$ for the two FRBs with RM sign change.} 
        \label{fig:temporal}
\end{figure*}

\textbf{FRB~20190303A}: 
This FRB has been localized to a merging galaxy pair \citep{Michilli2023}.
Five baseband events were recorded between June 2021 and September 2023. Only 2 bursts with significant RM detections are shown in Table~\ref{tab:RN1-2pol}. Comparing these new measurements to the ones published by \citet{Mckinven2023b} (see the left panel column in Fig.~\ref{fig:temporal}), FRB~20190303A shows a secular temporal RM trend where the RM values increase several hundreds of units. Given that the Galactic RM contribution at this position is $+$21$\pm$5\,rad\,m$^{-2}$, we can be certain that a sign change (from negative to positive) of the RM took place between 2022 and 2023.

\textbf{FRB~20191106C}: This is a relatively active repeater with 11 baseband events between November 2019 and May 2022. Four events were not recorded properly hence only seven are presented in Table~\ref{tab:RN3pol} and  Fig.~\ref{fig:temporal}. There are significant RM variations of several hundred RM units, showing a clear secular trend. 
There is a low level of linear polarization (red line) in all available waterfalls. 

\textbf{FRB~20200120E}: This is the FRB in a globular cluster associated with the galaxy M81 \citep{Bhardwaj2021,Kirsten2021}. We have 8 events with baseband data recorded between January 2022 and October 2023. Only 3 bursts with significant RM detections are shown in Table~\ref{tab:RN1-2pol}. One of these was previously announced by \citet{Bhardwaj2021} but not with complete polarization information. We note that the RMs of this source do not vary a lot ($\sigma$(RM)=2.0\,rad\,m$^{-2}$) throughout the few years of data we have, which is consistent with the expectation from the sparse ISM environment of a globular cluster.

\textbf{FRB~20200929C}: All seven bursts from this repeater have low $|$RM$|$ values, although given the relatively high $L/I$ fraction, the detected RM is unlikely to be (purely) instrumental. Given that the Galactic RM contribution at this position is $-$18$\pm$3\,rad\,m$^{-2}$, we can say that there are at least two RM sign changes in the corrected, extragalactic RM component between November 2020 and February 2022 (see third panel column in Fig.~\ref{fig:temporal}). 

\textbf{FRB~20201124A}: 15 events with baseband data were recorded for this repeater between March and September 2021. Some of these events were described by \citet{Lanman2022} although polarization information was not included. 
Only the three bursts with significant RM detections are shown in Table~\ref{tab:RN1-2pol}. CHIME/FRB has not detected this repeater after September 2021 despite regular operations, but if we combine the CHIME/FRB measurements with those in the literature \citep{Kumar2022,Hilmarsson2021b,Xu2022}, we see that the CHIME/FRB measurements are consistent with those of the other telescopes, showing secular temporal variations where the RMs fluctuate for hundreds of units (see right panel column of Fig.~\ref{fig:temporal}).
The CHIME/FRB data seem to have a lower $L/I$ as compared to the other telescopes, which might be because these other facilities observe at higher frequency than CHIME/FRB and their data suffer less Faraday depolarization.
We also note that burst~\#3 (TNS name 20210526D) shows possibly varying $L/I$ as a function of time during the burst (last panel in Fig.~\ref{fig:waterfalls4}).

\textbf{FRB~20201130A}: Six baseband events were recorded between December 2020 and March 2021, although one of them was not saved properly and thus cannot be analyzed. The RM of this source is stable at around 180$-$190\,rad\,m$^{-2}$ within this period. A range of morphologies was observed. 
All 5 events shown in Fig.~\ref{fig:waterfalls} have a high degree of linear polarization (red line). 

\section{Discussion} \label{sec:discussion}
In this work, we have presented a polarization analysis of 75 bursts from 28 repeating FRB sources, of which 63 bursts have significant RM and $L/I$ measurements. 
At the time of writing, combining these new results with those in the literature, there is a total of 26 repeating FRBs that have more than one RM measurement and we can see the RM vs time plot of all these sources in Fig.~\ref{fig:AllFRBs} and in Appendix Fig.~\ref{fig:AllFRBs2}.
From our results alone, we do not see bursts with extreme RM magnitudes, with the highest $|$RM$|$ value being 1348.7(3)\,rad\,m$^{-2}$ from FRB~20201114A. 

Similar to previous findings \citep[see, e.g.][]{Mckinven2023b}, we do not observe correlated variations in DM, $L/I$ or any other parameters that resemble the RM trends.
The $L/I$ measurement distribution is largely consistent with that previously reported by \citet{Mckinven2023b}, showing a wide range of values that suggest significant intrinsic variations per repeater source. 

We observe some variations in the PA as a function of time during the burst, but these are typically small fluctuations of less than 30$^{\circ}$. 
There are no events that show a significant S-swing in the PA profile, like the one reported by  \citet{Mckinven2024}. 

\begin{figure*}[ht]
    \centering
    \includegraphics[width=0.98\textwidth]{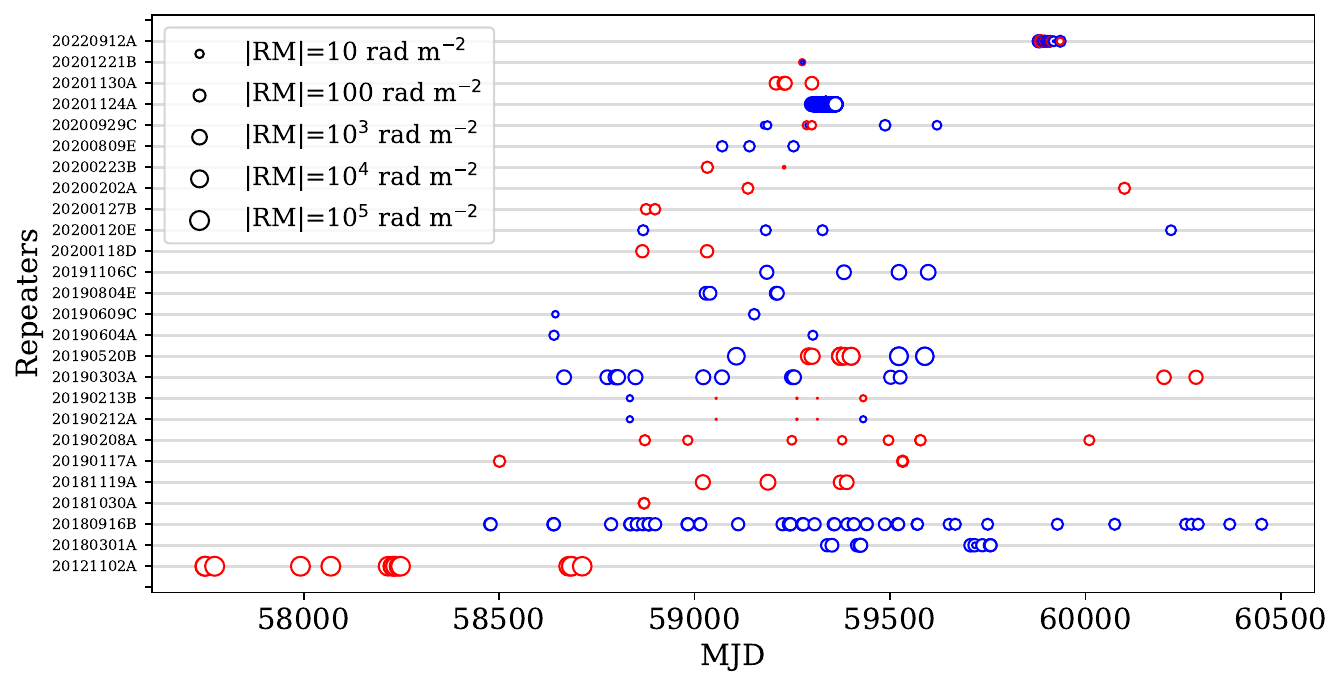}
    \caption{Temporal RM variations for 26 repeating FRBs with more than one RM. The size of the circles is proportional to the magnitude of the RM values, whereas red colour indicates positive RMs and blue negative RMs. }
        \label{fig:AllFRBs}
\end{figure*}

\subsection{Temporal RM variations} \label{sec:RMtemporal}
We further study the RM distribution of all 26 repeaters with more than one RM value. 
In Fig.~\ref{fig:RMstd}, we compare the standard deviation of the RM excess with the maximum time span of measurements (the time of the newest data point minus the oldest), as well as the averaged RM excess magnitudes. 
From the left panel, we can see that the majority of these repeaters have at least 1 year of data, albeit with varying degrees of cadence (refer to Fig.~\ref{fig:AllFRBs}). 
The amount of RM variations ($\sigma$(RM)) does not show a simple linear correlation with the time span. 
A number of FRB repeaters have small RM variations of a few tens of units irrespective of the time span (bottom half of the plot area). These are likely the repeaters associated with stochastic RM variations and the observed behaviour suggests a uniform environment with limited turbulence.
On the top half of the plot area, a number of exceptions show much higher levels of RM variations. 
This division is perhaps more apparent from the right panel of Fig.~\ref{fig:RMstd}, where these repeaters seem to separate into two categories. 
Those to the right of the dotted line, with a low ratio of $\sigma$(RM)/$|\overline{\mathrm{RM}}|$, likely come from stable magneto-ionic environments. 
On the other side (to the left of the dotted line), these sources have a high ratio of $\sigma$(RM)/$|\overline{\mathrm{RM}}|$ and likely reside in dynamic RM environments. 
The cases of FRBs~20190520B, 20121102A,  20181119A, 20190303A and 20201124A have already been discussed in the literature \citep{Mckinven2023b,AnnaThomas2023,Michilli2018,Luo2020}. 
This work adds FRB~20191106C to this subgroup of FRBs with high $|\overline{\mathrm{RM}}|$ which are thought to reside in dynamic RM environments, where the extreme circum-burst environment contributes to the high-level RM variations. 
It is also remarkable that the top boundary of the trend in the right panel is very close to a 1-1 relation between $\sigma$(RM) and $|\overline{\mathrm{RM}}|$.
This is additional evidence that for these repeaters in dynamic RM environments, the majority of the RM variations are likely due to contributions from the local environment (which we expect to be time variable), instead of from the host galaxy ISM (which we expect to be stable with time). 
We could also conclude that the sources located close to the 1-1 relation must have a low host galaxy ISM fractional contribution to the RM temporal variation, as the majority of the temporal variations are already accounted for by the evolving environment near the FRB source.

In general, practically all FRB repeaters demonstrate temporal RM variations that have larger amplitudes than those found for Galactic pulsars, which is typically no more than a few rad\,m$^{-2}$ \citep{Wahl2022}. 
There are certain pulsar systems that show comparably high magnitude of RM variations (see right panel of Fig.~\ref{fig:RMstd}). Examples include pulsars near the Galactic centre close to Sagittarius A* \citep[e.g. PSRs~J1745$-$2900, J1746$-$2849, J1746$-$2850, J1746$-$2856, and J1745$-$2912;][]{Desvignes2018,Abbate2023}, `windy'\footnote{A `windy' pulsar binary implies a massive star companion that is emanating strong stellar winds.} pulsar binaries \citep[e.g. PSRs~J2051$-$0827 and B1259$-$63;][]{Wang2023,Connors2002}, and a redback\footnote{A redback pulsar system typically consists of a low-mass companion like a red dwarf and the system usually has short orbital period of order of hours.} system \citep[PSR~B1744$-$24A;][]{Li2023}.
The FRB repeaters from evolving RM environments appear to agree quite well with the pulsars from (eclipsing) binary systems, but not so much with the solitary ones near the Galactic centre.
Given the similarities, it might be suggestive to say that 
binary stellar companion coronal winds are a potential explanation of the circum-burst environments of these FRB systems with a high ratio of $\sigma$(RM)/$|\overline{\mathrm{RM}}|$, and that these repeating FRBs probably did not come from the Galactic centre of their host galaxies, assuming similar host ISM properties to our Milky Way.
This is indeed the case for the three high $\sigma$(RM)/$|\overline{\mathrm{RM}}|$ sources which have VLBI localizations, namely FRB~20190520B at 5\,kpc from its host galaxy centre \citep{Niu2021}, FRB~20180916A with a 4.7\,kpc offset \citep{Marcote2020}, and FRB~20220912A with a 0.8\,kpc offset \citep{Hewitt2024}.
In addition, in the cases of the Galactic centre pulsars, large variations in DM were also reported \citep{Abbate2023} which is not the case for the FRB repeaters.

Beyond stochastic variations, some sources show deterministic trends that are not captured in Fig.~\ref{fig:RMstd} and that can cause an overestimation of stochastic variation. To investigate this, we have defined four simple empirical models for the RM evolution: 1) a constant RM (one parameter), 2) a linear trend (one parameter), 3) a model with constant RM episodes connected by a linear trend (four parameters, motivated by FRB~20180916B) and 4) a periodic model (four parameters, motivated by FRB~20190520B). We have used least-squares fitting to fit all four models to the 19 sources with $>2$ RM measurements and selected the best-fit model using the Akaike information criterion \citep[see Section~2.4 in ][]{Burnham2002}, modified for small sample sizes. 

In this framework, twelve FRBs are best fit by a constant RM. FRBs~20180301A and 20191106C are best fit by a linear trend with slopes of 0.24 and $-$2.0\,rad\,m$^{-2}$\,day$^{-1}$ over $\geq$418 and $\geq$413 days, respectively. FRBs~20121102A, 20180916B and 20190303A are best fit by model with constant episodes connected by a linear trend with slopes of $-$90, 0.19, 0.72\,rad\,m$^{-2}$ day$^{-1}$ over 378, 323 and $\geq$1275 days, respectively. FRB~20190520B is best fit by a periodic model with a period of 350 days (note that this does not imply it is in a orbit) and FRB~20201124A shows too much temporal variation to produce a meaningful fit using these four simple models. 

Indeed, the standard deviation of the residual RM after removing the best-fit models go down and all sources show stochastic changes $\sigma$(RM) $\lesssim$ 80\,rad\,m$^{-2}$, except for FRBs~20121102A, 20181119A and 20190520B that still show $\sigma$(RM) $\mathcal{O}(10^3)$ rad m$^{-2}$. Interestingly, the five best-fit linear trends normalized by the median RM of the respective sources all five fall within 0.1--0.3\% day$^{-1}$. The time evolution could be further quantified using correlation or structure functions, but we defer this investigation to future work.

\subsection{RM sign change}
\begin{figure*}[ht]
    \centering
    \includegraphics[width=\textwidth]{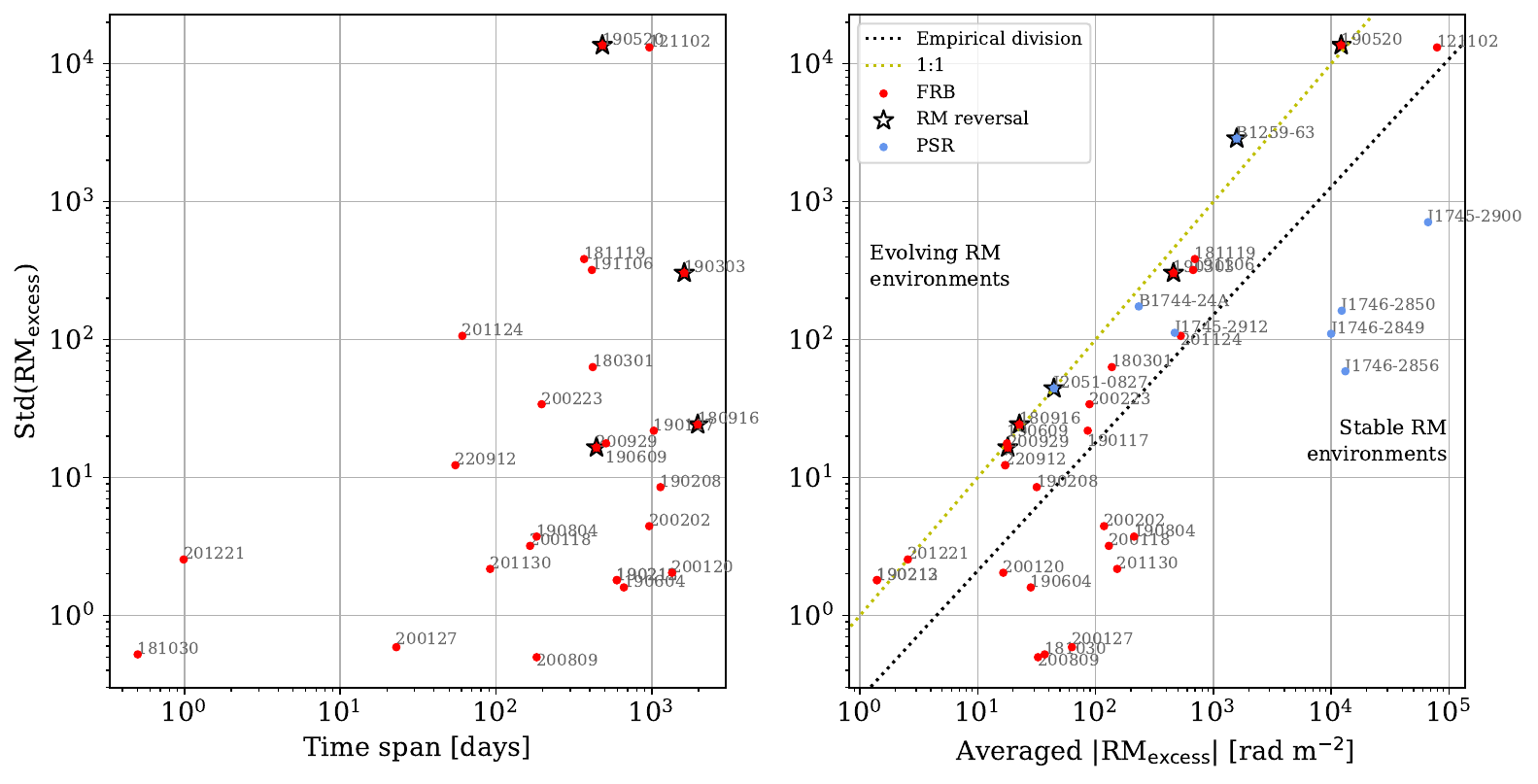}
    \caption{(Left) RM excess standard deviation vs  variability timescale (maximum time span of available data) for all repeating FRBs (red dots) with more than one RM measurement. (Right) RM standard deviation vs the averaged RM excess magnitude. A few notable pulsar systems with peculiar RM properties are included as blue dots.
    Sources that exhibit an RM sign change are represented by stars. The yellow dotted line indicate 1:1 ratio of $\sigma$(RM)/$|\overline{\mathrm{RM}}|$, where as the black dotted line is an empirical division of the two types of RM environments.}
        \label{fig:RMstd}
\end{figure*}

In this work, we report the change of signs in RMs for two FRBs, namely FRBs~20200929C and 20190303A. 
Together with one other previously published case of FRB~20190520B \citep{Dai2022,AnnaThomas2023}, there are now three examples (see red star symbols in Fig.~\ref{fig:RMstd}) in this sub-group.
A possible 4th source is FRB~20180916B (see Fig.~\ref{fig:temporalR3}), whose low Galactic latitude of 3.7$^{\circ}$ implies high uncertainty in its Milky Way RM contributions ($RM_{\mathrm{MW}}=-94\pm45$\,rad\,m$^{-2}$) that may or may not indicate the RM variations we see in FRB~20180916B fluctuates around zero. 
We note that these sources have the highest ratios of $\sigma$(RM)/$|\overline{\mathrm{RM}}|$ which are all very close to 1:1 (yellow line), occupying the further left region in the right panel of Fig.~\ref{fig:RMstd}.

This sub-sample of FRB repeaters with RM sign changes is a stark contrast to the Galactic pulsar population where out of the over 3000 known pulsars, an RM sign change has only been reported in two systems with windy binaries, namely the black widow binary PSR~J2051$-$0827 \citep{Wang2023} and the binary system PSR~B1259$-$63 which has a Be-star companion \citep{Connors2002,Johnston2005}. 
In both pulsar systems, the RM sign change is attributed to the change in magnetic field strength along the line of sight due to binary orbital motions, and indeed the RM temporal variations of these two pulsar systems show orbital modulations that match the orbital periods. 
The fact that RM sign changes have not been seen more commonly in other windy pulsar binaries might be mostly due to the detection limit of the telescope. \citet{Li2023} have shown that RM changes of thousands of  rad\,m$^{-2}$ can take place rapidly in these binary systems, and so depolarization could occur in these pulsar observations where an integration time of tens of seconds was typically used. 
In contrast, this is less of a problem for the much brighter FRB signals where the RM can be measured from each individual burst. 
Nonetheless, the four repeating FRBs with RM sign changes do not (yet) show any orbital modulations in their RM variations that resemble those observed in the windy pulsars. Further monitoring of the RM variations of these four repeaters would be insightful in understanding any potential relationship between them and the pulsar systems. 

\subsection{Comparing repeaters and non-repeaters} \label{sec:Ayush}
Recently, \cite{Pandhi2024} compared the polarization properties of $118$ non-repeating FRBs ($89$ polarized and $29$ unpolarized) to $13$ repeating sources. They find that the linear polarization fractions and rotation measure distributions of repeaters and non-repeaters are consistent with being drawn from the same population, however, they present only marginal evidence for repeaters originating from more magnetized environments. In this section, we update their analysis comparing repeaters and non-repeaters by adding the 75 bursts from 28 repeating FRBs presented in this work. For brevity, we do not detail the full methodology and statistical techniques involved, instead, we refer the reader to \cite{Pandhi2024} for this information.

\begin{figure}[ht]
    \includegraphics[width=0.47\textwidth]{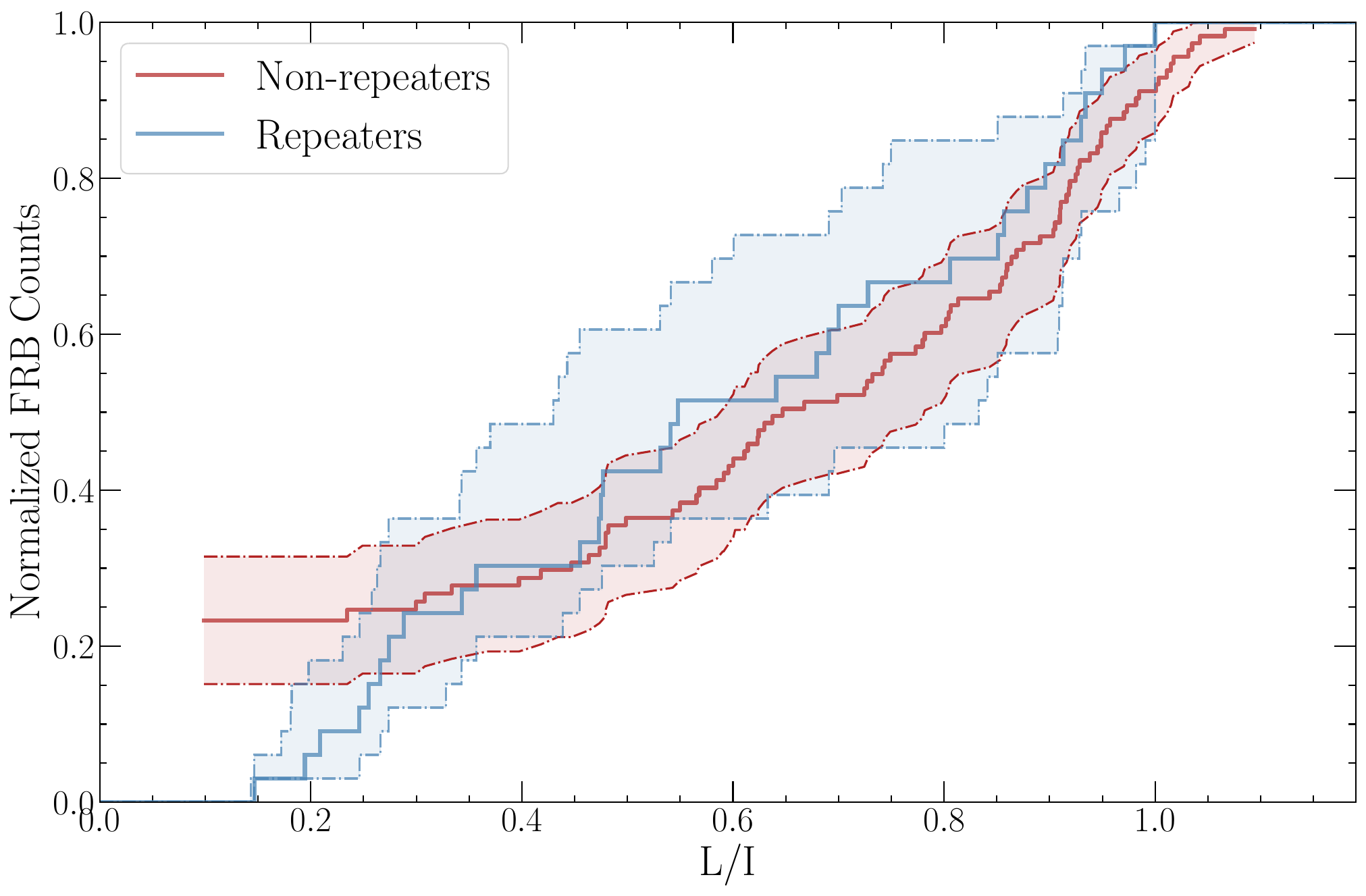}
    \includegraphics[width=0.47\textwidth]{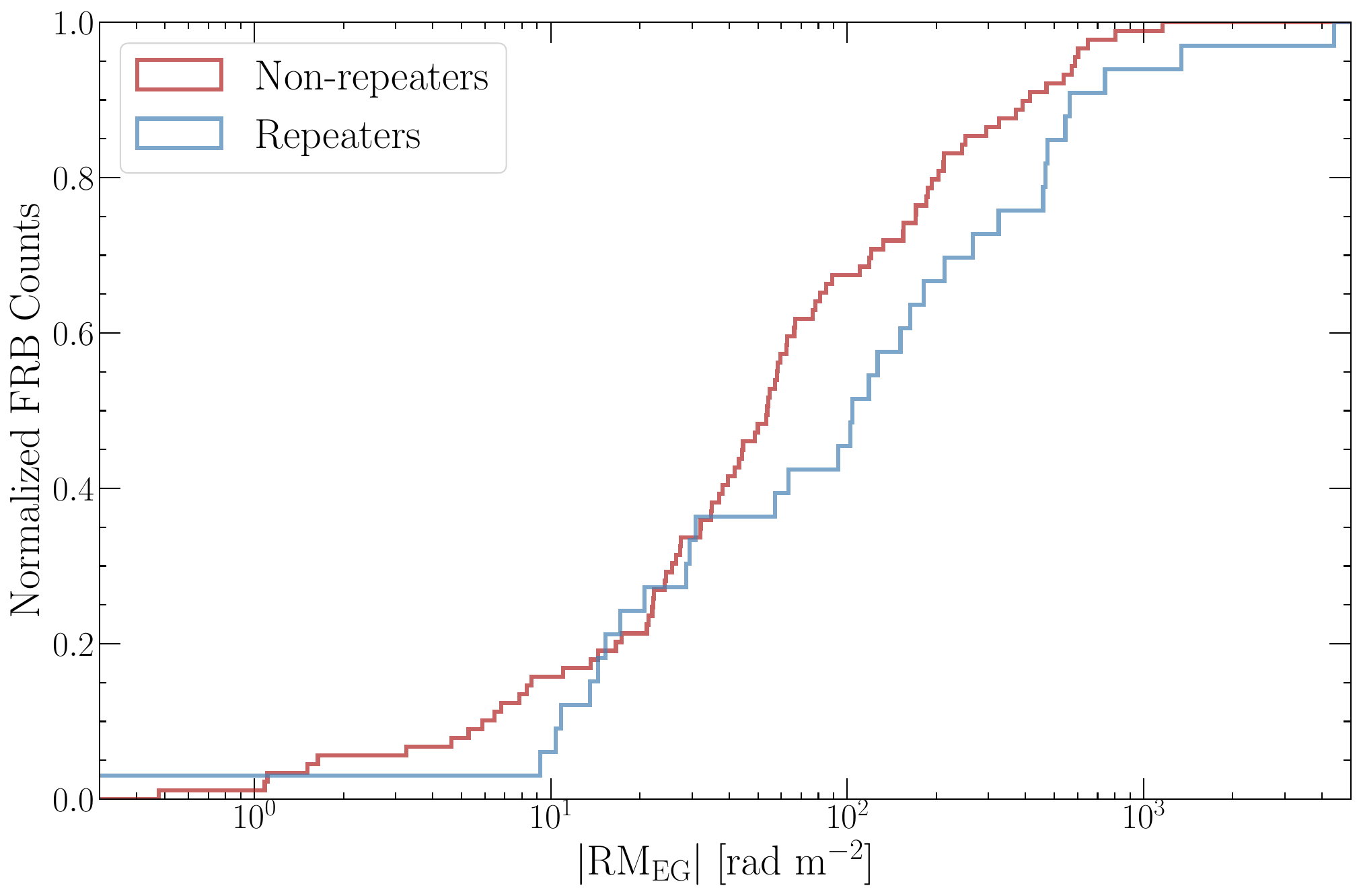}
    \includegraphics[width=0.47\textwidth]{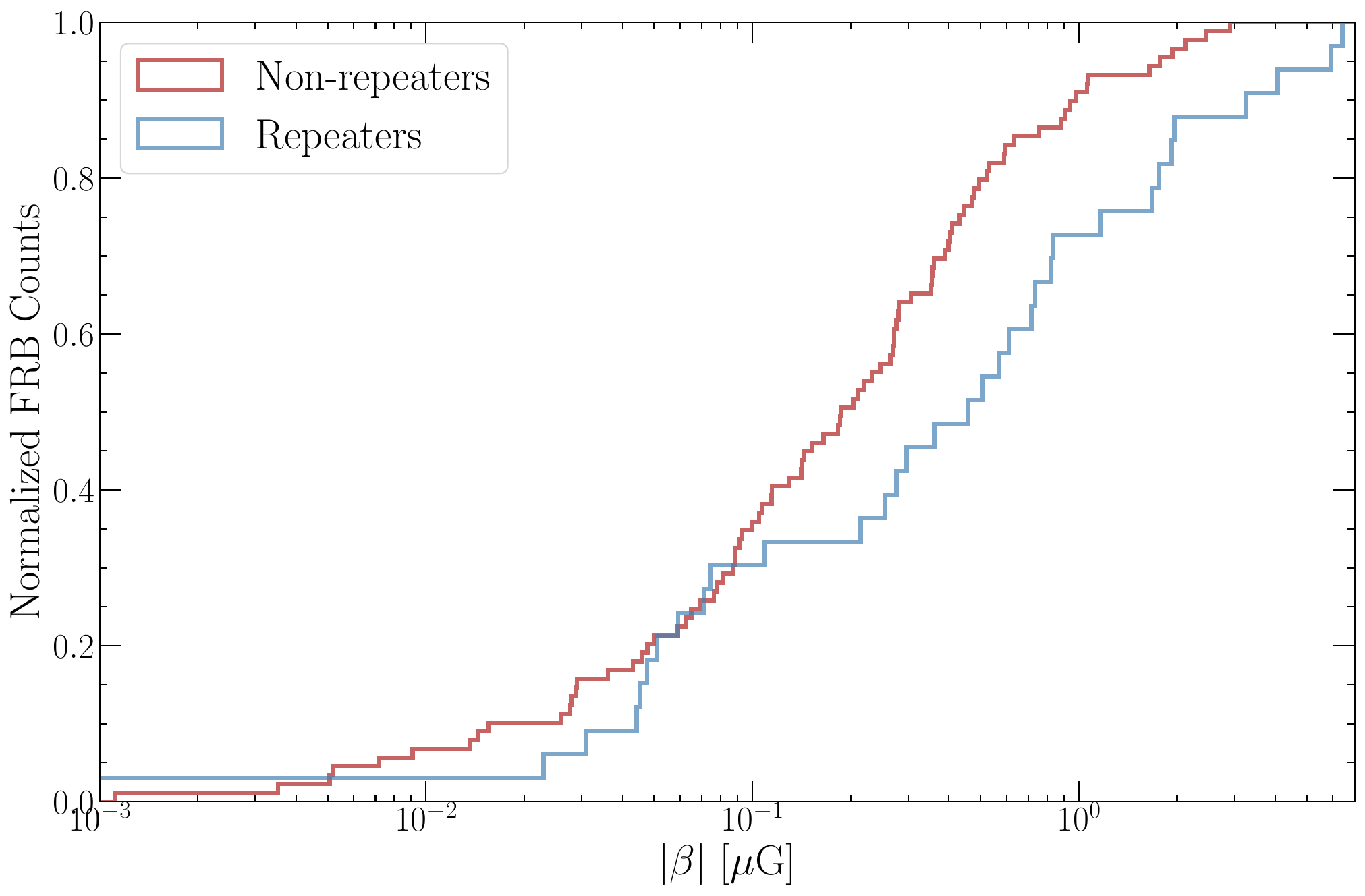}
    \caption{Comparison of the polarization properties of repeaters \citep[in blue;][and this work]{Bhardwaj2021, Mckinven2023b} and non-repeaters \citep[in red;][]{Pandhi2024} observed by CHIME/FRB. The panels, from top to bottom, show the CDF of $L/I$ (including upper limits for the unpolarized non-repeaters), the magnitude of the foreground subtracted rotation measure $|\mathrm{RM}_\mathrm{EG}|$, and the observer frame lower limit of the line of sight magnetic field strength $|\beta|$.}
    \label{fig:rep_nonrep_comparison}
\end{figure}

In the top panel of Fig.~\ref{fig:rep_nonrep_comparison}, we show the cumulative distribution function (CDF) of the linear polarization fraction for repeaters in blue and non-repeaters in red. The CDF for non-repeating sources is computed using the Kaplan-Meier method \citep{kaplan-meier1958}, which also accounts for upper limits in $L/I$ for unpolarized bursts, and the red shaded region is the $95$\% confidence interval. The blue-shaded region shows the full range of $L/I$ measurements across bursts for each repeating source. The middle panel of Fig.~\ref{fig:rep_nonrep_comparison} shows the CDFs for the Milky Way foreground subtracted RM ($|\mathrm{RM}_\mathrm{EG}|$) for repeaters and non-repeaters, respectively. Finally, the bottom panel of Fig.~\ref{fig:rep_nonrep_comparison} depicts the CDFs for the observer frame lower limit on the average line of sight magnetic field strength \citep[$|\beta|$; for details, see Section~2.6 by][]{Pandhi2024}.

Applying the Kolmogorov-Smirnov \citep[KS;][]{smirnov1948, kolmogorov1956} and Anderson-Darling \citep[AD;][]{ADtest} tests to each pair of CDFs presented in Fig.~\ref{fig:rep_nonrep_comparison}, we find no evidence for a dichotomy between repeating and non-repeating FRBs in terms of their $L/I$ using Kendall's $\tau$ test \citep{kendall1938} or their $|\mathrm{RM}_\mathrm{EG}|$ using the KS and AD tests. We find marginal evidence for a dichotomy in the repeater and non-repeater $|\beta|$ distributions (with $p$ values of $0.033$ and $0.015$ with the KS and AD tests, respectively), consistent with the results from \citet{Pandhi2024} who used a smaller sample of repeating FRBs.

\begin{figure}[ht]
    \includegraphics[width=0.47\textwidth]{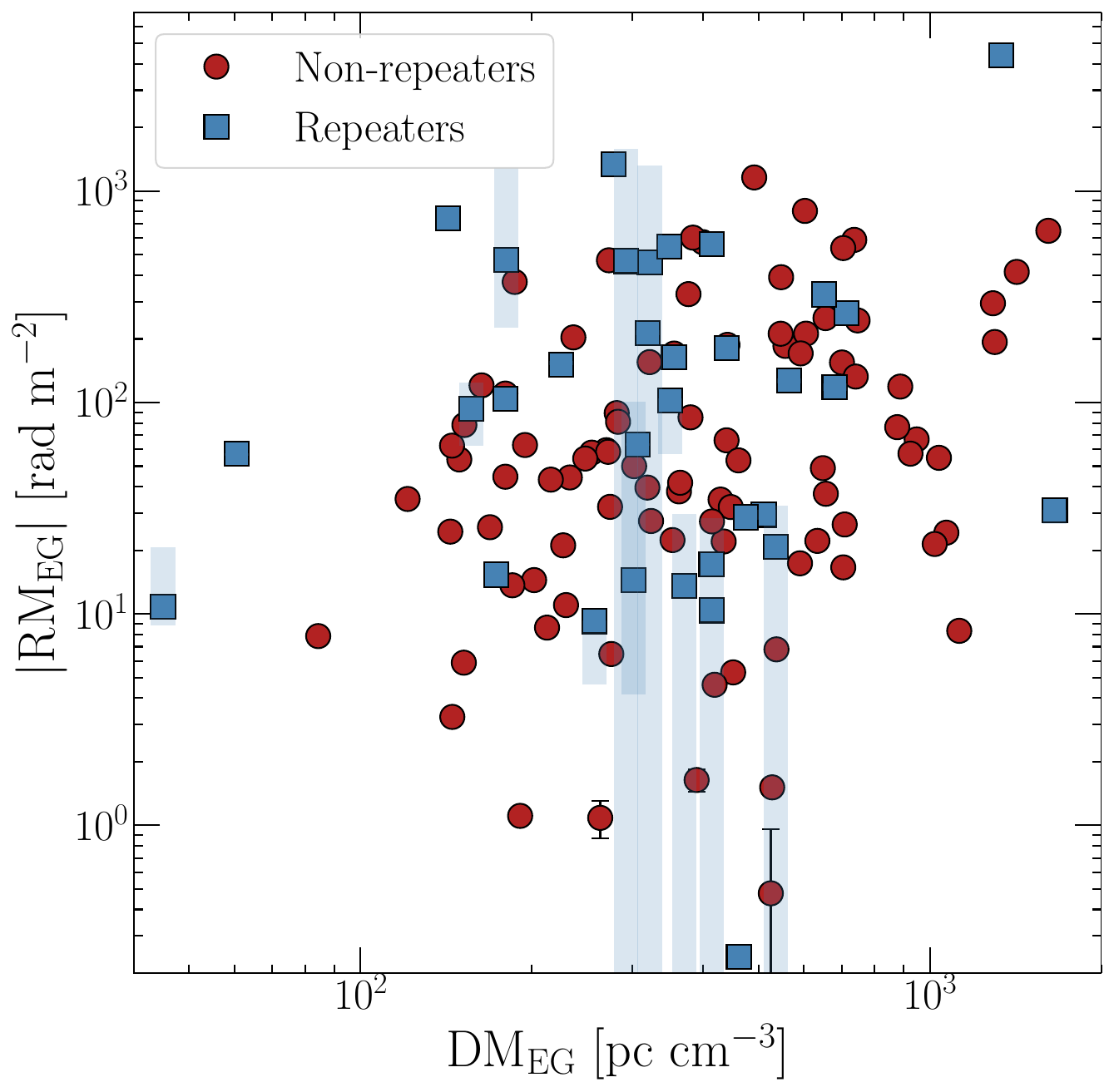}
    \caption{The foreground subtracted $|\mathrm{RM}_\mathrm{EG}|$ plotted as a function of the foreground subtracted DM, $\mathrm{DM}_\mathrm{EG}$ for all CHIME/FRB repeating (blue squares) and non-repeating (red circles) FRBs. Measurement uncertainties are plotted as black error bars and blue shaded regions represent the intrinsic variability across repeat bursts for each repeating source.}
    \label{fig:dm_rm_rn3}
\end{figure}

To test whether the DMs and RMs in our FRBs originate from the same type of media along the line of sight, we analyze the correlation between the Milky Way foreground subtracted RM ($|\mathrm{RM}_\mathrm{EG}|$) and the foreground subtracted DM ($\mathrm{DM}_\mathrm{EG}$). 
A plot of the $|\mathrm{RM}_\mathrm{EG}|$ versus $\mathrm{DM}_\mathrm{EG}$ for all CHIME/FRB observed repeaters and non-repeaters with polarimetric measurements to date, including the 28 repeating sources in this paper, is presented in Fig.~\ref{fig:dm_rm_rn3}. 
Mirroring the analysis by \citet{Pandhi2024}, we use Spearman's rank correlation coefficient test \citep{spearman1904} to gauge the extent of a possible correlations between $\mathrm{log}_{10}(\mathrm{DM}_\mathrm{EG})$ and $\mathrm{log}_{10}(|\mathrm{RM}_\mathrm{EG}|)$ in our expanded repeater data set. We find a Spearman rank $p$-value of $0.804$ (i.e., no statistical significance) when evaluating the $\mathrm{log}_{10}(\mathrm{DM}_\mathrm{EG}) - \mathrm{log}_{10}(|\mathrm{RM}_\mathrm{EG}|)$ correlation across all polarized repeating FRBs detected with CHIME/FRB. 
Interestingly, \cite{Pandhi2024} and \cite{Sherman2024} both find a marginal correlation between $\mathrm{DM}_\mathrm{EG}$ and $|\mathrm{RM}_\mathrm{EG}|$ in their respective non-repeating FRB samples. One possible explanation for the lack of a  $\mathrm{log}_{10}(\mathrm{DM}_\mathrm{EG}) - \mathrm{log}_{10}(|\mathrm{RM}_\mathrm{EG}|)$ correlation in repeating FRBs, and for the marginal dichotomy in $|\beta|$, is that the media contributing the bulk of the RM may be distinct from the media primarily contributing to the DM for repeating FRBs. 
As suggested by \cite{Pandhi2024}, repeating FRBs might be embedded in more strongly magnetized local environments (which contribute significantly more to the total RM than to the total DM) than non-repeating FRBs.

We note that in this analysis, we have restricted the comparison to CHIME/FRB-detected bursts only in order to limit bias due to different telescope instrumental properties. This means that our sample does not include some of the highest $\sigma$(RM)/$|\overline{\mathrm{RM}}|$ ratio repeaters (e.g. FRBs~20190520B and 20121102A, and only very few bursts from FRB~20201124A).
Intra-channel depolarization could be a reason why CHIME/FRB is less likely to detect high $\sigma$(RM)/$|\overline{\mathrm{RM}}|$ sources, compared to telescopes that operate at higher observing frequencies (e.g. FAST and ASKAP).
Another possible reason why CHIME/FRB might not have detected these sources is scattering, although based on the scattering measurements from \citet{Curtin2024}, we do not observe any correlation between $\sigma$(RM)/$|\overline{\mathrm{RM}}|$ and the scattering timescale among the events presented in this paper. 
As suggested in Section~\ref{sec:RMtemporal}, if indeed there are two categories of magneto-ionic environments and that the high $\sigma$(RM)/$|\overline{\mathrm{RM}}|$ repeaters are primarily the ones from more dynamic environments, the actual $|\beta|$ and DM-RM dichotomy between non-repeaters and repeaters might have been more significant. 

\section{Conclusion} \label{sec:conclusion}
We have presented a total of 75 bursts from 28 repeating FRB sources recorded between January 2019 and May 2024, including 63 new polarization measurements.
We present RM values ranging from $-$1044 to $+$1348\,rad\,m$^{-2}$, along with a wide range of $L/I$ from less than 0.2 to almost 1, while PA variations are typically small and no bursts in this paper exhibit clear S-swing in the PA profile.
We observe temporal RM variations that are higher than those associated with Galactic pulsars, consistent with previous findings.
Repeating FRBs appear to separate into two categories of dynamic vs stable RM environments that are differentiated by the ratios of $\sigma$(RM)/$|\overline{\mathrm{RM}}|$.
FRB~20191106C shows high-level RM changes (hundreds of rad\,m$^{-2}$ on months timescales) that could potentially point to dynamic magneto-ionic environments, similar to the cases of FRBs~20190520B, 20121102A,  20181119A, 20190303A and  20201124A.
From the last 5 years of observations of FRB~20180918B, we witness the RM variations change from stochastic to secular and back to stochastic, with no turning point that could suggest an orbital effect from a binary system, although we do not exclude the possibility that FRB~20180916B is in a binary.
We highlight two more repeaters with RM sign changes that imply a magnetic field sign change along those lines of sight.
Finally, we present an updated comparison of polarization properties between repeaters and non-repeaters, and our results 
show only marginal dichotonmy in the magnetic field strength distribution between the two groups, consistent with the findings of \citet{Pandhi2024}.

\begin{acknowledgements}
We acknowledge that CHIME is located on the traditional, ancestral, and unceded territory of the Syilx/Okanagan people. We are grateful to the staff of the Dominion Radio Astrophysical Observatory, which is operated by the National Research Council of Canada. CHIME is funded by a grant from the Canada Foundation for Innovation (CFI) 2012 Leading Edge Fund (Project 31170) and by contributions from the provinces of British Columbia, Qu\'{e}bec and Ontario. The CHIME/FRB Project is funded by a grant from the CFI 2015 Innovation Fund (Project 33213) and by contributions from the provinces of British Columbia and Qu\'{e}bec, and by the Dunlap Institute for Astronomy and Astrophysics at the University of Toronto (funded through an endowment established by the David Dunlap family and the University of Toronto). Additional support was provided by the Canadian Institute for Advanced Research (CIFAR), McGill University and the Trottier Space Institute thanks to the Trottier Family.
B.M.G. is supported by an NSERC Discovery Grant (RGPIN-2015-05948), and by the Canada Research Chairs (CRC) program. 
R.M. recognizes support from the Queen Elizabeth II Graduate Scholarship and the Lachlan Gilchrist Fellowship. 
V.M.K. holds the Lorne Trottier Chair in Astrophysics $\&$ Cosmology, a Distinguished James McGill Professorship, and receives support from an NSERC Discovery grant (RGPIN 228738-13), from an R. Howard Webster Foundation Fellowship from CIFAR, and from the FRQNT CRAQ.
K.W.M. is supported by an NSF Grant (2008031). 
D.Z.L is a Lyman Spitzer, Jr. Fellow.
A.B.P. is a Banting Fellow, a McGill Space Institute~(MSI) Fellow, and a Fonds de Recherche du Quebec -- Nature et Technologies~(FRQNT) postdoctoral fellow.
M.B is a McWilliams fellow and an International Astronomical Union Gruber fellow. M.B. also receives support from the McWilliams seed grant.
D.L.J. is a KIPAC Kavli Fellow.
K.S. is supported by the NSF Graduate Research Fellowship Program. 
The polarization analysis presented here makes use of the {\tt RMtools} package\footnote{https://github.com/CIRADA-Tools/RM-Tools}\citep{Purcell2020} written by Cormac Purcell, and maintained by Cameron Van Eck.
We thank Grégory Desvignes for sharing the RM data of PSR~J1745$-$2900. 

\end{acknowledgements}

\bibliographystyle{aasjournal}
\bibliography{references}

\appendix

\begin{figure*}[ht]
    \centering
    \includegraphics[width=0.98\textwidth]{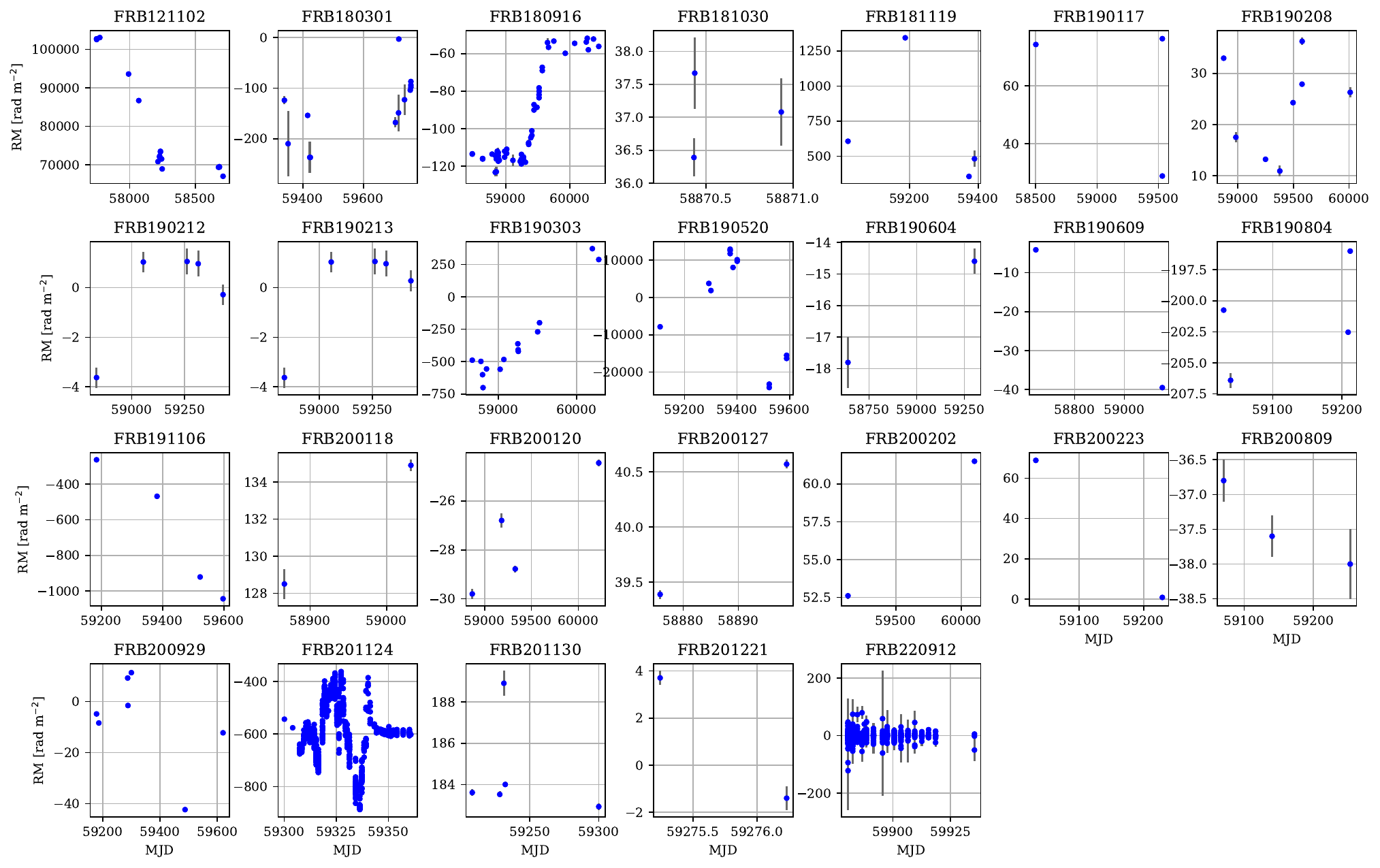}
    \caption{Temporal RM variations for 26 repeating FRBs with more than one RMs, including the new results from this work as well as data from the literature.}
        \label{fig:AllFRBs2}
\end{figure*}





\end{document}